\documentclass[twoside,preprint,11pt]{article}
\usepackage{xcolor}
\usepackage{blindtext}

\usepackage{jmlr2e}
\usepackage{amsmath}
\usepackage{amsbsy}  
\usepackage{amsfonts}
\usepackage{amssymb, bbm}
\usepackage{enumitem}
\usepackage{multirow}
\usepackage{algorithm}
\usepackage{algpseudocode} 
\usepackage{booktabs}
\usepackage{times}
\usepackage{subcaption}

\DeclareMathOperator*{\argmin}{\arg\!\min}

\newcommand{\ex}{\mathbb{E}}

\newcommand{\tr}{\operatorname{tr}}

\newcommand{\reals}{\mathbb{R}}

\newcommand{\var}{\mathbb{V}}

\newcommand{\E}{\mathbf{E}}

\newcommand{\R}{\mathbf{R}}

\newcommand{\X}{\mathbf{X}}

\newcommand{\cD}{{\cal D}}

\newcommand{\cF}{{\cal F}}

\newcommand{\cM}{{\cal M}}

\newcommand{\cO}{{\cal O}}

\newcommand{\cR}{{\cal R}}

\newcommand{\cX}{{\cal X}}

\newcommand{\hgamma}{\hat{\gamma}}
\newcommand{\hdelta}{\hat{\delta}}

\newcommand{\hmu}{\hat{\mu}}

\newcommand{\htau}{\hat{\tau}}

\usepackage{lastpage}
\jmlrheading{26}{2026}{1-\pageref{LastPage}}{6/25; Revised 12/25}{1/26}{21-0000}{Author One and Author Two}

\ShortHeadings{Improving Precision of RCT-Based CATE using Data Borrowing}{Asiaee, Di Gravio, Beck, Mei, Pal, and Huling}
\firstpageno{1}

\begin{document}

\title{Improving Precision of RCT-Based CATE Estimation \\using Data Borrowing with Double Calibration}

\author{
        \name \hspace{-0.4em} Amir Asiaee$^1$         \email{amir.asiaeetaheri@vumc.org} 
        \AND
        \name Chiara Di Gravio$^2$
        \email{c.di-gravio@imperial.ac.uk} 
        \AND
        \name Cole Beck$^1$
        \email{cole.beck@vumc.org} 
        \AND
        \name Yuting Mei$^1$ 
        \email{yutingmei.vu@gmail.com} 
        \AND
        \name Samhita Pal$^1$ 
        \email{samhita.pal@vumc.org} 
        \AND
        \name Jared D.~Huling $^3$
        \email{huling@umn.edu} 
        \vspace{4pt}
        \\
        \addr $^1$Department of Biostatistics, 
        Vanderbilt University Medical Center, 
        Nashville, Tennessee, U.S.A. \\
        \addr $^2$Department of Epidemiology and Biostatistics, 
        Imperial College London, 
        London SW7 2AZ, U.K. \\
        \addr $^3$Division of Biostatistics and Health Data Science, 
        University of Minnesota, 
        Minneapolis, Minnesota, U.S.A. 
}
\editor{}

\maketitle

\begin{abstract}
Understanding how treatment effects vary across patient characteristics is essential for personalized medicine, yet randomized controlled trials (RCTs) are often underpowered to detect heterogeneous treatment effects (HTEs). We propose a framework that improves the efficiency of conditional average treatment effect (CATE) estimation in RCTs by leveraging large observational studies (OS) while preserving the unbiasedness of RCT estimates. By framing CATE estimation as a supervised learning problem, we show that estimation variance is minimized using the counterfactual mean outcome (CMO) as an augmentation function. We derive finite-sample error bounds and establish conditions under which OS data improves CMO estimation, and thus CATE efficiency, even in the presence of confounding in the OS or outcome distribution shifts between populations. We introduce R-OSCAR (Robust Observational Studies for CMO-Augmented RCT), a two-stage estimator that calibrates OS outcome predictions to the RCT population and corrects residual biases through regularized regression. For any OS-derived nuisance, R-OSCAR is consistent for the RCT-population CATE, and it is efficient relative to RCT-only estimators when the RCT--OS outcome mean discrepancy is estimable from the RCT with lower complexity than the full RCT outcome model. A cross-fitted RCT diagnostic determines, from observable data alone, whether borrowing from a particular OS is supported. Simulations show that R-OSCAR can reduce the RCT sample size needed for HTE detection by up to 75\%, maintaining robustness to model misspecification. The empirical validation covers two real-world case studies: a semi-synthetic analysis of the Tennessee STAR study with constructed observational confounding, and the Greenlight Plus pediatric-obesity trial linked with external electronic-health-record controls, where borrowing improves control-arm estimation for small trials and the diagnostic certifies it only where the records cover the trial population. Our framework offers a principled approach to integrating observational and experimental data using tools from statistical learning and transfer learning.
\end{abstract}


\section{Introduction}
Understanding how treatment effects vary with patient characteristics is essential for tailoring interventions \citep{kosorok2019precision}. 
While randomized controlled trials (\textbf{RCTs}) are the gold standard for causal inference and estimating heterogeneous treatment effects (\textbf{HTEs}), they are often constrained by cost, time, and limited sample sizes, making it difficult to estimate HTEs across diverse covariates. In contrast, large-scale observational studies (\textbf{OS}) offer richer data and broader population coverage, but raise confounding concerns, making causal inference reliant on untestable assumptions such as unconfoundedness. As a result, recent work has focused on integrating RCT and OS data to mitigate bias in OS, improve the efficiency of RCT estimates, or enhance RCT finding's generalizability.

Most work on RCT–OS integration has focused on generalizing findings from RCTs to broader OS populations ($\cR \rightarrow \cO$); see \citet{colnet2024causal} and \citet{degtiar2023review} for overviews, and \citet{asiaee2026sharp, asiaee2026omitted} for recent developments. Our focus differs: we aim to improve estimation efficiency within the RCT population using OS data ($\cO \rightarrow \cR$).

Data integration to improve RCT precision has a long history, starting with \citet{Pocock1976}, who proposed augmenting control arms with data from previous RCTs or OS, treating these as \textit{historical controls}. Approaches fall into two categories: static borrowing \citep{li2019, Liu2022}, which treats external data as part of the RCT, and dynamic borrowing \citep{Ibrahim2000, Neuenschwander2010, Kaizer2018, Kotalik2021}, which weights external data based on its similarity or agreement with the current RCT data.

While prior work has mostly focused on borrowing control arms to reduce variance of the treatment effect, less attention has been given to leveraging both treatment arms from OS for improving CATE estimation. Moreover, existing methods often assume CATEs are transportable between populations, a strong assumption that can fail when unmeasured systematic differences exist between populations \citep{Hernan2020CausalInferenceWhat}. New methods that relax this transportability assumption are needed that reliably leverage both treatment arms of OS data when integrating with RCTs for more efficient CATE estimation.

\textbf{Our contribution.}
We propose a framework that efficiently leverages OS data to improve CATE estimation in RCTs, without assuming CATE transportability. Our key contributions include:

\begin{enumerate}[leftmargin=12pt, itemsep=0pt]
    \item developing an $\cO \rightarrow \cR$ framework that borrows both treatment arms from OS, not just controls;
    \item relaxing the assumption of CATE or outcome mean exchangeability across populations;
    \item using learning theory to derive non-asymptotic error bounds, showing how CATE estimation depends on an augmentation functional we call the Counterfactual Mean Outcome (CMO);
    \item identifying conditions under which borrowing improves CMO and thus CATE estimation;
    \item providing a practical implementation via regularized regressions in settings with linear outcome models and sparse population differences;
    \item and validating our method in extensive simulations and real-world data, showing improved efficiency and power for HTE detection.
\end{enumerate}

Our approach has potentials to enhance personalized treatment strategies by making more efficient use of available data while preserving the unbiasedness guarantees of RCTs.
The remainder of this introduction sets up the necessary notation and presents an overview of our results. 

\subsection{Mathematical notation}
Let $Y^{\text{obs}}$ be the observed outcome, $X \in \mathcal{X}^s$ the observed covariates, and $A \in \mathcal{A} = \{a_1, \ldots, a_T\}$ the treatment in population $s \in \{r, o\}$, where $r$ denotes an RCT and $o$ an OS. Let $Y(a)$ denote the potential outcome under treatment $a$, and assume consistency: $Y = Y(a)$ when $A = a$. Then the joint density decomposes as:
\begin{align} \label{decomp}
    \mathbb{P}^s(Y^{\text{obs}}, X=x, A=a)
    &=\mathbb{P}^s(Y(a), X=x, A=a) \\ \nonumber
    &=\mathbb{P}^s(Y(a) \!\mid\! X=x, A=a)\, \mathbb{P}^s(A=a \!\mid\! X=x)\, \mathbb{P}^s(X=x).
\end{align}

Throughout, superscripts $s\in\{r,o\}$ mark the population for distributions, expectations, outcome means, and propensities; we therefore omit redundant ``$\mid S=s$'' conditioning on the population index.
When $\mathbb{P}^r(Y(a), X, A) = \mathbb{P}^o(Y(a), X, A)$, both datasets share the same distribution. Otherwise, any of the three terms in \eqref{decomp} may differ. In machine learning, $\mathbb{P}^r(X) \neq \mathbb{P}^o(X)$ is called covariate shift, while $\mathbb{P}^r(Y(a) \!\mid\! X) \neq \mathbb{P}^o(Y(a) \!\mid\! X)$ is known as outcome (concept or label) shift \citep{kouw2018introduction}.

A summary of key notation appears in Table~\ref{tab-notation}. We assume binary treatment $A \in \{-1, +1\}$. Our goal is to use OS data $(s = o)$ to improve the efficiency of CATE estimation in the RCT $(s = r)$, where $\tau^r(x) = \ex^r[Y(+1) - Y(-1) \!\mid\! X = x]$. We denote the OS dataset as $\{(X_i^o, A_i^o, Y_i^o)\}_{i=1}^{n_o}$ and the RCT dataset as $\{(X_i^r, A_i^r, Y_i^r)\}_{i=1}^{n_r}$, with $n_r \ll n_o$ typically. A key quantity in our framework is the \textit{counterfactual mean outcome} (CMO), $\mu^s(x)$, defined as the average of $\mu_{+1}^s(x)$ and $\mu_{-1}^s(x)$ under swapped treatment assignment probabilities. Intuitively, the CMO represents the expected outcome at covariates $x$ under a hypothetical reversal of the observed treatment assignment.

\begin{table}[h!]
\centering
\caption{\small Notation and definitions used throughout the manuscript.}
\label{tab-notation}
\renewcommand{\arraystretch}{1.3}  
\small
\begin{tabular*}{.92\columnwidth}{@{}l@{\extracolsep{\fill}}|p{.87\columnwidth}@{}}
\hline
\textbf{Symbol} & \textbf{Description} \\
\hline
$A$ & Treatment indicator; $A \in \{-1, 1\}$ where $1$ is for treatment \\
\hline
$Y$ & Observed outcome of interest \\
\hline
$S$ & Study indicator; $S \in \{r, o\}$ with $r$: RCT, $o$: observational \\
\hline
$p$ & Number of measured covariates in both studies \\
\hline
$X^s$ & Covariates observed in study $S=s$; $X^s \in \mathcal{X}^s \subseteq \mathbb{R}^p$ \\
\hline
$n^s$ & Sample size of study $S=s$ \\
\hline
$\tau^s$ & Average treatment effect (ATE) in population $s$; $\tau^s \equiv \ex^s[Y(+1)-Y(-1)]$ \\
\hline
$\mu_a^s(x)$ & Conditional mean potential outcome in study $s$; $\mu_a^s(x) \equiv \ex^s[Y(a) \mid X=x]$ \\
\hline
$\mu^s(x, a)$ & Regression function (outcome mean) in study $s$;  $\mu^s(x, a) \equiv \ex^s[Y \mid X=x, A=a]$ \\
\hline
$\tau^s(x)$ & Conditional average treatment effect (CATE); $\tau^s(x) \equiv \ex^s[Y(+1)-Y(-1) \mid X=x]$ \\
\hline
$\pi_a^s(x)$ & Treatment assignment probability in study $s$; $\pi_a^s(x) \equiv \mathbb{P}^s(A=a \mid X=x)$ \\
\hline
$\mu^s(x)$ & Counterfactual mean outcome (CMO) in population $s$; $\mu^s(x) \equiv \sum_a \pi_{-a}^s(x) \mu_a^s(x)$ \\
\hline
$\delta_a(x)$ & Mean discrepancy between populations $r$ and $o$; $\delta_a(x) \equiv \mu_a^r(x) - \mu^o(x, a)$ \\
\hline
$\Delta_2(f,g)$ & $L^2$ distance (RMSE) between $f$ and ground truth $g$; $\Delta^2_2(f, g) \equiv \ex^s_X[(f(X) - g(X))^2]$ \\
\hline
$\Delta_\infty(f,g)$ & $L^\infty$ distance between function $f$ and ground truth $g$; $\Delta_\infty(f, g) \equiv \sup_X |f(X) - g(X)|$ \\
\hline
$\Delta_2^2(\cF, g)$ & Approximation error; $\Delta_2^2(\cF, g) \equiv \inf_{f \in \cF} \ex^s_X [(f(X) - g(X))^2]$ \\
\hline
\end{tabular*}
\end{table}

\subsection{Technical overview of our framework and results}

Our goal is to \emph{improve the efficiency} of CATE estimation in an RCT population by leveraging OS data.
We frame CATE estimation as a supervised learning task: given observed triplets \((X, A, Y)\) and an \emph{augmentation function} \(m(X)\), we define the pseudo-outcome 
\(
\tau_m(X, A, Y) \equiv \frac{A\,(Y - m(X))}{\pi_A(X)}.
\)
We estimate $\hat{\tau}(x)$ as the conditional mean of pseudo-outcomes by minimizing squared loss over a function class $\mathcal{F}$. Specifically, the CATE for the RCT is estimated as:
\begin{equation} \label{cate-opt}
\hat{\tau}^r(\cdot) ~=~ 
\arg\min_{f \in \mathcal{F}}
\frac{1}{n^r}\sum_{i=1}^{n^r} \left(\tau_m(X^r_i, A^r_i, Y^r_i) - f(X^r_i)\right)^2.
\end{equation}

Crucially, the estimator’s risk depends on the choice of \(m\). Our key insight is that the risk-minimizing choice is the \emph{counterfactual mean outcome} (CMO):
{\small \(
\mu^r(x)
~\equiv~
\pi_{-1}^r(x)\,\mu^r_{+1}(x)
~+~
\pi_{+1}^r(x)\,\mu^r_{-1}(x).
\)}

Under standard assumptions, our analysis yields two key results. \textbf{First}, with high probability:
\[
\Delta_2^2\left(\hat{\tau}^r, \tau^r\right)
~\lesssim~
\left(1 + \Delta_2(m, \mu^r)\right)\, \mathcal{R}_{n^r}(\mathcal{F}),
\]
where $\Delta_2(\cdot, \cdot)$ is the $L^2$ error and $\mathcal{R}_{n^r}(\mathcal{F})$ is the Rademacher complexity. This shows that a more accurate CMO (as $m$) reduces CATE estimation error.
\textbf{Second}, the prediction risk is bounded as:
\[
R(\hat{\tau}^r)
\equiv
\mathbb{E}^r\left[\tau_m(X, A, Y) - \hat{\tau}^r(X)\right]^2
~\lesssim~
\bar\sigma^2 
+ \Delta_2^2(m, \mu^r)
+ \Delta_2^2(\hat{\tau}^r, \tau^r),
\]
capturing irreducible noise ($\bar\sigma^2$), CMO, and CATE mean square errors (\textbf{MSE}). These bounds motivate improving CMO estimates to reduce both prediction and estimation error.

Since the CMO is a weighted average of $\mu^r_{+1}(x)$ and $\mu^r_{-1}(x)$, improving its estimation requires accurate estimation of these per-arm outcomes. Intuitively, estimating $\mu^r_a(x)$ using only RCT data yields unbiased but high-variance estimates, while using OS data can reduce variance at the cost of bias. We propose methods that combine RCT and OS data to reduce the overall error in estimating $\mu^r_a(x)$, and thus the CMO, without introducing bias in the CATE. 

To relax the CATE and outcome mean exchangeability assumptions, often required for data integration, we adopt an \emph{outcome-shift} framework in which conditional outcome models differ across populations. Let $\mu^o(x, a) \equiv \ex^o[Y \mid X = x, A = a]$ and $\mu^r_a(x) \equiv \ex^r[Y(a) \mid X = x]$. We define the discrepancy
\(
\delta_a(x) \equiv \mu^r_a(x) - \mu^o(x, a),
\)
and adopt as a working hypothesis that $\delta_a(x)$ belongs to a structured class (e.g., sparse or low-dimensional).
This formulation captures the relationship between the two populations without requiring full exchangeability and aligns with the concept-shift framework in transfer learning \citep{zhuang2020comprehensive}. It allows us to improve estimates of $\mu^r_a(x)$, and thus the CMO, by appropriately correcting for systematic differences between OS and RCT data.

We propose, a doubly-calibrated OS-augmented procedure for estimating CATE in the RCT, R-OSCAR. Algorithm \ref{alg:doubly-calibrated} implements R-OSCAR through four steps: estimate outcome models $\hat{\mu}^o(\cdot,a)$ (line 3) using the OS, estimate their discrepancies $\hat{\delta}_a(\cdot)$ from the RCT data to calibrate for population differences (line 4), construct the CMO $m(\cdot)$ (line 6), then estimate the \textit{CATE} discrepancy $\hat{\delta}(\cdot)$ using pseudo-outcomes. 

\begin{algorithm}[h]
{\small
\caption{R-OSCAR: Robust OS for CMO-Augmented RCT}
\label{alg:doubly-calibrated}
\begin{algorithmic}[1]
\State \textbf{Input:} OS data $\{(X_i^o, A_i^o, Y_i^o)\}_{i=1}^{n^o}$, RCT data $\{(X_i^r, A_i^r, Y_i^r)\}_{i=1}^{n^r}$
\For{$a \in \{-1, +1\}$}
    \State \textbf{OS outcome modeling: } $\hat{\mu}^o(\cdot,a) = \argmin_f \frac{1}{n_a^o}\sum_{i: A_i^o = a} [ Y_i^o - f(X^o_i) ]^2 + R^o_a(f)$
    \State \textbf{Outcome calibration to RCT: }$\hat{\delta}_{a}(\cdot) = \argmin_{d} \frac{1}{n_{a}^r} \sum_{i: A_i^r = a}[Y_{i}^r - \hat{\mu}^o(X_{i}^r,a) - d(X_i^r)]^{2} + R^r_a(d)$
\EndFor
\State \textbf{CMO construction: } $m(\cdot) = \sum_{a} \pi_{a}^r(\cdot) [\hat{\mu}^o(\cdot, a) + \hat{\delta}_a(\cdot)]$
\State \textbf{CATE calibration:} 
\Statex \quad
$\hat{\delta}(\cdot) = \argmin_{d} \frac{1}{n^r} \sum_{i=1}^{n^r}[\tau_m(X_{i}^r, A_{i}^r, Y_{i}^r) - \sum_{a} a (\hat{\mu}^o(X_i^r,a) + \hat{\delta}_a(X_i^r)) - d(X_{i}^r)]^{2} + R(d)$
\State \textbf{Output:} $\hat{\tau}^r(\cdot) = \sum_{a} a [\hat{\mu}^o(\cdot,a) + \hat{\delta}_a(\cdot)] + \hat{\delta}(\cdot)$
\end{algorithmic}
}
\end{algorithm}


All $R$ terms represent suitable function regularizers (e.g., ridge or lasso penalties). Without regularization in outcome calibration, estimated discrepancies may ignore information from the OS, so the penalties $R^r_a$ ensure effective ``knowledge transfer.'' Separate arm regularization can introduce regularization bias, where independently regularized outcome functions yield poor CATE estimates when contrasted. The CATE calibration step corrects this bias while its penalty term helps borrow information from the preliminary CATE estimate. For ease of analysis, our theoretical results use constrained formulations of all steps, while practical implementations employ regularized versions that maintain equivalent solution sets and theoretical guarantees.

For theoretical validity, outcome and CATE calibration require independent RCT samples. In practice, we use $K$-fold sample splitting: partition RCT data into $K$ folds, use $K-1$ folds for outcome calibration and one fold for CATE calibration in each iteration. The final estimator averages across folds: $\hat{\tau}^r(x) = \sum_{a} a [\hat{\mu}^o(x,a) + \frac{1}{K} \sum_{k=1}^K \hat{\delta}^k_a(x)] + \frac{1}{K} \sum_{k=1}^K \hat{\delta}^k(x)$. Empirically, sample-split results closely match the full-data approach.

{
\subsection{Related Work} \label{relwork}

\subsubsection{CATE Estimation} 
The field of heterogeneous treatment effect estimation has seen significant methodological development in recent years, particularly through the integration of machine learning techniques \citep{jacob2021cate}. Current methodological approaches in CATE estimation can be broadly categorized into two groups: meta-learners that leverage off-the-shelf machine learning methods \citep{kunzel2019metalearners} and specialized algorithms specifically designed for causal inference.

Meta-learners represent a flexible framework that adapts standard machine learning algorithms for estimating required functions such as outcome means and propensities uscausal inference tasks. These include the \textit{S-learner}, which models a single function including the treatment as a feature, and the \textit{T-learner}, which estimates separate conditional mean functions for treated and control groups. In our notation, the T-learner estimator becomes $\sum_{a} a \hat{\mu}_{a}^r(x)$. However, separate regularized estimation of outcome models may lead to different regularization patterns, potentially introducing artifacts in the final CATE estimates known as regularization bias \citep{wager2024causal}. In contrast, the estimator in \eqref{instance} provides greater robustness by directly targeting CATE estimation, making it less susceptible to such estimation artifacts. Other meta-learners include the \textit{doubly-robust (DR) learner}, which combines regression adjustment with inverse probability weighting \citep{kennedy2023towards}; the \textit{R-learner}, which employs orthogonalization techniques \citep{nie2021quasi}; and the \textit{X-learner}, which addresses treatment effect heterogeneity in imbalanced experimental settings \citep{kunzel2019metalearners}. These approaches leverage various machine learning algorithms such as random forests, gradient boosting, neural networks, and LASSO regression to tailor estimation to specific dataset characteristics.

In parallel, specialized algorithms have been developed to directly estimate heterogeneous treatment effects, including causal forests \citep{athey_generalized_2019}, causal BART \citep{hahn2020bayesian}, and causal boosting \citep{powers2018some}. These methods modify traditional machine learning approaches to specifically target treatment effect estimation, often incorporating sample-splitting and cross-fitting techniques to improve robustness and reduce overfitting. As the field continues to evolve, research increasingly focuses on understanding the statistical properties of these estimators, developing appropriate inference techniques, and addressing challenges in high-dimensional settings \citep{fan2022estimation}.

A closely related concept is that of \textit{individualized treatment rules} (ITRs), which aim to recommend the optimal treatment for each individual based on their covariates \citep{qian2011performance}. For binary treatments, ITRs are often derived by thresholding the estimated CATE: treatment is assigned when the estimated CATE is positive. Recent advances have also focused on directly estimating ITRs using methods such as outcome-weighted learning \citep{zhao2012estimating}, policy learning \citep{athey2017efficient}, and doubly robust policy evaluation \citep{dudik2011doubly}. Generalizing ITRs to target populations under covariate shift has recently been addressed by \citet{chen2024robust}. These methods are typically evaluated using the value function, which quantifies the expected outcome under the learned policy. While conceptually distinct, ITR estimation is closely tied to CATE estimation, as accurate CATEs support more effective individualized decision-making.

The closest analogue to our setting in the ITR literature is the variance-reduction result of \citet{laber2015tree} (Lemma~1): among augmented inverse-probability-weighted estimators of the policy value $V(\pi) = \ex[Y(\pi(X))]$ for a fixed rule $\pi$, the variance is minimized when the augmentation function equals $\ex[Y \mid X=x, A=\pi(x)]$, the outcome mean evaluated at the rule's recommended treatment. The CMO has the same variance-control role for the CATE pseudo-outcome used to estimate $\hat\tau^r(x)$ as a function on the covariate space, but in our setting this augmentation is learned from a separately collected observational study and enters a Rademacher-complexity-based finite-sample CATE-prediction risk bound.

\subsubsection{RCT and OS Data Integration Assumptions}  
As mentioned earlier, most work in data integration follows the $\cR \rightarrow \cO$ direction, commonly known as generalization or transportability. While the exact definitions and their corresponding assumptions vary between potential outcome \citep{dahabreh2020extending}
and structural causal model \citep{bareinboim2016causal} frameworks, we broadly refer to these as generalizability methods where the goal is to generalize the average treatment effect from RCT to OS; recent developments include sharp identification bounds under outcome distribution shift \citep{asiaee2026sharp} and omitted-variable sensitivity analysis \citep{asiaee2026omitted}. Although our work studies the $\cO \rightarrow \cR$ direction, understanding the assumptions of generalizability methods remains relevant. These methods primarily address shifts in covariate distribution ($\mathbb{P}^r(x) \neq \mathbb{P}^o(x)$) while assuming some aspect of the outcome model remains invariant across populations. This invariance typically relies on one or more of the following assumptions in descending order of strength \citep{colnet2024causal}: \textbf{1}) CATE invariance ($\tau^r(x)=\tau^o(x)$), \textbf{2}) mean exchangeability ($\forall a: \mu^r_a(x) = \mu^o_a(x)$), or \textbf{3}) trial participation ignorability ($Y(a), Y(a^{\prime}) \perp S \!\mid\! X$). The strongest assumption, trial participation ignorability, is equivalent to the ``no outcome shift'' assumption, which states that $\mathbb{P}^r(Y(a) \!\mid\! X=x)=\mathbb{P}^o(Y(a) \!\mid\! X=x)$.

These invariance assumptions have been criticized as unrealistic in practical applications. \citet{Hernan2020CausalInferenceWhat} identifies three primary ways the trial participation ignorability assumption can be violated: shifts in the distribution of unobserved effect modifiers across populations, inconsistencies in treatment versions between populations, and altered interference patterns of units across different populations. We address these concerns by explicitly modeling and accounting for outcome shifts between the RCT and OS populations rather than assuming mean exchangeability.

\subsubsection{Data Integration for CATE Estimation in RCT}
Recent work has explored combining OS and RCT data to improve CATE estimation for the RCT population. These approaches can be categorized into two main methodological frameworks, each with distinct assumptions and strategies.

\noindent \textbf{Weighted Combination Approaches.} The first category focuses on combining CATE estimates from both data sources through optimal weighting schemes. The approach of \citet{cheng2021adaptive} uses a weighted combination of preliminary CATE estimates from both RCT and OS datasets, where the weight is determined adaptively based on the mean squared error of the combined estimator relative to a proxy derived from out-of-fold RCT estimates. This allows efficient borrowing of information from the observational data while controlling for potential bias. A related line of work by \citet{oberst2022understanding} studies the optimal convex combination of an unbiased and a possibly biased estimator in a general setting, providing sharp MSE-based bias thresholds to guide integration. While not specific to CATE, their framework is directly applicable to combining RCT and OS estimators for causal inference.

Another related family uses James--Stein-type data-adaptive weights to combine an unbiased experimental estimator with a possibly biased observational one, building on the Stein-type shrinkage tradition \citep{strawderman1971proper}. \citet{rosenman2023combining} apply this idea to stratum-specific CATE estimates, shrinking the RCT-based estimator toward its OS-based counterpart with a weight derived from a generalized unbiased risk estimator. These methods, like the convex-combination approaches above, borrow at the final-estimator level and require an explicit bias model for CATE. Our framework instead borrows through the outcome-mean nuisance function, so the final CATE pseudo-outcome regression remains anchored entirely in the RCT.

\noindent \textbf{Confounding Function Approaches.} The second category, which is closest in spirit to our approach, relaxes the strong CATE transportability assumption by explicitly modeling the bias between populations. This category comprises only two recent works: \citep{wu_integrative_2022, yang_data_2025}, which integrate OS and RCT data to improve CATE estimation efficiency for the RCT population. Like our work, they relax stringent transportability conditions, though they focus on relaxing the CATE invariance assumption by introducing a confounding function that captures hidden biases specific to the OS data. In contrast, our approach relaxes the mean exchangeability assumption. Building on semiparametric modeling techniques, \citet{wu_integrative_2022} propose orthogonalized loss functions, while \citet{yang_data_2025} derive semiparametric efficient scores and construct estimators that achieve the efficiency bound.

\subsubsection{Relationship to Other Learning Paradigms}

Transfer learning can be viewed as the predictive analog of data integration for causal effect estimation, where the goal is to use source datasets to improve prediction accuracy—rather than estimate causal effects—in a target population. This area is well-studied in the machine learning literature \citep{weiss2016survey, zhuang2020comprehensive}. Within transfer learning, domain adaptation methods have received the most attention. These approaches primarily aim to address covariate shift, where the input distribution differs between source and target domains \citep{ben2010theory}, while outcome shift has been comparatively less explored \citep{kouw2018introduction}.

While transfer learning generally assumes a sequential relationship between source and target tasks, multi-task learning instead seeks to jointly train models across multiple related tasks, enabling the sharing of structure and inductive bias \citep{zhang2021survey}. The tasks in transfer and multi-task learning methods may involve predicting related but distinct outcomes \citep{li2022deep, zhili2020multi}, predicting the same outcome across different populations \citep{suresh2019learning, steingrimsson2023transporting} or treatment groups \citep{wang2022molecular, strauch2024improving}, or addressing temporal distributional changes \citep{nguyen2020temporal, helli2024drift}.

Our framework enables the integration of transfer learning techniques to estimate conditional mean outcomes in the RCT population, using OS data as the source domain. These transferred outcome models are then used to construct CMO estimates in the RCT population, which can be safely incorporated into CATE estimation without introducing bias. Our outcome discrepancy formulation provides a principled way to implement this idea, drawing on prior work in high-dimensional multi-task learning under sparse outcome shift assumptions \citep{gross2016data, asiaee2018high}.

To the best of our knowledge, our work is the first to explicitly connect transfer learning theory to CATE estimation across populations. The proposed method offers a flexible framework that accommodates a wide range of transfer learning strategies for outcome modeling, enhancing the efficiency of CATE estimation while preserving consistency guarantees.




The remainder of the manuscript is organized as follows. Section~\ref{theory} introduces our theoretical framework, formulating CATE estimation as a supervised learning problem and establishing how its efficiency depends on the MSE between the augmentation function ($m$) and the true CMO ($\mu$)  (Section~\ref{the-framework}). We then show how to leverage OS data to improve CMO estimation (Section~\ref{cmo-learning}), including a special case based on sparse shifts in linear outcome models (Section~\ref{on-outcome-shift}). Section~\ref{simulation-results} presents simulation studies, and Section~\ref{real-results} validates our method on real-world data. Appendix \ref{all-proofs} contain all proofs.

\section{A general framework for improving efficiency of CATE estimation} \label{theory}
This section presents our general framework for improving the efficiency of CATE estimation by incorporating OS data. We begin by showing that CATE estimation can be formulated as supervised learning with a pseudo-outcome whose variance is minimized when using the true CMO as the augmentation function (Section~\ref{the-framework}). We then propose strategies for estimating the CMO using both RCT and OS data under outcome-shift assumptions (Section~\ref{cmo-learning}). Finally, we illustrate the framework using linear models with sparse differences across populations (Section~\ref{on-outcome-shift}).

\subsection{Improved CATE estimation via augmentation: general prediction and estimation risk bounds} 
\label{the-framework}

Here, we reformulate CATE estimation as an empirical risk minimization problem, first demonstrating how efficient pointwise CATE estimation is achievable through a careful choice of the augmentation function. We then show how this formulation facilitates the systematic reduction of prediction risk and estimation error of the CATE. For clarity, in this section, we present our results without study population indicators, as the results hold under appropriate assumptions in any study population. While these assumptions (except A1) are typically satisfied by design in RCTs, they become explicit requirements for observational studies. Importantly, since our ultimate goal is to estimate CATE for the RCT population, these assumptions are only required to hold for the RCT data; their validity in the observational study is not necessary. 

\subsubsection{More efficient pointwise CATE estimation via augmentation}
The following proposition formalizes the supervised learning perspective for CATE estimation by showing how it can be framed as a risk minimization problem using transformed outcomes.

\begin{proposition} 
\label{prop1-cate-opt}
Consider any study population where the following hold:
\begin{enumerate}[label=(A\arabic*), nosep]
\item (SUTVA) $Y_{i} = Y_{i}(A_{i})$
\item (Conditional Ignorability) $(Y(-1), Y(+1)) \perp A \!\mid\! X$ 
\item (Weak Positivity) $\forall X, A: 0 < \pi_{A}(X) < 1$
\end{enumerate}

Let $m(X)$ be any function of $X$. For any given sample $(X, A, Y)$, define the following transformation: $\tau_{m}(X,A,Y) \equiv \frac{A (Y - m({X}))}{\pi_A({X})}$. 
Then, $\tau_{m}(X,A,Y)$ is an unbiased estimator of the CATE at $X$, that is: $\ex_{A, Y}({\tau}_{m}(x,A,Y)\mid X=x) = \tau(x)$.
Moreover, for any $x$, the true CATE $\tau(x)$ minimizes the following conditional risk:
\begin{align} 
    \label{eq-cate}
    \tau(x)=\argmin_f \ex_{Y, A}\left[\left(\tau_{m}(x,A,Y)-f(x)\right)^{2} \middle| X=x \right],
\end{align}
\end{proposition}
%

The special case of the transformation $\tau_0(X, A, Y) \equiv 2AY$ with equal treatment assignment probability $\pi_A(X) = 1/2$ and augmentation function $m(X) = 0$ was  introduced by \citet{tian2014simple}, where it was referred to as the ``modified outcome.'' Subsequently, \citet{athey2015machine} extended this idea to general propensity functions $\pi_A(X)$, still with $m(X) = 0$, calling it the ``CATE-generating transformation.'' , we further generalize this transformation by allowing an arbitrary augmentation function $m(X)$. We denote this general form by $\tau_m(X, A, Y)$ and refer to it as the ``pseudo-outcome,'' emphasizing its use within supervised learning frameworks.

Since $m(X)$ is arbitrary, we have the flexibility to choose it in a way that minimizes the variance of the CATE estimator $\tau_m(X, A, Y)$ at $X=x$. The following theorem demonstrates that minimum conditional variance for $\tau_m(X, A, Y)$ is achieved by choosing $m(X)$ to be the CMO, $\mu(X)$.

\begin{theorem}
\label{theo1-var-reduce}
Assuming assumptions (A1)-(A3) of Proposition \ref{prop1-cate-opt} hold. Then, setting $m(X) := \mu(X)$ in \eqref{eq-cate} results in the minimum variance CATE estimator. In other words, the solution of the following is the CMO: $\mu(x) = \argmin_{m} \var(\tau_{m}(x,A,Y) \!\mid\! X=x)$.
\end{theorem}

One can show (see Proposition~\ref{prop2-decomp} in  Appendix \ref{all-proofs}) that the mean potential outcome decomposes into the sum of the CMO and the CATE. This decomposition offers an intuitive interpretation of Theorem~\ref{theo1-var-reduce}: subtracting the CMO from outcome yields the most efficient estimator of the CATE. The function $m(X)$ can be viewed as either a \textit{nuisance function} that removes baseline variation without introducing bias, or as an \textit{augmentation function} that improves efficiency when properly chosen. We adopt the term \textbf{augmentation function}, following the augmented inverse probability weighting (AIPW) literature \citep{kurz2022augmented}, as it better captures the specific role of $m(X)$ in enhancing CATE estimation efficiency.

\subsubsection{Bounding and minimizing the expected CATE prediction error}
Use of the true CMO is optimal for variance reduction with respect to the population loss function. In practice an empirical loss function is minimized using the observed data.
Furthermore, the true CMO is never available and must be estimated from finite samples, introducing an additional layer of estimation error. 
This raises critical questions about the practical implications of outcome transformation: how does the selected $m(X)$ impact CATE estimation across all covariates $x$ when the corresponding plug-in estimator of \eqref{eq-cate} is used? and how does the error in estimating the CMO influence the efficiency of the resulting CATE estimates? 

To address these questions, we leverage the standard concept of risk, or expected squared prediction error, from supervised learning. In a typical supervised learning task, the risk of a function $f$ is defined and decomposes as:
$
    R(f) \equiv \ex_{X, Y} [(Y - f(X))^2] 
    = \ex_X \var(Y \!\mid\! X) + \Delta^2_2(f, g),
$
where $(X, Y)$ is the predictor-outcome pair, $\Delta^2_2(f, g) = \ex_X [f(X) - g(X)]^2$, and $g(X) \equiv \ex[Y \!\mid\! X]$ is the Bayes estimator, minimizing the risk in the MSE sense. This decomposition reveals two components:
1) The \textbf{irreducible error}, $\ex_X \var(Y \!\mid\! X)$, representing the variance of $Y$ given $X$.
2) The \textbf{estimation error}, $\Delta^2_2(f, g) = \ex_X [f(X) - g(X)]^2$, quantifying the MSE of $f$ with respect to $g$.
For homoskedastic additive noise, the risk simplifies to $R(f) = \sigma^2 + \Delta^2_2(f, g)$, where $\sigma^2$ represents the irreducible noise variance in the model $Y = g(X) + \epsilon$ with $\ex[\epsilon|X]=0$.

Analogously, in the supervised learning formulation of CATE estimation, the transformed outcome $\tau_m(X, A, Y)$ serves as the analog of $Y$, and the true CATE $\tau(X) = \ex[\tau_m \!\mid\! X]$ plays the role of $g(X)$. Then for the risk of any CATE estimator we have:
\begin{align} \label{eq-risk-decompose}
    R_m(\htau) = \ex_{X, A, Y} [\tau_m(X, A, Y) - \htau(X)]^2 
    = 
    \ex_X \var(\tau_m \!\mid\! X) + \Delta^2_2(\htau, \tau).
\end{align}
\paragraph{Bounding and minimizing irreducible error.\,} 
Here, unlike a typical supervised learning setting, the \textbf{irreducible error} $\ex_X \var(\tau_m \!\mid\! X)$ depends on the choice of $m(X)$. As shown in Theorem \ref{theo1-var-reduce}, this term is minimized when $m(X) = \mu(X)$, the true CMO. The ``irreducible error'' is a misnomer in this context because it can be reduced by improving the choice of $m(X)$. However, given its established usage in regression, we retain the term here for consistency. Proposition \ref{prop3-variance-reduction} summarizes our discussion. 

\begin{proposition}[Risk Reduction with the Optimal Augmentation Function]
\label{prop3-variance-reduction}
The variance of the transformed outcome $\tau_m$ determines the irreducible error in CATE estimation. Selecting the augmentation function $m(X)$ to minimize $\ex_X \var(\tau_m \!\mid\! X)$ reduces the risk of ANY estimator $\htau(X)$. The optimal augmentation function, $m(X)=\mu(X)$, achieves the smallest possible irreducible error.
\end{proposition}
\vspace{6pt}
Proposition \ref{prop3-variance-reduction} underscores the importance of quantifying the relationship between the quality of $m(X)$ and the variance reduction achieved. In the following theorem, we establish bounds on the mean irreducible variance $\ex_X \var(\tau_m \!\mid\! X)$ as a function of the MSE of $m(X)$ relative to the true CMO $\mu(X)$, $\Delta_2^2(m, \mu)$. These bounds provide a practical criterion for determining when an estimated CMO, subject to estimation error and possible misspecification, improves the risk over simpler alternatives, such as setting $m(X) = 0$.

\begin{theorem} \label{theo2-sandwich}
Under assumptions (A1)-(A3) of Proposition \ref{prop1-cate-opt} and:
\begin{enumerate}[label=(A\arabic*), nosep, start=4]
\item (Strong Positivity) $\exists \rho \in (0,1/2]$ such that $\rho \leq \pi_a(X) \leq 1-\rho$ for $a \in \{-1, 1\}$;
\end{enumerate}
Let $\mu$ be the true CMO and $m$ be any function used to generate pseudo-outcome $\tau_m$ introduced in Proposition \ref{prop1-cate-opt}. Define $\Delta_2^2(m, \mu) = \ex_X[(m(X) - \mu(X))^2]$ as the mean squared error between $m(X)$ and $\mu(X)$. The excess irreducible error from using $m$ instead of $\mu$ is bounded as:
\begin{equation} \label{eq-sandwitch}
\frac{\Delta_2^2(m, \mu)}{(1-\rho)^2}
\leq
\ex[\var(\tau_{m} \!\mid\! X)] - \ex[\var(\tau_{\mu} \!\mid\! X)]
\leq
\frac{\Delta_2^2(m, \mu)}{\rho^2},
\end{equation}
where $\var(\tau_{\mu} \!\mid\! X)$ is the variance of the oracle estimator using the true CMO.
\end{theorem}
\vspace{6pt}
For the special case of 
$\pi^r_{-1} = \pi^r_{+1} = 1/2$, the inqualities simplify to: $\ex[\var({\tau}_{m}\mid X)] = \ex[\var({\tau}_{\mu}\mid X)] + 4\Delta_2^2(m, \mu).$ Note that $\var(\tau_{\mu} \!\mid\! X=x)$ represents the minimum achievable variance at $X=x$ according to Theorem \ref{theo1-var-reduce} and is therefore truly irreducible. By combining \eqref{eq-risk-decompose} and \eqref{eq-sandwitch}, we establish how the MSE between $m(x)$ and the CMO $\mu(x)$ controls the prediction risk:

\begin{corollary}
\label{irred-error}
Let $\bar\sigma^2 \equiv \sup_{x \in \mathcal{X}} \var\left(\tau_\mu \!\mid\! X=x\right)$ denote the supremum of the conditional variance of the optimal transformed outcome. 
Then, under the assumptions of Theorem \ref{theo2-sandwich}, the prediction risk admits the following bound:
$
    R_m(\hat{\tau}) \leq \bar\sigma^2 + \frac{\Delta_2^2(m, \mu)}{\rho^2} + \Delta^2_2(\hat{\tau}, \tau).
$
\end{corollary}
This risk decomposition shows that, beyond the irreducible variance term $\bar\sigma^2$ and the CATE estimation error $\Delta^2_2(\hat{\tau}, \tau)$, the prediction risk is further governed by $\Delta_2^2(m, \mu)$, the quality of the CMO estimate. In the next section, we demonstrate that this same term also influences the upper bound on $\Delta^2_2(\hat{\tau}, \tau)$, highlighting the central role of accurate CMO estimation.


\paragraph{Bounding and minimizing the CATE estimation error.\,}
Now we turn to the second term of the risk \eqref{eq-risk-decompose}, which is the \textbf{CATE estimation error}. Here we focus on the estimation error of the empirical risk minimizer (ERM), i.e., the plug-in estimator corresponding to the population-level objective introduced in \eqref{eq-cate}. We restrict $f$ to a function class $\cF$ to ensure tractability and avoid overfitting in finite sample settings. For instance, in Section \ref{on-outcome-shift}, we consider the class of linear functions with bounded $\ell_1$ norm coefficients, which is the commonly-used lasso-penalized linear model.
We show that, similar to the irreducible error part of the prediction risk, the CATE's non-asymptotic estimation error bound also relies on the choice of the augmentation function. Using tools from statistical learning theory, Theorem \ref{theo3-risk-bound} quantifies how the choice of both the function class $\cF$ and $m(X)$ influence the risk of the ERM estimator, establishing that accurately estimating the CMO and using it as $m(X)$ can significantly reduce the CATE estimation error.

\begin{theorem}[Non-asymptotic MSE bound for an empirical risk minimizer]
\label{theo3-risk-bound}
Suppose Assumptions \textup{(A1)--(A4)} hold, and let $\cF$ be a function class for approximating the CATE $\tau(x)$. Consider a given augmentation function $m(x)$ used to form the pseudo-outcome $\tau_m(X_i, A_i, Y_i)$ from an i.i.d.\ dataset $\{(X_i, A_i, Y_i)\}_{i=1}^n$. Define the empirical risk minimizer as
\begin{equation*}    
\htau_n(\cdot) 
=
\arg\min_{f \in \cF}
\frac{1}{n}\sum_{i=1}^n\left(\tau_m(X_i,A_i,Y_i) - f(X_i)\right)^2.
\end{equation*}
Then, with probability at least $1 - 2\varepsilon$, the mean squared error of $\htau_n$ satisfies
\begin{equation}  \label{eq-risk-bound} 
   \Delta_2^2(\htau_n, \tau)
   \leq
   \Delta_2^2(\cF, \tau) 
   + 2C(m,\varepsilon)\cR_{n}(\cF)
   +
   C^2(m,\varepsilon)
   \sqrt{\frac{\log\left(2/\varepsilon\right)}{n}},
\end{equation}

where the quantities are defined as follows:
\begin{itemize}[leftmargin=12pt, itemsep=0pt]
    \item $\Delta_2^2(\cF, \tau) = \inf_{f \in \cF} \ex_X\left[(f(X) - \tau(X))^2\right]$ is the approximation error of $\cF$, i.e., the minimal achievable error within the function class $\mathcal{F}$, capturing the inherent bias due to model misspecification, 
    \item $\cR_n(\cF) = \ex_{\epsilon, X} \left[\sup_{f\in \cF} \frac{1}{n} \sum_{i=1}^{n} \epsilon_i f(X_i)\right]$ is the Rademacher complexity of $\cF$ based on $n$ samples and $\epsilon_i \stackrel{\text{i.i.d.}}{\sim} \mathrm{Unif}\{-1,1\}$,
    \item $B$ is a constant such that $|\tau(x)|\leq B$ and $|f(x)| \leq B$ for all $f \in \cF$,
    \item $C(m, \varepsilon) = \left(2B + \frac{\bar{\sigma}}{\sqrt{\varepsilon}} + \frac{\Delta_\infty(m, \mu)}{\rho}\right)$ depends on the uniform distance of $m(x)$ from $\mu(x)$, $\Delta_\infty(m, \mu)$.
\end{itemize}
\end{theorem}
%
%


\vspace{6pt}
Rademacher complexity is a well-studied, unifying measure of function class capacity that leads to tight generalization bounds.  
Tables~\ref{tab:lin_convex}--\ref{tab:nonparam} in Appendix~\ref{r-complexity-bounds} collect sharp bounds for many classes used in statistical learning, illustrating how structural constraints such as sparsity, smoothness, or norm constraints govern learnability.  
For instance, for the class of linear $s$-sparse predictors
\(
\mathcal{F}_{s}
  \;=\;
  \bigl\{x \mapsto w^{\top}x: x \in \reals^p, \; \|w\|_{0}\le s,\; \|x\|_{2}\le 1\bigr\},
\)
the empirical process theory ~\citep{bartlett2006local,koltchinskii2011oracle} gives the complexity bound  
\(
\mathcal{R}_{n}(\mathcal{F}_{s})
  \;\lesssim\;
  \sqrt{\frac{s\log(p/s)}{n}},
\)
showing that the burden of high dimensionality enters only through a benign $\log(p/s)$ factor. 

{
These Rademacher bounds can be converted into minimax squared-error rates by the standard \emph{localization argument} \citep{bartlett2006local}.  Localization requires that the class $\mathcal{F}$ be \emph{star-shaped about the oracle}  
$f^{\star}=\arg\min_{f\in\mathcal{F}}\mathbb{E}\,\ell(f)$; that is, for every $f\in\mathcal{F}$ and $\lambda\in[0,1]$, the convex combination $\lambda f^{\star}+(1-\lambda)f$ also belongs to $\mathcal{F}$.  
Since every norm ball (e.g.\ $\ell_{1}$, $\ell_{2}$, RKHS, Hölder, or Sobolev) is convex, this condition is automatically satisfied for all classes listed in the tables.  
Consequently, a bound $\mathcal{R}_{n}(\mathcal{F})\lesssim\varepsilon_{n}$ implies the minimax rate
\(
\inf_{\hat{f}}\sup_{f\in\mathcal{F}}\mathbb{E}\lVert\hat{f}-f\rVert^{2}
  \;\lesssim\;
  \varepsilon_{n}^{2}.
\)
This connection unifies the Rademacher-complexity viewpoint with classical minimax theory \citep{bartlett2006local,wainwright2019high}.  
For example, the bound $(1/n)^{\alpha/(2\alpha+p)}$ for Hölder-smooth functions yields the familiar nonparametric rate $(1/n)^{2\alpha/(2\alpha+p)}$ \citep{tsybakov_introduction_2008}, while the bound $\sqrt{s\log(p/s)/n}$ for $s$-sparse linear functions produces the optimal high-dimensional linear-regression rate $s\log(p/s)/n$ \citep{wainwright2019high}.
}

Theorem~\ref{theo3-risk-bound} shows that the excess MSE of $\hat\tau_n$ decomposes into three familiar pieces: approximation, estimation, and concentration components. 
Both stochastic terms are multiplied by the factor  
\(
C(m,\varepsilon)=2B+\frac{\bar\sigma}{\sqrt{\varepsilon}}
               +\frac{\Delta_\infty(m,\mu)}{\rho},
\)
so the quality of the augmentation $m$ directly modulates the sample complexity.  
If the oracle augmentation $m=\mu$ were available, then $\Delta_\infty(m,\mu)=0$ and we recover the classical Rademacher bound. 
In practice, however, we plug in an estimate $\hat\mu$, and the uniform error
$\Delta_\infty(\hat\mu,\mu)$ inflates both the estimation and concentration terms.  
Thus, accurate estimation of the CMO is indispensable for fully realizing the benefits of low-complexity CATE classes.  
The next section quantifies how errors in the per-arm mean functions drive $\Delta_\infty(\hat\mu,\mu)$ and, consequently, the overall precision of CATE estimation.

\subsubsection{Controlling the CMO error via treatment-arm error bounds}

The preceding analysis has demonstrated that both prediction and estimation risks are governed by the accuracy of our CMO estimate, measured through MSE (Theorem \ref{theo2-sandwich}) and uniform distance (Theorem \ref{theo3-risk-bound}), respectively. 
In what follows, we establish that controlling the mean squared estimation error for each treatment arm separately suffices for reliable control of the uniform and MSE estimation error of the CMO jointly.
While uniform convergence generally implies convergence in MSE, the converse does not always hold. However, under mild conditions, convergence in MSE \emph{does} imply uniform convergence. We show below how the per-arm rate of convergence in MSE simultaneously controls both the uniform and convergence in MSE of the CMO.
\begin{proposition}[MSE and Uniform Convergence of the CMO via Separate-Arm Estimation]
\label{prop4-cmo-convergence}
Let $\cD' = \{(X_j, A_j, Y_j)\}_{j=1}^{n'}$ be an i.i.d.\ dataset of size $n'$, reserved for estimating the CMO, $\mu$.  
Partition $\cD'$ by treatment arm $A_j \in \{+1,-1\}$ into two subsets
$
  \cD'_{+1} = \{(X_j, Y_j) : A_j = +1\}, \quad \cD'_{-1} = \{(X_j, Y_j) : A_j = -1\},
$
with sizes 
$
  n'_{+1} = |\cD'_{+1}|
$
and
$
n'_{-1} = |\cD'_{-1}|,
$
so that $n'_{+1} + n'_{-1} = n'$. Suppose there exists a constant $\eta>0$ such that 
$
n'_{+1} \ge \eta n'
$
and 
$ 
n'_{-1} \ge \eta n'.
$

For each arm $a\in\{+1,-1\}$, let $\hat{\mu}_a$ be an estimator of the outcome mean of treatment $a$, $\mu_a : \cX \to \reals$, and assume that the per-arm MSE satisfies
$
   \Delta_2^2\big(\hat{\mu}_{a}, \mu_{a}\big)
   = \cO_p(r^2(n'_a)).
$
Define the true and estimated CMO functions as
$
  \mu(x) \equiv \pi_{-1}(x)\,\mu_{+1}(x) + \pi_{+1}(x)\,\mu_{-1}(x)
$
and
$
  \hat{\mu}(x) \equiv \pi_{-1}(x)\,\hat{\mu}_{+1}(x) + \pi_{+1}(x)\,\hat{\mu}_{-1}(x)
$, respectively.
Then, the following hold:
\begin{enumerate}
    \item[\textbf{(i)}] 
    The MSE of the CMO estimator satisfies
    $
       \Delta_2^2(\hat{\mu}, \mu)
       :=
       \ex_{X}[(\hat{\mu}(X) - \mu(X))^2]
       = \cO_p(r^2(n')).
    $
    
    \item[\textbf{(ii)}] 
    If, in addition, each $\mu_a$ is $L$-Lipschitz and the domain $\cX\subset\reals^d$ is compact, then the uniform error of the CMO estimator satisfies
    $
       \Delta_{\infty}\big(\hat{\mu}, \mu\big)
       :=
       \sup_{x\in \cX} |\hat{\mu}(x) - \mu(x)|
       = \cO_p(r(n')).
    $
\end{enumerate}
\end{proposition}
\vspace{6pt}

This proposition reveals a fundamental connection between per-arm estimation rates and the overall CMO convergence,  enabling us to leverage standard statistical learning results for individual outcome estimation to control both prediction risk and estimation error of CATE. The following corollary further elucidates this relationship.

\begin{corollary}[Simplified Risk Bounds]
\label{corollary-main-simplified}
Under the conditions of Proposition~\ref{prop4-cmo-convergence}, the following simplified risk bounds hold:
\begin{enumerate}
    \item With high probability, the MSE of the CATE estimator satisfies (up to a standard additive concentration term)
    $
    \Delta_{2}^2\bigl(\hat{\tau}_n, \tau\bigr)
    \;\lesssim\;
    \Delta_{2}^2\bigl(\cF, \tau\bigr)
    \;+\;
    (1 + \sum_{a} \Delta_{2}\bigl(\hat{\mu}_{a}, \mu_{a}\bigr)) \cdot \cR_n(\cF).
    $
    
    \item The overall prediction risk satisfies
    $
    R(\hat{\tau}_n) \;\lesssim\; \sigma^2 \;+\; \sum_{a} \Delta_{2}^2\bigl(\hat{\mu}_{a}, \mu_{a}\bigr) \;+\; \Delta_{2}^2\bigl(\hat{\tau},\tau\bigr).
    $
\end{enumerate}
\end{corollary}

We have now completed presenting our general CATE estimation framework. So far, the discussion has been limited to a single data source, where we partitioned the dataset into two parts: one for estimating CATE under assumed conditions of Proposition \ref{prop1-cate-opt} using the variational form in \eqref{eq-cate}, and another for estimating per-arm outcome means to construct an optimal augmentation function for minimizing the risk of the estimated CATE.
In the subsequent sections, we explore various CATE estimators for \textit{the RCT population}, focusing on strategies to enhance per-arm outcome mean estimation by leveraging large auxiliary OS data to construct the augmentation function.


\subsection{Enhancing CATE estimation efficiency via CMO estimation from OS and RCT data}
\label{cmo-learning}
We now develop approaches to integrate OS with RCT data for improved CATE estimation. Our methods focus on enhancing CMO estimation, which directly improves CATE estimates via our established risk bounds. Importantly, in this approach, leveraging potentially confounded OS data for CMO estimation does not introduce bias into the RCT-derived CATE estimate.

By design, all the assumptions required for Proposition \ref{prop1-cate-opt} and Theorems \ref{theo1-var-reduce} and \ref{theo2-sandwich} are satisfied in the RCT setting, with the exception of SUTVA (A1). Specifically, randomization ensures conditional ignorability (A2), while both weak (A3) and strong (A4) positivity are guaranteed by the trial protocol. Therefore, under the additional assumptions of SUTVA and correct specification of the CATE model, we establish the consistency of the ERM-based CATE estimator for the RCT setting, as formalized in the following proposition.

\begin{proposition}[Consistency of ERM in RCTs] \label{prop3-rct-est}
Assume SUTVA and
\begin{enumerate}[label=(A\arabic*), nosep, start=5]
\item (Correct specification of the CATE model): that $\tau^r \in \cF$
\end{enumerate}
Then, the empirical risk minimizer
\begin{align} \label{instance}
\htau^r(\cdot)=\argmin_{f\in \cF} \frac{1}{n^r} \sum_{i=1}^{n^r}\left[\frac{A_{i}^r\left(Y_{i}^r-m\left(X_{i}^r\right)\right)}{\pi_{A_{i}^r}^r\left(X_{i}^r\right)}-f\left(X_{i}^r\right)\right]^{2}
\end{align}
is consistent for the true CATE $\tau^r(\cdot)$ as $n^r \to \infty$.
\end{proposition}
While $\cF$ could theoretically include all measurable functions, in practice, we must control its complexity through constraints or regularization. 



\subsubsection{Baseline: naive CATE estimator} \label{naive-est}
We refer to the most basic version of estimator \eqref{instance}, obtained by setting $m(X):=0$, as the \textit{Naive} CATE estimator. This approach corresponds to the ``transformed outcome'' method explored in \citep{athey2015machine}. However, using $m(X) = 0$ is known to result in an inefficient estimator with high variance \citep{athey2016recursive}.


\subsubsection{Baseline: RCT-augmented CATE estimator} \label{racer-est}


To reduce the variance of the CATE estimate in \eqref{instance}, as guided by Section \ref{the-framework}, we use part of the RCT data to estimate the CMO:
\begin{align} \label{car-m}
m(x) := \hmu^r(x) = \sum_{a} \pi_{-a}^r(x) \hmu_{a}^r(x), \quad \hmu_{a}^r(x) \in \cM_a^r,
\end{align}
where $\hmu_{a}^r(x)$ denotes the estimated regression function from RCT data, constrained to lie in the function class $\cM_a^r$. The function class $\cM_a^r$ can be specified, for instance, as the set of linear functions with $\ell_1$ norm of coefficients bounded.
We refer to the resulting CATE estimator as \textbf{\textit{RACER}} (RCT Augmented by CMO Estimated from RCT). While RACER achieves improved efficiency over the naive estimator, its performance can be constrained by the high variance of the estimated outcome means due to RCT's limited sample size. 

\subsubsection{Proposed: OS-augmented CATE estimator with RCT calibration}
To address the high variance of estimated outcome means due to the sample size limitations inherent in RCT data, we propose utilizing OS data to estimate the CMO. A direct but potentially suboptimal approach would set the augmentation function to $m(x) := \sum_a \pi_{-a}^r(x) \hmu^o(x, a)$, where $\hmu^o(x, a)$ is an outcome mean estimate derived from OS data and applied to covariates from the RCT population. Ideally, if both ignorability in the OS population ($\mu^o(x, a) = \mu_a^o(x)$) and  mean exchangeability ($\forall a: \mu^r_a(x) = \mu^o_a(x)$) hold, then $\hmu^o(x, a)$ would be a valid substitute for $\hmu^r_a(x)$, enabling unbiased estimation of the RCT's potential outcome means from the OS.

However, in practice, OS data may suffer from unmeasured confounding and/or outcome shifts are often observed across populations, implying that $\forall a, x: \mu^r_a(x) \neq \mu^o(x, a)$. To account for this bias, we introduce an outcome mean discrepancy and shifted outcome mean functions defined as:
\begin{alignat}{3}
    &\text{outcome mean discrepancy:} \quad &\delta_a(x) &\equiv \mu^r_a(x) - \mu^o(x, a), \text{ and} \\ \label{shifted-outcome-means}
    &\text{shifted outcome mean:} \quad &\tilde{\mu}^o_a(x; d_a) &\equiv \hmu^o(x, a) + d_a(x), \quad \hmu^o(x, a) \in \cM_a^o,
\end{alignat}
where $\delta_a(x)$ represents the \textbf{true discrepancy} that will be estimated in the following steps and $\tilde{\mu}^o_a(x; d_a)$ is defined for an arbitrary discrepancy function $d_{a}(x)$ with $\hmu^o(x, a)$ being estimated from the OS data. Here, $\cM_a^o$ denote the function classes used for estimating the OS outcome means. The discrepancy functions are then estimated (using RCT data) so the overall estimated shifted outcome means best predict the observed outcomes in the RCT data. A key issue to note is that if the discrepancies $\delta_a(x)$ are estimated without constraining their function class, minimal to no information would be borrowed from the OS. Therefore, we constrain $d_{a}(x)$ to lie in the function class $\cD^j_a$ so that the estimated shifted outcome mean is shrunk towards the preliminary estimates $\hmu^o(x, a)$ of the CMO from the OS.
In summary, the discrepancy functions are estimated by solving the following constrained optimization problem:
\begin{equation} \label{delta-estimation}
    (\hdelta^j_{-1}, \hdelta^j_{+1}) = 
    \argmin_{d_{-1} \in \cD_{-1}^j, d_{+1} \in \cD_{+1}^j} \frac{1}{n^r} \sum_{i=1}^{n^r} \left[ \frac{A_i^r}{\pi^r_{A_i^r}(X_i^r)} \left(Y_i^r - \tilde{\mu}^o(X_i^r; d_{+1}, d_{-1})\right) - \tilde{\tau}^o(X_i^r; d_{+1}, d_{-1}) \right]^2. 
\end{equation}
Here, we used CMO and CATE \textit{template functions} using the shifted outcome means from \eqref{shifted-outcome-means} as 
\(
\tilde{\mu}^o(x; d_{+1}, d_{-1}) \equiv \sum_a \pi_{-a}^r(x)\, \tilde{\mu}^o_a(x; d_{a}) \text{ and }
\tilde{\tau}^o(x; d_{+1}, d_{-1}) \equiv \sum_a a\, \tilde{\mu}^o_a(x; d_{a}).
\)
These template functions serve as placeholders in the objective of \eqref{delta-estimation}, where the discrepancy functions $d_a$ are optimized.
The superscript $j$ on the estimated discrepancies $\hdelta^j_{a}$ indicates that these functions are estimated jointly, in the sense that both the calibrated CMO and the resulting CATE depend on the same discrepancy functions optimized within the joint risk-minimization framework.

Once the optimal discrepancies are obtained, the final \textit{calibrated CMO} is $\hmu^{o, j}(x) 
\equiv \tilde{\mu}^o(x; \hdelta^j_{+1}, \hdelta^j_{-1}) 
= \sum_a \pi_{-a}^r(x) \left[\hmu^o(x, a) + \hdelta_a^j(x)\right]$ and our proposed \textbf{OSCAR} (Observational Studies for CMO-Augmented RCT) estimator for the CATE will be (Table \ref{tab-estimator}):

\begin{align} \label{cate-oscar}
\htau_{\text{OSCAR}}(x) 
&\equiv \tilde{\tau}^o(x; \hdelta^j_{+1}, \hdelta^j_{-1}) 
= \sum_a a \left[\hmu^o(x, a) + \hdelta_a^j(x)\right].
\end{align}

As an example of function class constraints in \eqref{delta-estimation}, suppose $d_a(x)$ is modeled as a linear function of covariates. Then, the constraint sets $\cD^j_a$ may be defined as an $\ell_1$ or $\ell_2$ norm ball on the coefficients, corresponding to LASSO or Ridge regularization, respectively.

\begin{table}
\centering
\caption{\small \textbf{Summary of four CATE estimators for the RCT population within our general framework.} The naive estimator serves as a baseline and does not utilize any augmentation functions. RACER uses RCT data alone to estimate outcome means, forming the counterfactual mean outcome (CMO) used as the augmentation function in CATE estimation. The proposed methods, OSCAR and R-OSCAR, leverage large observational study (OS) samples to estimate outcome means. These are calibrated to match the RCT’s outcome means, yielding an estimator for the RCT's CMO. OSCAR performs joint estimation of CMO and CATE. R-OSCAR follows a two-stage procedure: it calibrates the outcome means to estimate the CMO, then revises the CATE using a correction term $\hdelta(x)$. 
}
\label{tab-estimator}
{\small 
\begin{tabular*}{\columnwidth}{@{}l|@{\extracolsep{\fill}}l}
\hline
\textbf{Selected Augmentation Function $m(x)$} & \textbf{Functional Form of the Corresponding CATE Estimator} \\
\hline
\hline
$0$ & $\htau_{\text{Naive}}(x) \in \cF$ \\
\hline
$\hmu^r(x)=\sum_{a} \pi_{-a}^r(x) \hmu_{a}^r(x)$ 
& $\htau_{\text{RACER}}(x) \in \cF$, $\hmu_{a}^r(x) \in \cM_a^r$ \\
\hline
$\hmu^{o, j}(x)=\sum_{a} \pi_{-a}^r(x)[\hmu^o(x, a)+\hdelta_a^j(x)]$
& $\htau_{\text{OSCAR}}(x) = \sum_a a \left[\hmu^o(x,a) + \hdelta_a^j(x)\right]$ \\
& $\hmu^o(x,a) \in \cM_a^o, \quad \hdelta_a^j(x) \in \cD_a^j$ \\
\hline
$\hmu^{o,t}(x)=\sum_{a} \pi_{-a}^r(x)[\hmu^o(x, a)+\hdelta_a^t(x)]$
& $\htau_{\text{R-OSCAR}}(x) = \sum_a a \left[\hmu^o(x,a) + \hdelta_a^t(x)\right] + \hdelta(x)$ \\ 
& $\hmu^o(x,a) \in \cM_a^o, \quad \hdelta_a^t(x) \in \cD_a^t, \quad \hdelta \in \cD$ \\
\hline
\end{tabular*}}
\end{table}

\subsubsection{Proposed: Doubly-calibrated OS-augmented CATE estimator} \label{robust-est}
The OSCAR estimator may suffer from two potential limitations: model misspecification in either the OS outcome regression means or mean discrepancy functions and regularization bias from the separate estimation of treatment arm outcomes \citep{wager2024causal}. To address these issues, we propose a robust variant called \textit{\textbf{R-OSCAR}} (Robust OSCAR) that employs a two-stage calibration process that decouples the CMO estimation from CATE estimation. First, it calibrates the OS regression means to the RCT data to reduce their biases and uses these calibrated estimates to construct both the CMO and a preliminary CATE estimate. Second, it applies an additional calibration step to the preliminary CATE estimate to mitigate  regularization and/or misspecification bias.

Specifically, we first calibrate the learned OS outcome means $\hmu^o(x, a)$ to RCT by estimating the discrepancy functions for each treatment arm:
\begin{align} \label{robust-delta-a}
\hdelta^t_a = \argmin_{d_a \in \cD_a^t} \frac{1}{n_a^r} \sum_{i:A_i^r=a} \|Y_i^r-\left[\hmu^o(X_i^r, a)+ d_a(X_i^r)\right]\|_{2}^2,
\end{align}
where $n_a^r$ denotes the number of RCT samples in treatment arm $a$ and $\hdelta^t_a$ is constrained to a function class $\cD_a^t$ and superscript $t$ emphasizes that this estimate is for the two-stage calibration process in contrast to the joint estimation in \eqref{delta-estimation}. {This step is inspired by the calibration of computer experiments \citep{kennedy2001bayesian, dai2018another}, which involves using observational data to refine the parameters of a computer model, making it better match real-world behavior. }

In the next step, we construct the calibrated CMO as (Table \ref{tab-estimator}): 
$$\hmu^{o, t}(x)\equiv \tilde{\mu}^o(x; \hdelta^t_{+1}, \hdelta^t_{-1})=\sum_{a} \pi_{-a}^r(x) \left[\hmu^{o}(x,a) + \hdelta^{t}_a(x)\right].$$ 
Then, we estimate an additional CATE discrepancy function $\delta(x)$ to refine the preliminary CATE estimate further:
{\small
\begin{align} \label{robust-delta}
\hdelta = \argmin_{d \in \cD} \frac{1}{n^r} \sum_{i=1}^{n^r} \left[ \frac{A_i^r}{\pi^r_{A_i^r}(X_i^r)} \left(Y_i^r - \tilde{\mu}^o(x; \hdelta^t_{+1}, \hdelta^t_{-1})\right) - \left[\tilde{\tau}^o(X_i^r; \hdelta^t_{+1}, \hdelta^t_{-1}) + d(X_i^r)\right] \right]^2,
\end{align}}
where $\tilde{\tau}^o(x; \hdelta^t_{+1}, \hdelta^t_{-1}) = \sum_a a \left[\hmu^o(x,a) + \hdelta^t_a(x)\right]$ is the preliminary CATE estimate and $\hdelta$ is constrained to a function class $\cD$.
The final R-OSCAR estimator combines these components:
\begin{align} \label{robust-final}
\htau_{\text{R-OSCAR}}(x) \equiv \tilde{\tau}^o(x; \hdelta^t_{+1}, \hdelta^t_{-1}) + \hdelta(x) = \sum_a a \left[\hmu^o(x,a) + \hdelta^t_a(x) \right] + \hdelta(x).
\end{align}
Compared to \eqref{cate-oscar}, R-OSCAR estimates the per-arm discrepancy functions separately in the first calibration step \eqref{robust-delta-a}, and incorporates an additional CATE calibration term $\hdelta(x)$ for robustness.


\subsubsection{CATE Identifiability and Estimator Consistency}
\label{on-bias}

\textbf{Identifiability.} Our framework leverages the key strength of RCTs: both conditional ignorability (A2) and positivity (A3–A4) hold by design. As a result, the identifiability of the CATE requires only the SUTVA (A1), making the assumption set minimal (Proposition~\ref{prop1-cate-opt}).

\noindent \textbf{Estimator Consistency.} The consistency result in Proposition~\ref{prop3-rct-est} requires correct specification of the CATE model (A5). In both OSCAR and R-OSCAR, CATE estimation involves combining estimated outcome means and discrepancy functions. Thus, consistency of these components must be connected to the overall consistency of the CATE estimator. We now examine the consistency behavior of OSCAR and R-OSCAR separately.

\begin{enumerate}[leftmargin=12pt]
\item \textbf{OSCAR Estimator.} When both the observational outcome means $\mu^o(x, a)$ and the discrepancy functions $\delta_a(x)$ are correctly specified, their sum $\hat{\mu}^o_a(x) + \hat{\delta}^j_a(x)$ consistently estimates the true RCT outcome mean $\mu^r_a(x)$. Consequently, their contrast $\hat{\tau}_{\text{OSCAR}}(x)$ is consistent for the RCT’s CATE. However, if either $\mu^o(x, a)$ or $\delta_a(x)$ is misspecified, then both CMO and CATE may be misspecified. While CMO misspecification alone affects only efficiency, misspecification of the CATE leads to bias and inconsistency in $\hat{\tau}_{\text{OSCAR}}(x)$.

\item \textbf{R-OSCAR Estimator.} To mitigate OSCAR’s reliance on correct specification of multiple functions, R-OSCAR introduces a separate CATE discrepancy function. This reduces the number of functions that must be correctly specified for consistent CATE estimation from four to one. In this framework, misspecification of $\mu^o(x, a)$ and $\delta^r_a(x)$ results in variance inflation through CMO misspecification, but does not compromise CATE consistency if the CATE estimation step is correctly specified. Here, two estimation strategies are possible:

\begin{itemize}
    \item \emph{Direct CATE Estimation.} One approach is to directly estimate CATE using a model class known to contain the true CATE function, as assumed in Proposition~\ref{prop3-rct-est}. This mirrors standard CATE estimation from RCTs, with the distinction that our method would incorporate calibrated OS outcome functions $\hat{\mu}^o_a(x) + \hat{\delta}^t_a(x)$ for augmentation.
    
    \item \emph{CATE Calibration (Our Approach).} Our preferred strategy is CATE calibration, in which we first form a preliminary CATE estimate using calibrated outcome means $\hat{\mu}^o_a(x) + \hat{\delta}^t_a(x)$, then refine this using a CATE discrepancy function learned from RCT data. Provided this discrepancy function is correctly specified, the resulting estimator is consistent (Table~\ref{tab-estimator}).
\end{itemize}
\end{enumerate}

In both approaches, CATE consistency depends on the correct specification of a single function: either the CATE itself or the CATE discrepancy function.

\noindent \textbf{Theoretical Advantages of R-OSCAR.} An additional benefit of R-OSCAR is its compatibility with sharper theoretical analysis. The prediction and estimation error bounds in Section~\ref{the-framework} require that the CMO and CATE be estimated using independent datasets. This assumption fails for OSCAR, where CMO and CATE are jointly learned, complicating rigorous analysis. In contrast, while R-OSCAR uses RCT data for both the outcome mean calibration and CATE calibration steps, theoretical guarantees can still be obtained through sample splitting and cross-fitting \citep{zeng2023efficient, chernozhukov2018double}. Specifically, the RCT dataset is divided into $K$ folds: one fold is used for calibration (Eq.~\eqref{e2}) and the remaining $K-1$ folds for CATE estimation (Eq.~\eqref{e3}). The final CATE estimator aggregates results across folds as
\[
\hat{\tau}^r(x) = \sum_a a \left[\hat{\mu}^o(x,a) + \frac{1}{K} \sum_{k=1}^K \hat{\delta}^k_a(x) \right] + \frac{1}{K} \sum_{k=1}^K \hat{\delta}^k(x).
\]

\paragraph{Single-arm OS as a special case.} The preceding construction uses OS observations from both treatment arms, because the OS outcome models $\hat\mu^o(\cdot,+1)$ and $\hat\mu^o(\cdot,-1)$ are estimated separately. In many applications the intervention exists only in the trial, leaving the OS with controls only. R-OSCAR specializes to this setting by replacing $\hat\mu^o(\cdot,+1)$ with the RCT-only treated outcome model $\hat\mu^r_{+1}(\cdot)$ (no calibration is needed since this estimate already lives on the RCT distribution), while retaining the OS-derived control model $\hat\mu^o(\cdot,-1)$ together with its calibration $\hat\delta_{-1}^t(\cdot)$. The CMO becomes
\[
m(x) ~=~ \pi^r_{-1}(x)\,\hat\mu^r_{+1}(x) ~+~ \pi^r_{+1}(x)\,\bigl[\hat\mu^o(x,-1) + \hat\delta_{-1}^t(x)\bigr],
\]
The final CATE-calibration step in line~7 of Algorithm~\ref{alg:doubly-calibrated} remains unchanged, and this single-arm adaptation is denoted \emph{R-OSCAR-1arm}. The construction matches hybrid-control and external-control designs \citep{Pocock1976,Liu2022,Kotalik2021}, which reduce the control-arm variance by borrowing from real-world controls; here, borrowing occurs through the control-arm outcome model, followed by R-OSCAR's CATE-calibration step. Consistency follows from the same argument as for R-OSCAR: by Proposition~\ref{prop1-cate-opt} the CATE pseudo-outcome is unbiased regardless of the choice of $m$, while a structured $\delta_{-1}$ produces efficiency gains over RACER.

\subsubsection{Theoretical justification of borrowing strength from an OS}\label{justif}
As shown in Section~\ref{the-framework}, improving the MSE of each treatment arm’s outcome mean reduces both prediction and estimation errors. Here, we justify how borrowing from OS data via our proposed method can lower this MSE relative to relying solely on RCT data. Specifically, we compare outcome mean estimation under RACER (using partial RCT data) and R-OSCAR (with two-stage calibration), focusing on per-arm MSE.

For clarity, assume correct model specification: each true function lies within the function class used to estimate it (e.g., $\mu^r_a(x) \in \mathcal{M}^r_a$). Under standard assumptions, the high-probability non-asymptotic MSE of an estimator $\hat{g}_n$ of $g \in \mathcal{G}$ with $n$ samples satisfies:
\(
\Delta^2_2(\hat{g}_n, g) \leq \mathcal{O}_p\left(\frac{c(\mathcal{G})}{n^\eta}\right),
\)
where $c(\mathcal{G})$ reflects function class complexity and $\eta$ the convergence rate. Such bounds typically arise from localized Rademacher complexity arguments. For example, in LASSO regression with sparsity $s$ and $p$ covariates, $c(\mathcal{G}) \approx s\log(p)$; in kernel methods, $c(\mathcal{G}) \approx B^2\text{tr}(K)$ for RKHS norm bound $B$ and kernel matrix $K$. 
The rate $\eta$ depends on the function class: $\eta = 1$ for parametric models, and $\eta = \frac{2\alpha}{2\alpha + p}$ for nonparametric Hölder classes with smoothness $\alpha$ in $p$ dimensions.

Assuming such bounds hold, we now state the following proposition:
\begin{proposition}[RACER vs. R-OSCAR]
\label{prop:two-step-vs-direct}
Assume standard regularity conditions and that the following non-asymptotic MSE bounds hold for per-arm outcome mean estimation:

\begin{enumerate}
  \item \textbf{Direct Estimation (RACER):} Using RCT data only,
  \(
  \Delta^2_2\left(\hat{\mu}_a^r,\, \mu_a^r\right) 
  \leq \mathcal{O}_p\left(\frac{c(\mathcal{M}_a^r)}{{n^r_a}^{\eta^r_a}}\right).
  \)

  \item \textbf{Two-Stage Calibration (R-OSCAR):} Using OS data with RCT-based calibration,
  \(
  \Delta^2_2\left(\hat{\mu}_a^{o,t},\, \mu_a^r\right) 
  \leq
  \mathcal{O}_p\left(\frac{c(\mathcal{M}_a^o)}{{n^o_a}^{\eta^o_a}}\right)
  +
  \mathcal{O}_p\left(\frac{c(\mathcal{D}_a^t)}{{n^r_a}^{\eta^t_a}}\right),
  \)
  where $\hat{\mu}_a^{o,t}(x) = \hat{\mu}^o(x, a) + \hat{\delta}^t_a(x)$.
\end{enumerate}

Then, borrowing from OS data will lead to CATE estimation improvement  when the following inequality holds for some constants $C_1, C_2, C_3 > 0$: 
\(
\forall a:\quad 
C_1 \frac{c(\mathcal{M}_a^o)}{{n^o_a}^{\eta^o_a}}
+
C_2 \frac{c(\mathcal{D}_a^t)}{{n^r_a}^{\eta^t_a}}
<
C_3 \frac{c(\mathcal{M}_a^r)}{{n^r_a}^{\eta^r_a}}.
\)
\end{proposition}
\vspace{6pt}
The result follows directly: if the inequality holds, then the per-arm outcome means are more accurately estimated under R-OSCAR than under RACER,
\(
\Delta^2_2\left(\hat{\mu}_a^{o,t},\, \mu_a^r\right) < 
\Delta^2_2\left(\hat{\mu}_a^r,\, \mu_a^r\right),
\)
leading to a more accurate estimate of the CMO and, in turn, improved CATE estimation.

To interpret the proposition, suppose $C_1 = C_2 = C_3$ and that the RCT and OS model classes have comparable complexity, i.e., $c(\mathcal{M}_a^o) \approx c(\mathcal{M}_a^r)$, with identical convergence rates $\eta^o_a = \eta^r_a = \eta^t_a = \eta$. The inequality then reduces to:
\(
\frac{c(\mathcal{M}_a^r)}{{n^o_a}^{\eta}}
+
\frac{c(\mathcal{D}_a^t)}{{n^r_a}^{\eta}}
< 
\frac{c(\mathcal{M}_a^r)}{{n^r_a}^{\eta}}.
\)
This condition is satisfied when (i) $n^o_a \gg n^r_a$, i.e., the OS model benefits from a much larger sample size, and (ii) $c(\mathcal{D}_a^t) < c(\mathcal{M}_a^r)$, meaning the discrepancy function is simpler than the full outcome model. That is, by estimating $\hat{\mu}^o(x,a)$ from abundant OS data and only calibrating a simpler discrepancy $\hat{\delta}^t_a(x)$ using the limited RCT data, R-OSCAR achieves lower error than directly estimating $\mu_a^r(x)$ from the RCT alone.

Crucially, this does not require $\mathcal{M}_a^o \subseteq \mathcal{M}_a^r$, the OS model class can be more complex (e.g., a neural network) so long as the OS sample size is large enough to keep its error term small, i.e., $c(\mathcal{M}_a^o)/{n^o_a}^{\eta} \lesssim c(\mathcal{M}_a^r)/{n^r_a}^{\eta}$. If the discrepancy function belongs to a much simpler class $\mathcal{D}_a^t$ such that $c(\mathcal{D}_a^t) \ll c(\mathcal{M}_a^r)$, the setup resembles transfer learning: a large source dataset establishes a strong baseline, and only a lightweight correction is learned in the target domain.

\begin{example}[Sparse Linear Model] \label{example}
Consider a setting where:
\(\mu_a^r(x) = x^\top \beta_a^r\), \(\mu^o(x,a) = x^\top \beta_a^o\),
\(\delta_a^t(x) = x^\top (\beta_a^r - \beta_a^o).\)
Assume $\beta_a^r, \beta_a^o \in \mathbb{R}^p$, and their difference $\beta_a^r - \beta_a^o$ is $s$-sparse (with $s \ll p$). Then, both $\mathcal{M}_a^r$ and $\mathcal{M}_a^o$ (linear functions in $\mathbb{R}^p $) have complexity $c(\mathcal{M}_a^r) \approx c(\mathcal{M}_a^o) \approx p \log p$, while $\mathcal{D}_a^t$ (all $s$-sparse linear functions) has complexity $c(\mathcal{D}_a^t) \approx s \log(p)$.
Using LASSO for estimation, the estimation errors are approximately:
$
\Delta^2_2(\hat{\mu}_a^r, \mu_a^r)
\approx
\frac{p \log(p)}{n_a^r}
$
and
$
\Delta^2_2(\hat{\mu}_a^{o,t}, \mu_a^r)
\approx
\frac{p \log(p)}{n_a^o}
+
\frac{s \log(p)}{n_a^r}.
$
When $n_a^o \gg n_a^r$ and $s \ll p$, the two-stage R-OSCAR estimator achieves a substantially smaller estimation error than the direct RACER estimator, confirming the benefit of borrowing strength from OS data. Specifically, if $\forall a: n_a^o \gtrsim (p/(p-s)) \cdot n_a^r$, then R-OSCAR will have lower estimation error for the CMO than RACER.
\end{example}

\paragraph{Behavior when the structured-discrepancy assumption fails.} Proposition~\ref{prop1-cate-opt} implies that, under the RCT's design (A1--A4), the CATE pseudo-outcome regression remains unbiased for $\tau^r(x)$ for \emph{any} fixed augmentation $m(\cdot)$, including one constructed from a severely biased or misspecified OS. Failure of the structured-discrepancy regime therefore affects efficiency, not bias. The relevant quantity is the discrepancy-class complexity, $c(\mathcal{D}_a^t)$, rather than the magnitude of $\delta_a$: a sparse-but-large discrepancy still satisfies $c(\mathcal{D}_a^t) \ll c(\mathcal{M}_a^r)$ and therefore remains in the regime where Proposition~\ref{prop:two-step-vs-direct}'s threshold inequality continues to hold, so the regime of Example~\ref{example} applies; a dense discrepancy with $c(\mathcal{D}_a^t) \approx c(\mathcal{M}_a^r)$ makes the inequality fail, so R-OSCAR's variance approaches RACER's rather than dropping below it. Section~\ref{sec:severe-bias} characterizes this regime empirically, and the diagnostic of Section~\ref{sec:diagnostic} defaults to RACER when the borrowing condition fails.

\subsection{A cross-fitted RCT diagnostic for deciding when to borrow from the OS}\label{sec:diagnostic}

The structured-discrepancy condition used in Section~\ref{robust-est} and quantified by Proposition~\ref{prop:two-step-vs-direct} is a sufficient condition under which R-OSCAR improves on the RCT-only RACER estimator, but it cannot be checked directly in an application. The operational question is therefore not whether $\delta_a$ has the posited structure, which is generally untestable; it is whether the OS-calibrated outcome models predict held-out RCT outcomes better than the RCT-only outcome models. The diagnostic formulates that question as a cross-fitted comparison within the RCT and evaluates it through a finite-sample paired hypothesis test.

\paragraph{Procedure.}
Partition the RCT index set $\{1,\dots,n^r\}$ into $K$ disjoint folds
$\mathcal{I}^1,\dots,\mathcal{I}^K$ (we use $K=5$), and let
$\mathcal{I}^{-k}=\{1,\dots,n^r\}\setminus\mathcal{I}^k$. For each fold
$k \in \{1,\dots,K\}$:
\begin{enumerate}[leftmargin=12pt, itemsep=0pt]
\item On the $K-1$ training folds, fit the RCT-only per-arm outcome model $\hat\mu^{r,-k}_a(x)$ used inside RACER, and the OS-calibrated per-arm outcome model $\hat\mu^{o,t,-k}_a(x)$ used inside R-OSCAR.
\item On the held-out fold $k$, compute the paired squared-error difference for each subject $i$:
\begin{equation} \label{eq:diagnostic}
\forall i \in \mathcal{I}^k: D_i^k = \left(Y_i^r - \hat\mu^{r,-k}_{A_i}(X_i^r)\right)^2 - \left(Y_i^r - \hat\mu^{o,t,-k}_{A_i}(X_i^r)\right)^2.
\end{equation}
\end{enumerate}
The statistic aggregates $\{D_i^k\}$ across folds, weights each subject as $D = \sum_{k=1}^K \sum_{i\in \mathcal{I}^{k}} \pi_{-A_i}^2 D^k_i$ so the diagnostic matches each arm's contribution to pseudo-outcome variance, and uses a one-sided $1-\alpha$ percentile bootstrap lower confidence bound for $\E[D]$. The decision rule is deliberately conservative: R-OSCAR is recommended as the primary estimator only when this lower bound exceeds zero; otherwise the diagnostic defaults to RACER and reports R-OSCAR as a sensitivity analysis.

The diagnostic is not a test of the structural assumption on $\delta_a$. It tests the predictive implication relevant for CATE estimation: on randomized held-out data, whether OS-calibrated outcome means predict RCT outcomes more accurately than RCT-only outcome means. That condition is the one under which Proposition~\ref{prop4-cmo-convergence} guarantees a smaller CMO error and, by Theorem~\ref{theo3-risk-bound}, a smaller CATE risk bound. If it fails, the diagnostic defaults to RACER; the pseudo-outcome unbiasedness in Proposition~\ref{prop1-cate-opt} rules out bias from the failed borrowing attempt beyond the cost of running the diagnostic. Section~\ref{sec:diagnostic-sim} evaluates this behavior in simulation.

\subsection{Using the general framework under linear outcome means with sparse discrepancy} \label{on-outcome-shift}
We now specialize our general framework to a parametric setting with linear potential outcome means and sparse linear mean discrepancies between populations. Specifically, we assume linearity of the RCT potential outcome means $\mu_{a}^{r}(x) = \langle \gamma^r_a, x \rangle$ for both treatments $a \in \{-1, +1\}$, which implies linearity of the CATE ($\tau^r(x)$), the CMO ($\mu^r(x)$), and the function class $\cF$ in \eqref{eq-cate}. Similarly, we assume the OS population's outcome means follow a linear form $\mu^o(x, a) = \langle \gamma^o_a, x \rangle$. The mean discrepancy function is defined as $\delta_a(x) = \mu_{a}^{r}(x) - \mu^o(x, a) = \langle \delta_a, x \rangle$, where $\delta_a \equiv \gamma^r_a - \gamma^o_a$ is a sparse vector. 
For simplicity, we assume the RCT employs full randomization, where $\pi_{a}^r(x^r)=\pi_{a}^r$. 
\subsubsection{The OSCAR for linear outcome mean with a sparse discrepancy parameterization} \label{no-calib}
The OSCAR estimation proceeds in two stages: first, estimating the OS outcome means $\hmu^o\left(\cdot, a\right)$ using LASSO regression:
\begin{align} \label{olin}
\hat{\gamma}_a^o = \argmin_\gamma \frac{1}{n_a^o}\sum_{i: A_i^o = a}\left[ Y_i^o - \gamma^T X^o_i \right]^2 + \lambda^o_a\left\|\gamma\right\|_{1},
\end{align}
where $\gamma \in \reals^{p}$ and the estimated outcome mean function is $\hmu^o\left(x, a\right) \equiv \langle \hat{\gamma}_a^o, x \rangle$. 
For the second step, we instantiate \eqref{delta-estimation} to learn CMO and CATE jointly through  the estimation of the sparse mean discrepancy coefficient vectors:
\begin{align} \nonumber
(\hdelta_{-1}^j, \hdelta_{+1}^j)= \argmin_{(d_{-1}, d_{+1})} 
& \frac{1}{n^r}\sum_{i=1}^{n^r}\left[ \frac{A_i^r}{\pi^r_{A_i^r}} \left(Y_i^r - \sum_{a} \pi^r_{-a} (\hat{\gamma}_a^o +d_{a})^T X^r_i \right) -\sum_{a} a (\hat{\gamma}_a^o +d_{a})^T X^r_i \right]^2 \\ \label{rlin}
&+\sum_{a} \lambda_a^j \|d_{a}\|_{1}.
\end{align}
The objective in \eqref{rlin} can be simplified by recognizing that it is separable in the $d_a$ terms:
\begin{align} \label{l1}
\hdelta_{a}^j = \argmin_{d_a} \frac{1}{n^r_a}\sum_{i: A_i^r = a} & \left[ \left( \frac{a}{\pi^r_{a}} Y_i^r - a\left(1 + \alpha^a\right) \hat{\gamma}_a^{o^T} X^r_i\right) -a \left(1 + \alpha^a\right) d_{a}^T X^r_i \right]^2 + \lambda_a^j \|d_{a}\|_{1},
\end{align}
where $\alpha \equiv \frac{\pi^r_{-1}}{\pi^r_{+1}}$. The optimization in \eqref{l1} represents two separate $\ell_1$-regularized regressions, each using data from the corresponding treatment arm $a$ to estimate $\delta_a$. In total, our approach requires solving four LASSO problems: one for each treatment arm in the OS data \eqref{olin} and one for each discrepancy term using RCT data in \eqref{l1}. 
All tuning parameters are chosen via cross validation.

\subsubsection{The Robust OSCAR for possibly misspecified models} \label{with-calib}
To enhance the robustness of OSCAR against model misspecification, we introduce a two-stage calibration process. 
First, we estimate the per-treatment-arm linear coefficient vectors from the OS data, following  \eqref{olin}. Then, we calibrate the OS outcome means to the RCT outcomes by estimating a sparse mean discrepancy vector for each treatment arm:
\begin{align} \label{e2}
    \hdelta_{a}^t
    = \underset{d_{a}}{\argmin} \frac{1}{n_{a}^r} \sum_{i: A_i^r = a}\left[Y_{i}^r - \left(\hat{\gamma}_{a}^o + d_{a}\right)^{T} X_{i}^r\right]^{2} 
    + \lambda_a^t\left\|d_{a}\right\|_{1}.
\end{align}
This step produces a CMO and a preliminary CATE estimate tailored for the RCT. In the second stage, we refine the preliminary CATE by estimating a sparse linear discrepancy function to correct any residual misspecification using the calibrated CMO as the augmentation function:
\begin{align} \label{e3}
    \hdelta 
    = \underset{d}{\argmin} \frac{1}{n^r} \sum_{i=1}^{n^r}\left[\frac{A_{i}^r}{\pi_{A_{i}^r}^r}\left(Y_{i}^r
    - \sum_{a} \pi_{-a}^r \hat{\gamma}_{a}^{o,t^{T}} X_{i}^r\right) 
    - \left(\hat{\gamma}_{+1}^{o,t} - \hat{\gamma}_{-1}^{o,t} + d \right)^{T} X_{i}^r\right]^{2} + \lambda \|d\|_{1},
\end{align}
where $\hgamma_a^{o,t} \equiv \hgamma_a^o + \hdelta_a^t$ represents the calibrated OS coefficients for arm $a$, and $\hmu^{o,t}(x) = \sum_a \pi_{-a}^r (\hgamma_{a}^o + \hdelta_a^t)^{T} x$
is the calibrated CMO. The $\ell_1$-norm regularization induces sparsity in the discrepancy estimates. The final CATE estimate is given by: $\htau_{\text{R-OSCAR}}(x) = \langle [(\hgamma_{+1}^o + \hdelta_{+1}^t) - (\hgamma_{-1}^o + \hdelta_{-1}^t)] + \hdelta, x \rangle$,
where $\hdelta_a^t$ is estimated during the outcome mean calibration step, and $\hdelta$ is estimated during the CATE calibration step. 
If the preliminary CATE estimate is already accurate, the calibration step shrinks all elements of $\hdelta$ to zero, avoiding unnecessary corrections.

\section{Numerical Experiments}
\label{simulation-results}

We conduct a comprehensive set of simulation studies to evaluate the performance of the proposed CATE estimators under diverse conditions, including covariate and outcome shifts, model misspecification, and unmeasured confounding. All simulations are based on a common generative setup described below, with specific variations introduced in each experiment.

\subsection{Simulation Setup}
\label{simulation-setup}

Each simulation generates paired OS and RCT datasets. Covariates \( x \in \mathbb{R}^{100} \) are drawn from a multivariate Gaussian distribution with zero mean and a dense covariance matrix (i.e., all off-diagonal entries are nonzero), inducing correlation among features. Potential outcomes are defined separately for each treatment arm, which allows us to compute the ground truth CATE before treatment assignment.

In the OS dataset, treatment is assigned via a logistic model using ten randomly selected covariates. The model includes an intercept term adjusted to ensure that at least one-third of samples receive the treatment. Treatment assignment in the RCT is randomized with probability \( \pi^r_1(x) = 0.5 \).
For each arm \( a \in \{-1, 1\} \) in the OS, the potential outcome is generated linearly as
$
y^o(a) = \langle \beta^o_a, x \rangle + \epsilon,
$
where the coefficient vector \( \beta^o_a \) has a support size of 10, with non-zero entries drawn from \( \text{Uniform}(-2/3, -1/3) \cup \text{Uniform}(1/3, 2/3) \). The noise term \( \epsilon \sim \mathcal{N}(0, 1/9) \).

To introduce domain shift in the RCT, we simulate both covariate and outcome shifts:
\begin{itemize}[leftmargin=12pt, itemsep=0pt]
    \item \textbf{Covariate shift:} Ten randomly selected covariates are shifted by adding a value sampled from \( \text{Uniform}(-1/2, -1/4) \cup \text{Uniform}(1/4, 1/2) \).
    \item \textbf{Outcome shift:} Two non-zero coefficients in each \( \beta^r_a \) are perturbed by adding values from \( \text{Uniform}(-1, -1/2) \cup \text{Uniform}(1/2, 1) \).
\end{itemize}

\begin{table}
\centering
\caption{\small \textbf{CATE estimation with data simulated from the proposed model.} At three-decimal precision, OSCAR surpasses all alternatives, with R-OSCAR a close second. Notably, OSCAR and R-OSCAR with $n_r = 250$ achieve comparable or better performance than the RCT-only methods (Naive and RACER) with $n_r = 1000$, demonstrating a 75\% reduction in required RCT sample size when leveraging OS data.}
\label{tab-outcome}

\begin{tabular*}{\columnwidth}{@{}l@{\extracolsep{\fill}}c@{\extracolsep{\fill}}c@{\extracolsep{\fill}}c@{}}
\hline
\multicolumn{4}{c}{\textbf{Number of Samples per Study}} \\
\hline
$n_r$ & 250 & 500 & 1000 \\
$n_o$ & 10,000 & 10,000 & 10,000 \\
\hline
\textbf{Method} & \multicolumn{3}{c}{\textbf{RMSE of CATE Estimation}} \\
\hline
Naive, (\ref{naive-est}) & {1.25 $\pm$ 0.29} & {1.03 $\pm$ 0.28} & {0.74 $\pm$ 0.19} \\
RACER, (\ref{racer-est}) & {0.31 $\pm$ 0.04} & {0.21 $\pm$ 0.03} & {0.15 $\pm$ 0.02} \\
OSCAR, (\ref{no-calib}) & \textbf{0.15 $\pm$ 0.03} & \textbf{0.11 $\pm$ 0.02} & {0.08 $\pm$ 0.02} \\
R-OSCAR, (\ref{with-calib}) & {0.16 $\pm$ 0.03} & \textbf{0.11 $\pm$ 0.02} & \textbf{0.08 $\pm$ 0.01} \\
R-OSCAR, Cross Fit & {0.18 $\pm$ 0.03} & {0.12 $\pm$ 0.02} & {0.09 $\pm$ 0.01} \\
\hline
\end{tabular*}
\vspace*{-12pt}
\end{table}

\subsection{Estimators Compared}
\label{estimators-compared}

We compare five estimators for the CATE. All methods use LASSO regression with tuning parameter chosen via 10-fold cross-validation via \texttt{cv.glmnet} for model fitting unless otherwise specified:
\begin{itemize}[leftmargin=12pt, itemsep=0pt]
    \item \textbf{Naive (\( \hat{\tau}_{\text{Naive}} \))}: Fits arm-specific LASSO regressions using only RCT data, then contrasts predictions to estimate CATE.
    \item \textbf{RACER (\( \hat{\tau}_{\text{RACER}} \))}: Estimates potential outcome means from RCT data and plugs them into the CMO expression in Equation~\eqref{instance}.
    \item \textbf{OSCAR (\( \hat{\tau}_{\text{OSCAR}} \))}: Uses OS data to estimate outcome models via Equation~\eqref{olin}, then solves for the CMO and CATE jointly using Equation~\eqref{l1}.
    \item \textbf{R-OSCAR (\( \hat{\tau}_{\text{R-OSCAR}} \))}: Calibrates OS-based outcome models to the RCT population using Equation~\eqref{e2}, then estimates CATE using Equation~\eqref{e3}.
    \item \textbf{Cross-Fitted R-OSCAR}: Applies five-fold cross-fitting to R-OSCAR: one fold is used for calibration, and the remaining for estimation; final results average across folds.
\end{itemize}

\subsection{Varying Sample Sizes Under Linear Outcome Models}

We begin by evaluating performance under a correctly specified linear outcome model, focusing on how RCT and OS sample sizes affect CATE estimation accuracy.

\paragraph{Varying RCT size.} In the first experiment, we fix the OS sample size at \( n^o = 10{,}000 \) and vary the RCT sample size \( n^r \) from 250 to 1,000. Table~\ref{tab-outcome} reports the RMSE of CATE estimation averaged over 100 replicates.  As expected, estimation accuracy improves with larger RCT sample sizes. However, both OSCAR and R-OSCAR consistently outperform Naive and RACER, even when the latter have access to four times more RCT data (compare the last column of Naive and RACER rows with the first column of OSCAR and R-OSCAR in Table~\ref{tab-outcome}). The cross-fitted R-OSCAR performs comparably to its non-cross-fitted counterpart but is computationally more expensive, so we omit it in later experiments.

\paragraph{Varying OS size and covariate dimension.} 
We next fix the RCT sample size at \(n^r = 300\) and systematically vary both the OS sample size \(n^o\) and the covariate dimension \(p\).  In each setting, we designate \(0.10p\) covariates as true effect modifiers in the OS population, and we allow \(s = 0.02p\) of modifiers to differ in the RCT model, which means, the number of effect modifiers in the RCT will be between \(\lceil0.10p\rceil\) and \(\lfloor0.12p\rfloor\) effect modifiers, so all of the OS modifiers are shared.  For small \(p\), this means only zero or one modifier differs (e.g.\ when \(p=50\), there are 5 OS modifiers and \(s=1\), so the RCT has 5 or 6 modifiers).

Figure~\ref{fig:overall}A plots the RMSE difference between R‑OSCAR and RACER as a function of \(n^o\) (averaged over 100 replicates), where negative values indicate superior performance by R‑OSCAR. The results align closely with the theoretical predictions from Example~\ref{example}, confirming two key behaviors:
\begin{enumerate}[leftmargin=12pt,itemsep=0pt]
  \item \textbf{Threshold behavior around \(n^o = 300\).} R‑OSCAR begins to outperform RACER once \(n_a^o \approx n_a^r\), matching the theoretical condition
  \(n_a^o \;\gtrsim\;\frac{p}{p - s}\,n_a^r.\)
  With \(s = 0.02p\), this simplifies to \(n_a^o \approx n_a^r\).
  \item \textbf{Increasing advantage with higher dimension.} For fixed sample sizes and \(s = 0.02p\), Example~\ref{example} shows that the MSE gap between RACER and R‑OSCAR grows linearly with \(p\), specifically in proportion to \({p}(0.98(1/n^r) - 1/n^o)\). As a result, the relative RMSE benefit of R‑OSCAR increases approximately with \(\sqrt{p}\), which is reflected in the widening performance gap at higher dimensions in the figure.
\end{enumerate}

\begin{table}
\centering
\caption{\small \textbf{CATE estimation under model misspecification.} RMSE (with standard deviation) is reported for each method based on 100 replicates. “Without RF” columns correspond to settings where all methods—including baselines—use LASSO regression. “With RF” columns show results when Naive and RACER use random forest to model outcomes, while OSCAR and R-OSCAR remain fixed with LASSO. In both misspecification settings, R-OSCAR achieves the lowest RMSE. Notably, baseline methods perform better under LASSO-regularized regression than random forests.}
\label{tab-nonlinear}
\begin{tabular*}{\columnwidth}{@{}l@{\extracolsep{\fill}}cc@{\extracolsep{\fill}}cc@{}}
\hline
& \multicolumn{2}{c}{\textbf{Quadratic}} & \multicolumn{2}{c}{\textbf{Sinusoidal}} \\
\textbf{Method} & Without RF & With RF & Without RF & With RF \\
\hline
Naive & 1.99 $\pm$ 0.25 & 5.06 $\pm$ 0.33 & 0.90 $\pm$ 0.24 & 2.23 $\pm$ 0.67 \\
RACER & 1.54 $\pm$ 0.16 & 1.87 $\pm$ 0.12 & 0.32 $\pm$ 0.03 & 1.24 $\pm$ 0.15 \\
OSCAR & 1.73 $\pm$ 0.19 & -- & 0.32 $\pm$ 0.03 & -- \\
R-OSCAR & 1.50 $\pm$ 0.19 & -- & 0.28 $\pm$ 0.03 & -- \\
\hline
\end{tabular*}
\vspace*{-10pt}
\end{table}

\subsection{Robustness to Model Misspecification}

In this experiment, we test robustness to nonlinear effects by adding misspecified terms to the linear outcome model. For each effect modifier \( x_j \in \text{supp}(\beta^s_a) \), we generate outcomes as:
\(
y^s_a = \sum_j \beta^s_a(j) x_j + m^s_a(j) g(x_j) + \epsilon,
\)
where \( g(x_j) \) is either a quadratic term \( x_j^2 \) or sinusoidal term \( \sin(x_j) \), and \( m^s_a(j) \sim \text{Uniform}(0.25, 0.5) \). 

We fix the number of RCT and OS samples at \( n_r = 1{,}000 \) and \( n_o = 10{,}000 \), respectively, and repeat each simulation setting 100 times. In these misspecified scenarios, we allow Naive and RACER to use random forests for outcome modeling, while OSCAR and R-OSCAR deliberately retain LASSO, remaining misspecified by design. As shown in Table~\ref{tab-nonlinear}, R-OSCAR consistently outperforms all baseline variants—including both LASSO- and random forest-based implementations—demonstrating strong robustness to model misspecification. OSCAR performs slightly worse than RACER in the quadratic setting and comparably in the sinusoidal setting. This underscores the value of R-OSCAR’s two-stage design: by learning the mean discrepancy functions \( \delta_a(x) \) during calibration and then estimating the final CATE discrepancy \( \delta(x) \) separately, R-OSCAR maintains resilience to nonlinearity even when outcome models are misspecified.

Interestingly, in both the quadratic and sinusoidal settings, the LASSO versions of Naive and RACER outperform their random forest counterparts. This somewhat counterintuitive result likely stems from the challenge tree-based models face in capturing smooth nonlinear transformations—such as sine functions—in high-dimensional settings \citep{breiman2001random}. In contrast, LASSO continues to provide stable performance through effective regularization and sparsity induction.

\begin{figure}[t]
    \centering
    \includegraphics[width=\textwidth]{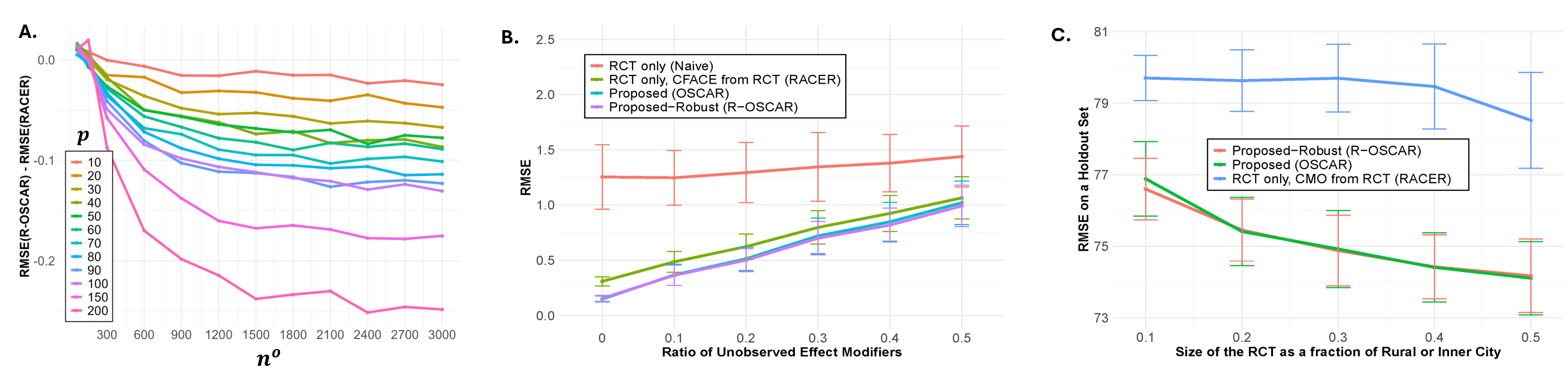}
    \caption{\small 
    \textbf{CATE estimation accuracy.}   
    \textbf{(A)} Difference in RMSE between R‑OSCAR and RACER as the OS sample size \(n^o\) increases (with fixed \(n^r=300\)).  Negative values indicate superior performance by R‑OSCAR.  Each curve corresponds to a different covariate dimension \(p\), where \(10\%\) of covariates are true effect modifiers and \(2\%\) of these modifiers differ between the RCT and OS models (so \(s=0.02p\)).  Two phenomena are noteworthy, both of which align with our theoretical results from Example~\ref{example}: (i) a clear threshold around \(n^o \approx n^r\), and (ii) an increasing advantage of R‑OSCAR at higher $p$. Points are averages over 100 replicates; error bars are omitted for clarity.
    \textbf{(B)} Sensitivity to unmeasured effect modifiers. To simulate latent confounding, a fraction of true effect modifiers is removed from the observed covariates. RMSE is reported for each method across 100 replicates. All methods degrade with increasing unobserved confounding, but R-OSCAR consistently achieves the lowest error.    
    \textbf{(C)} Results on the STAR dataset with varying RCT cohort sizes. Each point shows average RMSE across 30 replicates. Both OSCAR and R-OSCAR outperform the RCT-only baseline (RACER). The naive estimator is omitted due to substantially worse performance.}
    \label{fig:overall}
    \vspace*{-12pt}
\end{figure}

\subsection{Effect of Unmeasured Effect Modifiers}\label{sec:unmeas-modifiers}

In this experiment, we investigate the impact of unmeasured effect modifiers on CATE estimation. We fix the RCT and OS sample sizes at 250 and 10,000, respectively. Starting from the base linear outcome model, we randomly remove 10\% to 50\% of the true effect modifiers from both datasets, corresponding to the removal of up to 8 covariates (out of approximately 15 total modifiers in each replication). 

As shown in Figure~\ref{fig:overall}B, estimation performance deteriorates across all methods as more modifiers are omitted. R-OSCAR consistently outperforms the baselines, though its margin of improvement narrows as the number of unmeasured effect modifiers increases. Notably, even the removal of a single covariate (roughly 10\% of effect modifiers) leads to a sharp decline in performance for methods that rely on CMO estimation. In contrast, the Naive method exhibits relatively stable performance, likely because it does not depend on modeling outcome means across datasets. 
These findings highlight both the promise and limitations of leveraging OS data: while borrowing information can substantially improve accuracy, its effectiveness hinges on the completeness of key effect modifiers.

\subsection{Severe OS bias and unmeasured confounding}\label{sec:severe-bias}

The previous subsection varies the fraction of unmeasured effect modifiers, but it does not cover settings where OS bias is large enough for $\delta_a$ to fall outside the assumed structured class. This regime distinguishes smooth efficiency loss from risk inflation. We study it with two experiments, each using $n^r=250$, $n^o=10{,}000$, $p=100$ and $100$ Monte Carlo replicates per cell.

\paragraph{Experiment 1: varying OS--RCT shift and unmeasured confounding strengths.} In the linear DGP of Section~\ref{simulation-setup} with $p=100$ covariates, we vary two parameters that control the distance between the RCT and OS outcome models. The first is the \emph{discrepancy support fraction} $s/p \in \{0.02, 0.10, 0.20, 0.40\}$, where $s = \lvert\mathrm{supp}(\beta_a^r - \beta_a^o)\rvert$ is the number of coordinates on which the RCT and OS linear outcome models differ, so larger $s/p$ moves $\delta_a$ from sparse toward dense. The second is the \emph{unmeasured-modifier fraction} $\eta \in \{0, 0.1, 0.3, 0.5, 0.7\}$, defined as the fraction of true effect-modifier covariates removed from $X$ before any estimator sees the data (the procedure used in isolation in Section~\ref{sec:unmeas-modifiers}, now varied jointly with $s/p$). The $4\times 5$ grid ranges from settings where the structured-discrepancy assumption is plausible ($s/p$ small, $\eta$ small) to settings where it fails ($s/p$ large, $\eta$ large). For each cell we report the median of $\text{RMSE}(\text{R-OSCAR})-\text{RMSE}(\text{RACER})$, the rank correlation of $\hat\tau$ with the true $\tau$, and the rate at which the diagnostic of Section~\ref{sec:diagnostic} selects R-OSCAR. Figure~\ref{fig:severe-bias}A,B summarizes the results. Across the grid, R-OSCAR's median RMSE almost never exceeds RACER's: the maximum overhead is approximately $0\%$ at $(s/p=0.4,\, \eta=0.5)$, where the two estimators are statistically indistinguishable. The diagnostic's selection rate declines monotonically from $1.00$ in the low-bias corner to $0.13$ in the most heavily biased corner.

\paragraph{Experiment 2: latent-confounder injection.} Experiment 1 targets the structural complexity of $\delta_a$; Experiment 2 isolates unmeasured confounding in the OS. Starting from the linear DGP of Section~\ref{simulation-setup}, each OS subject $i$ receives a scalar latent variable $U_i \sim \mathcal{N}(0, 1)$ that is omitted from the covariates available to the estimators. The OS treatment-assignment logit becomes $\mathrm{logit}(\pi^o_{+1}(x_i, U_i)) = \mathrm{logit}(\pi^o_{+1}(x_i)) + \gamma\, U_i$, and the OS potential outcomes become $y_i^o(a) = \langle \beta_a^o, x_i\rangle + \kappa\, U_i + \epsilon$ for $a \in \{-1, +1\}$, so $U$ is a classical unmeasured confounder: when $\gamma \neq 0$ and $\kappa \neq 0$ it shifts both $A$ and $Y$ in the OS, making the OS-fitted outcome means $\hat\mu^o(x,a)$ biased for the population conditional means. Conditioning on the OS arm $\{A^o=a\}$ and applying Bayes within the selection logit, the arm-wise OS outcome mean is $\mu_a^o(x) = \langle \beta_a^o, x\rangle + \kappa\, g_\gamma^a(x)$ with $g_\gamma^a(x) := \mathbb{E}^o[U \mid X=x, A^o=a]$ a smooth nonlinear function of $x$, so the induced discrepancy $\delta_a(x) = \langle \beta_a^r - \beta_a^o, x\rangle - \kappa\, g_\gamma^a(x)$ (for $a \in \{-1,+1\}$) acquires a nonlinear, confounding-induced piece whose magnitude scales with $\kappa$. Increasing $\kappa$ therefore strengthens the violation of $\delta_a$'s assumed sparse-linear structure, and we track that strength by varying $\kappa$. The RCT is unchanged from Section~\ref{simulation-setup}. We fix $\gamma = 1$ and vary $\kappa \in \{0, 0.25, 0.5, 1, 2, 4\}$ to trace efficiency loss as confounding strengthens. Figure~\ref{fig:severe-bias}C,D and Table~\ref{tab:latent-u} show the pattern implied by the construction: RACER's RMSE is invariant in $\kappa$ at $0.330$ across all six values, because the RCT is unmodified, whereas R-OSCAR's RMSE rises smoothly from $0.204$ at $\kappa=0$ to $0.300$ at $\kappa=4$ without diverging. The diagnostic shifts from selecting R-OSCAR with probability $1.00$ at $\kappa\le 0.5$ to selecting RACER at $\kappa=4$ ($P(\text{R-OSCAR}) = 0.19$).

\begin{figure}[!tbp]
    \centering
    \begin{subfigure}[t]{0.48\textwidth}
        \centering
        \includegraphics[width=\textwidth]{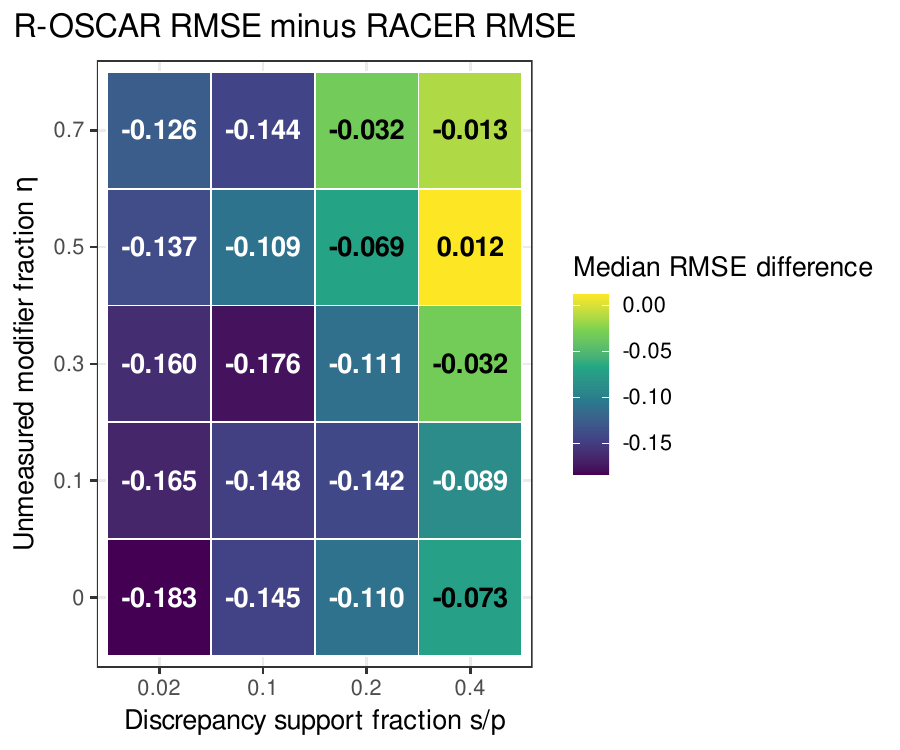}
        \caption{\small Phase 1: median difference in $\text{RMSE}(\text{R-OSCAR})-\text{RMSE}(\text{RACER})$ over the $4{\times}5$ bias grid.}
        \label{fig:severe-bias-A}
    \end{subfigure}\hfill
    \begin{subfigure}[t]{0.48\textwidth}
        \centering
        \includegraphics[width=\textwidth]{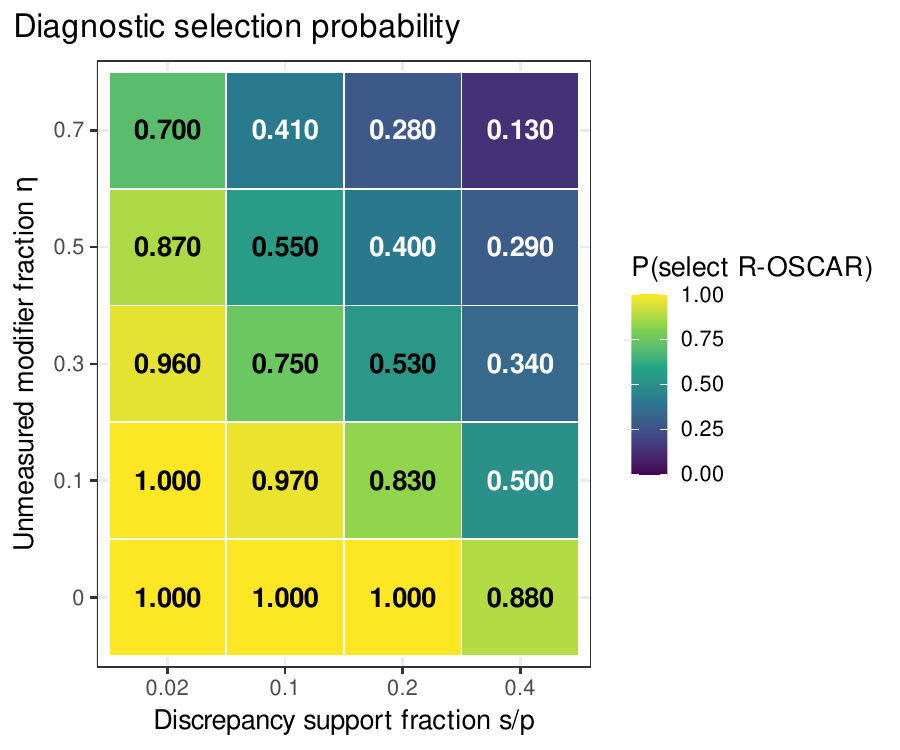}
        \caption{\small Phase 1: $P(\text{diagnostic selects R-OSCAR})$ over the same grid.}
        \label{fig:severe-bias-B}
    \end{subfigure}\\[6pt]
    \begin{subfigure}[t]{0.48\textwidth}
        \centering
        \includegraphics[width=\textwidth]{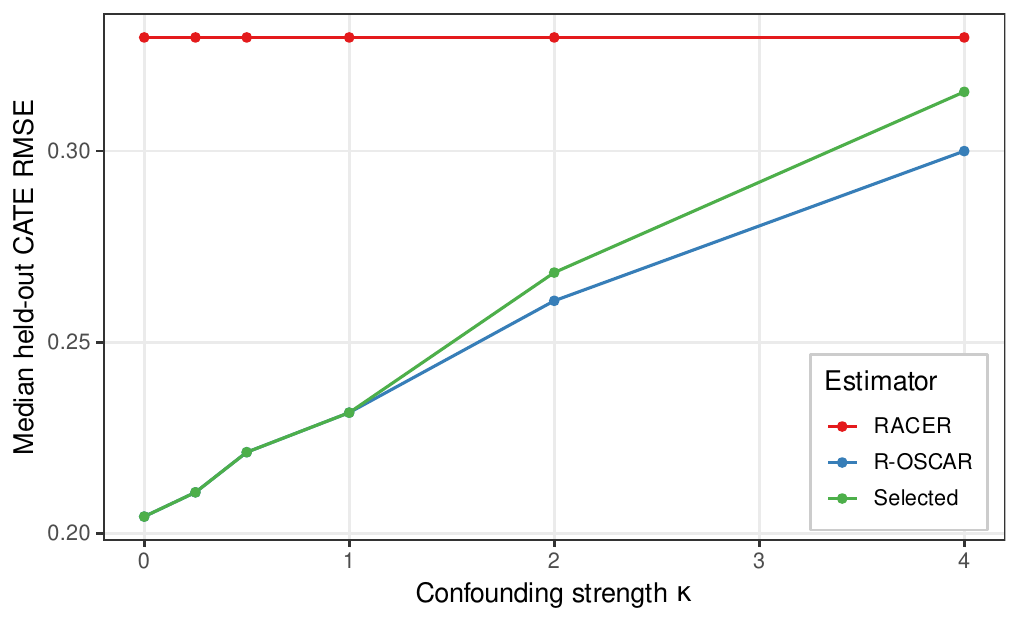}
        \caption{\small Phase 2: RMSE as a function of latent-confounder strength $\kappa$ for RACER, R-OSCAR, and the diagnostic-selected estimator.}
        \label{fig:severe-bias-C}
    \end{subfigure}\hfill
    \begin{subfigure}[t]{0.48\textwidth}
        \centering
        \includegraphics[width=\textwidth]{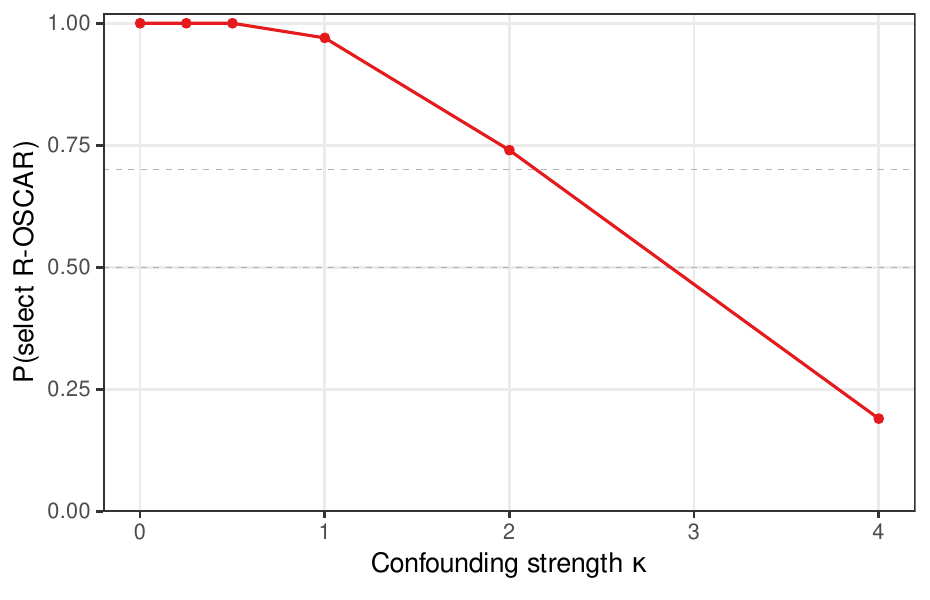}
        \caption{\small Phase 2: $P(\text{diagnostic selects R-OSCAR})$ as a function of $\kappa$.}
        \label{fig:severe-bias-D}
    \end{subfigure}
    \caption{\small \textbf{Severe OS bias and unmeasured confounding.} Phase 1 varies the structured-discrepancy class boundary in a 2D grid (panels A--B); Phase 2 adds a latent confounder into the OS only and traces RMSE and diagnostic behavior as a function of confounding strength $\kappa$ (panels C--D). R-OSCAR's median RMSE remains essentially at or below RACER's, and the diagnostic selection rate decreases monotonically with bias. All cells use $n^r=250$, $n^o=10{,}000$, $p=100$, and $100$ Monte Carlo replicates.}
    \label{fig:severe-bias}
\end{figure}

\begin{table}[t]
\centering
\caption{\small \textbf{Latent-confounder $\kappa$ curve (Phase 2).} Median RMSE of RACER (RCT-only, unaffected by $\kappa$), R-OSCAR, and the diagnostic-selected estimator, together with the diagnostic's R-OSCAR selection probability, as the OS-only confounding strength $\kappa$ varies. RACER's invariance follows from leaving the RCT unmodified; R-OSCAR degrades smoothly; the diagnostic transitions from full R-OSCAR selection at low $\kappa$ to RACER at high $\kappa$.}
\label{tab:latent-u}
\small
\begin{tabular*}{0.7\columnwidth}{@{\extracolsep{\fill}}ccccc@{}}
\hline
& \multicolumn{3}{c}{Median CATE RMSE} & \\
\cline{2-4}
$\kappa$ & RACER & R-OSCAR & Selected & $P(\text{R-OSCAR})$ \\
\hline
$0$    & $0.330$ & $0.204$ & $0.204$ & $1.00$ \\
$0.25$ & $0.330$ & $0.211$ & $0.211$ & $1.00$ \\
$0.5$  & $0.330$ & $0.221$ & $0.221$ & $1.00$ \\
$1$    & $0.330$ & $0.232$ & $0.232$ & $0.97$ \\
$2$    & $0.330$ & $0.261$ & $0.268$ & $0.74$ \\
$4$    & $0.330$ & $0.300$ & $0.316$ & $0.19$ \\
\hline
\end{tabular*}
\end{table}

\subsection{Empirical validation of the cross-fitted RCT diagnostic}\label{sec:diagnostic-sim}

The diagnostic is evaluated on the linear DGP of Section~\ref{simulation-setup}, varying the discrepancy support fraction over $s/p \in \{0,0.02,0.05,0.10,0.25,0.50,1.0\}$ with $p\in\{50,100,200\}$, $n^r=300$, $n^o\in\{60,150,300,600,1500,3000,10{,}000\}$, and $100$ Monte Carlo replicates per cell. In each replicate, the diagnostic is run on the RCT alone, with no access to the test set, and its decision selects between RACER and R-OSCAR. The selected estimator's CATE RMSE is compared with that of the \emph{oracle CATE estimator}, defined within each replicate as the member of $\{$Naive, RACER, R-OSCAR$\}$ with the smallest held-out RMSE against the true $\tau$.

Figure~\ref{fig:diagnostic} summarizes the diagnostic behavior. Panel A reports the diagnostic's R-OSCAR pick rate as a function of $s/p$, faceted by covariate dimension $p$. At the largest OS sample size $n^o = 10{,}000$, in the \emph{sparse-shift} regime $s/p \le 0.05$ the diagnostic selects R-OSCAR with mean probability $0.97$; in the \emph{dense-shift} regime $s/p \ge 0.5$ the mean probability is $0.40$. The pick rate increases with $p$ at fixed $s/p$, matching the theoretical pattern of Example~\ref{example}: the population RMSE gap $\mathrm{RMSE}(\text{RACER}) - \mathrm{RMSE}(\text{R-OSCAR})$ scales like $(p-s)\log p / n^r - p\log p / n^o$, which widens with $p$ at fixed $s/p$ and improves the diagnostic's signal. Within the dense-shift regime, the diagnostic separates partially-dense from fully-dense settings: at $s/p=0.5$ R-OSCAR's median RMSE is below RACER's in all three $p$ cells at $n^o=10{,}000$ (for example $0.540$ vs $0.582$ at $p=100$) and the diagnostic's pick rate is correspondingly elevated (mean $0.52$ across the three cells); at $s/p=1.0$, where every OS coefficient differs from its RCT counterpart, RACER wins on RMSE in all three $p$ cells and the diagnostic defaults to RACER (mean $P(\text{R-OSCAR})=0.23$). The resulting recommendation follows empirical loss rather than the asymptotic class boundary.

Panel B shows that the median-RMSE curve of the diagnostic-selected estimator nearly coincides with the oracle's across the bias grid at $n^o = 10{,}000$; the largest residual gap is at $p = 50$ in the intermediate $s/p$ range, under $7.3\%$ of the oracle RMSE at that cell. Across the full $147$-cell grid, the per-replicate median \emph{regret} $\mathrm{RMSE}(\text{selected}) - \mathrm{RMSE}(\text{oracle})$ is exactly $0$ in $114$ cells and stays below $10\%$ of the oracle RMSE at the same cell in the remaining $33$ (max absolute regret $0.025$). The \emph{false-borrow rate}, defined as the fraction of replicates in which the diagnostic recommends R-OSCAR but R-OSCAR underperforms RACER on the held-out test, averages $0.19$ over the dense-shift cells; these cases are concentrated in cells where R-OSCAR's median RMSE is lower than RACER's, indicating finite-sample variation around the population-level signal rather than systematic over-borrowing.

\begin{figure}[!htbp]
    \centering
    \begin{subfigure}[t]{0.85\textwidth}
        \centering
        \includegraphics[width=\textwidth]{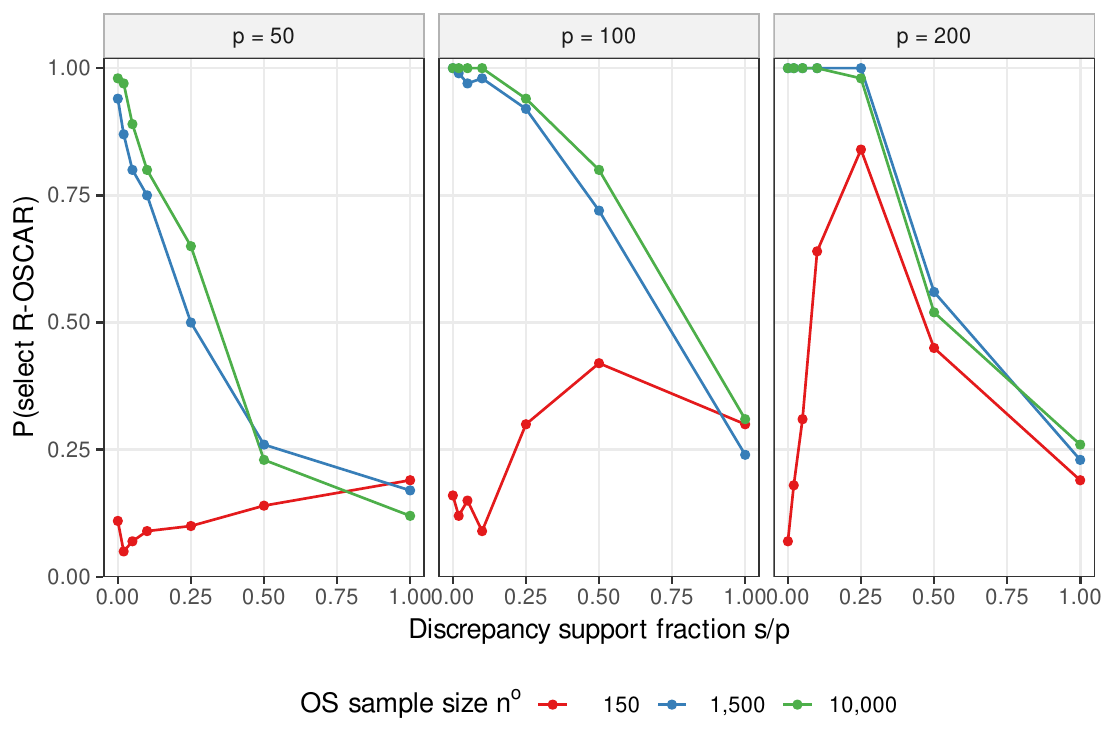}
        \caption{\small $P(\text{diagnostic selects R-OSCAR})$ as a function of the discrepancy support fraction $s/p$, faceted by covariate dimension $p$ and colored by OS sample size $n^o$.}
        \label{fig:diagnostic-A}
    \end{subfigure}\\[8pt]
    \begin{subfigure}[t]{0.85\textwidth}
        \centering
        \includegraphics[width=\textwidth]{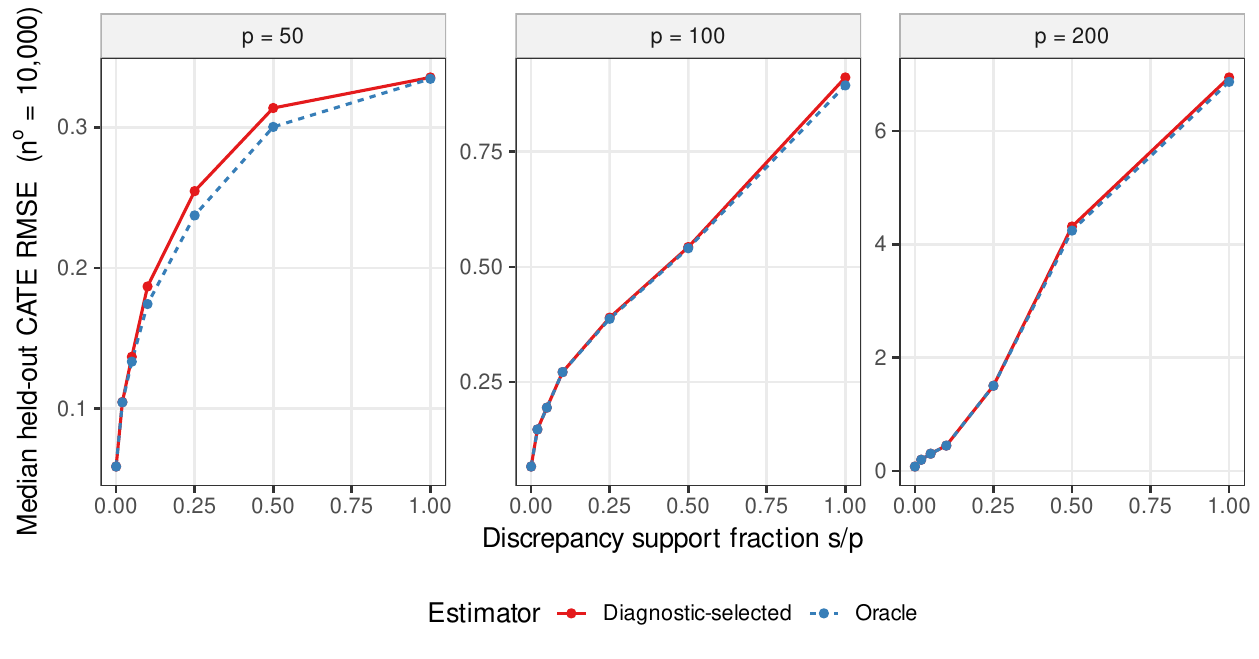}
        \caption{\small Median RMSE for the diagnostic-selected estimator and the oracle best-of-three, faceted by covariate dimension $p$, at $n^o = 10{,}000$. The curves nearly coincide across the bias grid; the small residual gaps at $p=50$ in the intermediate $s/p$ range reflect minority-replicate disagreements between the diagnostic and the oracle rather than systematic bias.}
        \label{fig:diagnostic-B}
    \end{subfigure}
    \caption{\small \textbf{Cross-fitted RCT diagnostic for OS borrowing.} The diagnostic selects R-OSCAR most often in sparse-shift regimes (panel A) and follows finite-sample RMSE as shifts become denser; the selected estimator nearly matches the oracle best-of-three on RMSE, with small residual gaps in a minority of cells (panel B). $n^r=300$, $100$ Monte Carlo replicates per cell.}
    \label{fig:diagnostic}
\end{figure}

\section{Real-World Experiments} \label{real-results}

The real-data evaluation has two components. Section~\ref{sec:star} re-analyzes the Tennessee STAR randomized study with a constructed observational cohort drawn from the same trial: the resulting \emph{semi-synthetic} setup follows \citet{kallus2018removing}, where confounding is induced by a known sampling rule and a ground-truth CATE is therefore available. STAR therefore tests whether R-OSCAR remains unbiased and recovers the known CATE. Section~\ref{sec:gps} is a fully external analysis of the Greenlight Plus pediatric-obesity trial (an RCT) paired with electronic-health-record data from Duke, UNC, and Vanderbilt that were collected separately under usual care. The EHR cohort contains controls only, prompting the one-arm adaptation R-OSCAR-1arm introduced in Section~\ref{robust-est}; the analysis is held-out RCT evaluation, so randomization continues to anchor the inference. The two analyses address distinct evidentiary questions: STAR evaluates recovery against ground truth; Greenlight Plus tests, on independently collected OS data, where borrowing helps the borrowed pathway and whether the diagnostic identifies the regimes in which it is and is not supported.

\subsection{Semi-synthetic STAR study with constructed observational confounding}\label{sec:star}

We evaluate our estimators using the Tennessee Student/Teacher Achievement Ratio (STAR) dataset, following the setup in \citet{kallus2018removing}. Their goal was to use RCT data to reduce bias in OS-based CATE estimates; ours is to reduce variance in RCT-based CATE estimation using OS data. While the inference targets differ, the population construction (see  below) is agnostic to this distinction, so we replicate their sampling strategy. This analysis is semi-synthetic because the observational cohort is drawn from the same underlying randomized trial rather than from an independently collected source; it tests R-OSCAR against a known ground-truth CATE.

STAR was a randomized study of class size effects on performance. Students were randomly assigned to small classes, regular classes, or regular classes with a teacher’s aide, and remained in these assignments through third grade. Performance was assessed via standardized tests in kindergarten, first, and third grade. We focus on 4,218 students with complete first-grade data, estimating the CATE of small (treatment) vs. regular (control) class assignment on average first-grade test scores, conditional on gender, race, birth date, free lunch status, and teacher ID.

To simulate limited RCT data, we randomly sampled a fraction $q \in \{0.1, 0.2, ..., 0.5\}$ of students from rural or inner-city schools (30 replicates per $q$). OS data was then constructed by including: (1) rural/inner-city students in regular classes not selected for the RCT; (2) rural/inner-city students in small classes not in the RCT whose scores were in the bottom 50\%; (3) all urban/suburban students in regular classes; and (4) urban/suburban students in small classes with scores in the bottom 50\%. This induces artificial confounding; see \citet{kallus2018removing}.

We compared four estimators of Table \ref{tab-estimator}. Following \citet{kallus2018removing}, we defined ground-truth CATE values based on full STAR data, but with respect to the RCT population.
%
Figure~\ref{fig:overall}C shows the results. $\htau_{\text{Naive}}$ performed substantially worse and is omitted from the plot for readability. Both OSCAR and R-OSCAR clearly outperform RACER, with nearly identical performance.

\subsection{Greenlight Plus trial with linked electronic-health-record data}\label{sec:gps}

Greenlight Plus provides the fully external counterpart to the semi-synthetic STAR analysis, pairing a randomized trial with a separately collected observational cohort. The Greenlight Plus trial \citep[GPS;][]{heerman2024digital} is a multi-site cluster-randomized study of a digital health-literacy intervention to prevent early-childhood obesity, conducted across six clinical sites with $n^r=860$ children randomized to clinic-only care versus clinic care plus a digital intervention. The observational source is a separately collected electronic-health-record (EHR) cohort of pediatric usual-care patients from three of the participating medical centers (Duke, UNC, and Vanderbilt). The EHR cohort is restricted to children with complete shared covariates, non-missing baseline WFLz (weight-for-length z-score at the earliest available well-child measurement, not birth weight), non-missing 24-month interpolated WFLz, and WFLz values in the physiologically plausible range $|z|\le 5$ (values outside this range are treated as data-entry errors and excluded). After these restrictions, the EHR cohort contains $n^o=8{,}867$ children (Duke: $6{,}604$; UNC: $433$; Vanderbilt: $1{,}830$). After applying the same filter to the RCT across all six sites, the analysis sample comprises $467$ randomized participants (Table \ref{tab:gps-samples}). The target estimand is the CATE of the digital intervention on 24-month WFLz, conditional on the shared baseline covariate vector
\[
X = \{\text{site},\,\text{sex},\,\text{race/ethnicity},\,\text{primary language},\,\text{insurance},\,\text{baseline WFLz}\},
\]
which is measured in both the trial and the EHR cohort. Table~\ref{tab:gps-samples} reports sample sizes after the plausibility filter.

The EHR cohort contains \emph{controls only} because the digital intervention exists only inside the trial, so the standard R-OSCAR procedure has no OS treated-arm outcome model to fit. The one-arm adaptation R-OSCAR-1arm from Section~\ref{robust-est} therefore uses the EHR cohort to supply $\hat\mu^o(x,-1)$, the RCT controls to calibrate $\hat\delta_{-1}^t$, and the RCT treated participants to supply $\hat\mu^r_{+1}(x)$ directly, and forms the CMO estimate as $m(x)=\pi^r_{-1}\hat\mu^r_{+1}(x)+\pi^r_{+1}[\hat\mu^o(x,-1)+\hat\delta_{-1}^t(x)]$, leaving the final CATE-calibration step unchanged.

\textbf{Setting.} We treat each clinical site as a small trial borrowing the pooled three-site EHR controls. We first ask whether borrowing reduces the mean held-out control-arm prediction error, $\bar D = 1/|n^{r}_{-1}| \sum_{k=1}^K \sum_{i\in \mathcal{I}^{k}} D_i^k$ as measured by the fold-specific loss difference $D_i^k$ in Equation~\eqref{eq:diagnostic}, and whether this gain increases as the trial size decreases. We then ask whether the diagnostic in Section~\ref{sec:diagnostic}, based on the weighted statistic
$D=\sum_{k=1}^K \sum_{i\in \mathcal{I}^{k}} \pi_{-A_i}^2 D_i^k$,
certifies borrowing. Finally, we study the influence of the EHR coverage status on both the unweighted prediction-error gain and the weighted diagnostic statistic, where coverage means that the EHR contains samples from the same clinical site as the target RCT. The full algorithms, tables, and per-individual analysis are in Appendix~\ref{app:gps-supp}.

\textbf{The borrowing improves the control-arm estimate for small trials whose populations are covered by the EHR.} Each of the six trial sites is treated in turn as the small RCT; the analysis borrows the pooled three-site EHR ($n^o=8{,}867$) and estimates the mean of the held-out control-arm loss reduction $\bar D$ of R-OSCAR-1arm over RACER by $K$-fold cross-fitting, with a paired bootstrap $95\%$ confidence interval (Appendix~\ref{app:gps-controlarm}). Figure~\ref{fig:gps}A plots $\bar D$ against trial size. Among the three sites whose populations are represented in the EHR (Duke, UNC, Vanderbilt), borrowing reduces control-arm error, and the reduction grows as the trial shrinks: $+4.6\%$ at Vanderbilt ($n^r=241$), $+16.9\%$ at Duke ($n^r=55$), and $+28.4\%$ at UNC ($n^r=34$), where the interval excludes zero ($\bar D=0.34$, CI $[0.10,0.62]$). Among the three sites absent from the EHR (Miami, NYU, Stanford), there is no reliable benefit; in particular NYU, the largest such trial ($n^r=86$) and the one with the most room to improve, shows essentially no gain ($\bar D=-0.02$). The pattern follows RCT--OS population coverage in addition to the difference in sample count of the two populations: a larger uncovered trial gains nothing while a smaller covered trial gains substantially.

\textbf{The diagnostic certifies borrowing only in covered sites.} The analysis varies the diagnostic's significance threshold and records, for each site, the one-sided level at which borrowing is certified (Figure~\ref{fig:gps}B; Appendix~\ref{app:gps-diagsens}). As the threshold is relaxed, certification appears exactly for the three covered sites (Duke, then UNC, then Vanderbilt), and never for an uncovered site: the loosest covered site reaches one-sided $p=0.12$ while the strictest uncovered site sits at $p=0.29$. Under this one-sided test, Duke and UNC are certified at the conventional $5\%$ level.  Interestingly, in our experiment, the diagnostic tracks the OS-coverage condition, endorsing borrowing where the EHR covers the trial population and withholding it otherwise.

\textbf{Borrowing helps because the EHR carries outcome signal, not because it supplies more samples.} To separate outcome signal from sample size, the UNC analysis is repeated after randomly permuting the EHR control outcomes while holding every sample size fixed. This destroys the covariate-outcome relationship while preserving all counts (Appendix~\ref{app:gps-negctrl}). The control-arm gain falls from $\bar D=0.34$ to $-0.02$ (averaged over $30$ permutations), and the diagnostic certifies borrowing in only $7\%$ of permutations.  Together with the overlap influence characterized above, this rules out a pure sample-size explanation: identical counts with the signal removed yield no gain, and a larger uncovered trial (NYU) gains nothing while a smaller covered one (UNC) gains substantially.

The implementation here does not fully exploit the complementarity between the RCT and OS data in studies such as Greenlight Plus. It operates only on covariates shared by the two datasets, so the borrowing benefit is confined to that shared subset and does not extend to non-shared (mismatched) covariates. Using this covariate-mismatch structure to improve the precision of CATE estimation is the subject of companion work \citep{pal2026covariate, asiaee2026improving, asiaee2026bcalmbiaslimitedbayesianborrowing}.

\begin{table}[t]
\centering
\caption{\small \textbf{Greenlight Plus and linked EHR sample sizes.} All six trial sites with their randomized counts after the plausibility filter (complete shared covariates, non-missing baseline and 24-month WFLz, $|\text{WFLz}|\le 5$); ``covered'' sites (Duke, UNC, Vanderbilt) appear in the linked EHR, which is pooled as a single shared external control across all sites.}
\label{tab:gps-samples}
\small
\begin{tabular*}{0.7\columnwidth}{@{\extracolsep{\fill}}lccc@{}}
\hline
Site & Covered? & RCT $n^r$ & EHR control $n^o$ \\
\hline
Duke         & yes & $55$  & $6{,}604$ \\
UNC          & yes & $34$  & $433$ \\
Vanderbilt   & yes & $241$ & $1{,}830$ \\
NYU      & no  & $86$ & --- \\
Miami    & no  & $28$ & --- \\
Stanford & no  & $23$ & --- \\
\hline
\textbf{Total} & & \textbf{467} & \textbf{8{,}867} \\
\hline
\end{tabular*}
\end{table}

\begin{figure}[!htbp]
    \centering
    \begin{subfigure}[t]{0.49\textwidth}
        \centering
        \includegraphics[width=\textwidth]{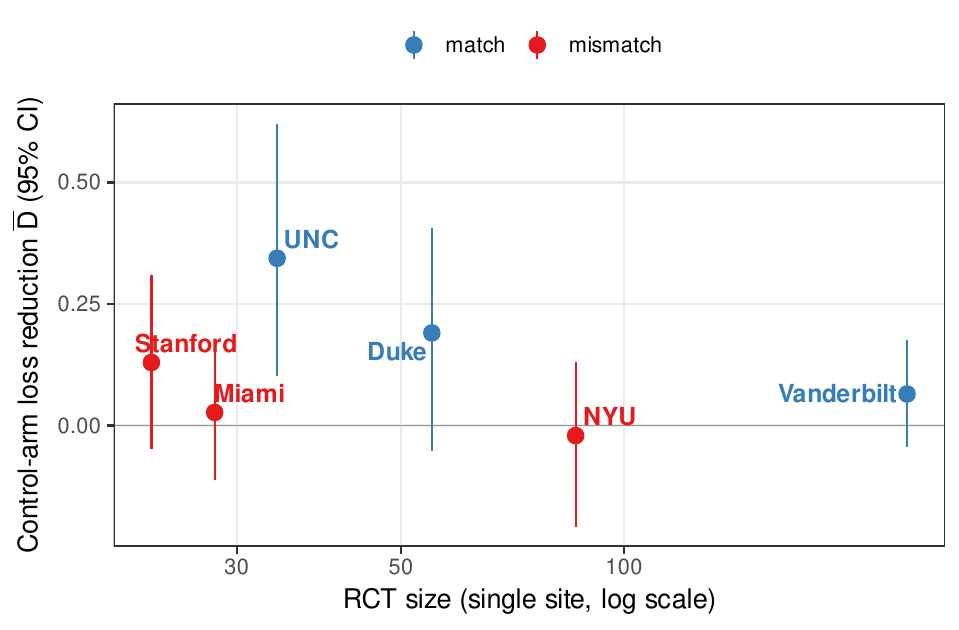}
        \caption{\small Per-site paired-bootstrap test for the held-out control-arm loss reduction $\bar D$ of R-OSCAR-1arm over the control arm of RACER (paired $95\%$ CI), by single-site trial size. Blue denotes a trial site covered by the EHR; red denotes a site absent from the EHR.}
        \label{fig:gps-A}
    \end{subfigure}\hfill
    \begin{subfigure}[t]{0.49\textwidth}
        \centering
        \includegraphics[width=\textwidth]{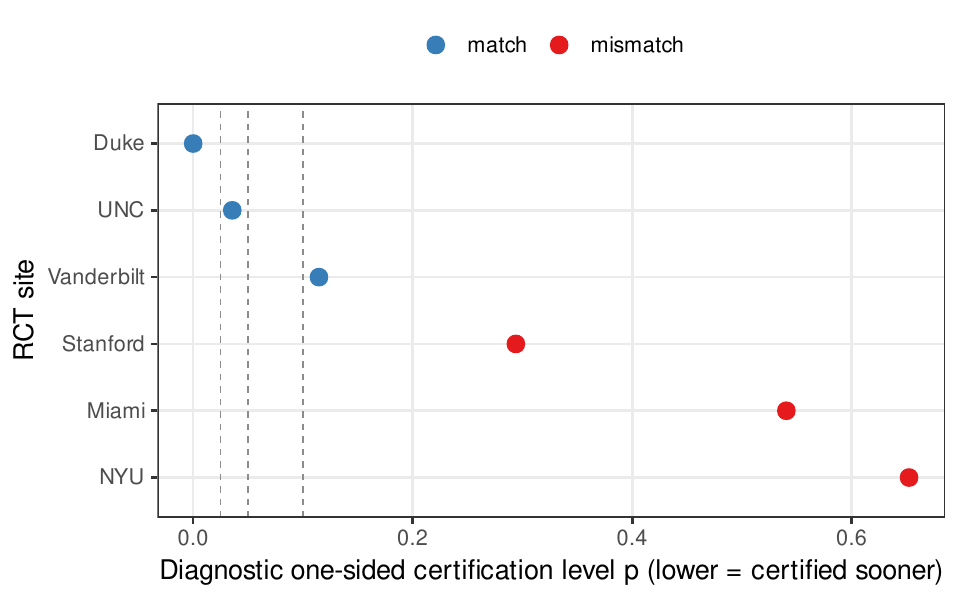}
        \caption{\small Cross-fitted RCT diagnostic from Section~\ref{sec:diagnostic} (cross-fitted, $\pi_{-A_i}^2$-weighted): one-sided certification level $p$ per site (lower $=$ certified sooner); dashed lines mark the two-sided $\alpha=0.05,0.10,0.20$ thresholds ($p<\alpha/2$). Covered sites certify; absent sites do not.}
        \label{fig:gps-B}
    \end{subfigure}
    \caption{\small \textbf{Greenlight Plus: EHR borrowing improves estimation for a small trial when the EHR covers it.} Two borrowing diagnostics are shown, one per panel. (A) Per-site paired-bootstrap test on the control-arm loss reduction $\bar D$. Within each of the covered and uncovered groups, the average $\bar D$ shrinks monotonically as the trial size $n^r$ grows; the covered curve sits above the uncovered curve at comparable $n^r$, and UNC ($n^r=34$) is the certifiable single-site case. (B) Cross-fitted RCT diagnostic from Section~\ref{sec:diagnostic}. Under threshold relaxation it certifies the three covered sites first (Duke, UNC, Vanderbilt in order of evidence strength, all at $p \le 0.12$); only at a much larger threshold ($\alpha > 0.29$) does Stanford join from the uncovered group; Miami and NYU require $\alpha > 0.5$ and are not certified at any reasonable level. The right-panel diagnostic is the formal recommendation test; the two tests address related but different questions and can rank sites within the covered group differently. 
    }
    \label{fig:gps}
\end{figure}

\section{Conclusion}
This work presents a novel framework for improving conditional average treatment effect (CATE) estimation in RCTs by leveraging data from an observational study. Our approach addresses a fundamental challenge in precision medicine: RCTs provide unbiased causal estimates but often lack the sample size needed for precise heterogeneous treatment effect estimation, while observational studies offer abundant data but suffer from confounding bias. By reformulating CATE estimation as a supervised learning problem with pseudo-outcomes, we demonstrate that the counterfactual mean outcome (CMO) serves as the optimal augmentation function for variance reduction. Our key insight is that accurate CMO estimation---achieved by borrowing strength from observational data while calibrating for population differences---directly translates to improved CATE estimation through established risk bounds. The proposed R-OSCAR estimator employs a two-stage calibration strategy that relaxes the strong transportability assumptions typically required in data integration, instead modeling outcome shifts through sparse discrepancy functions that accommodate differences in the data generation process between RCT and observational study.

For any OS-derived nuisance, R-OSCAR remains consistent for the RCT-population CATE. It improves efficiency relative to RCT-only estimators when the RCT--OS outcome discrepancy is estimable from the trial at lower complexity than the full outcome model; severe OS bias or a dense discrepancy affects efficiency rather than bias, and the cross-fitted RCT diagnostic uses observed data alone to determine whether borrowing from a given OS is certifiable. In simulations, R-OSCAR often matches an RCT-only estimator with four times the sample size, and the Greenlight Plus analysis shows the same borrowing logic in a fully external control-arm setting: EHR controls improve held-out control-arm estimation for small trial sites when the diagnostic certifies support, and not for sites absent from the EHR. Together, these results show how observational databases can serve as a precision source for trials while keeping the final treatment-effect analysis anchored in randomized data.





\acks{
This work was supported in part by the Patient-Centered Outcomes Research Institute (PCORI) under award ME-2023C1-32148. A.A. acknowledges support from the National Human Genome Research Institute of the National Institutes of Health under award R00HG011367. }

\section*{Reproducibility}

All experiments in this paper can be replicated with the code at \url{https://github.com/AsiaeeLab/r-oscar}, which implements R-OSCAR and the full analysis pipeline for the simulations of Section~\ref{simulation-results} and the real-data studies of Section~\ref{real-results}. The Tennessee STAR data of Section~\ref{sec:star} are publicly available. The Greenlight Plus trial data and the linked electronic-health-record controls of Section~\ref{sec:gps} are governed by data use agreements and cannot be redistributed.



\newpage

\appendix

\section{Technical Proofs}
\label{all-proofs}
\subsection{Proof of Proposition 1}
\label{proof-prop1-cate-opt}
\begin{proof}[Proof of Proposition \ref{prop1-cate-opt}] 
The first-order optimality condition requires implies that
$f^*(x) = \ex\left[\tau_m(x, A, Y) \!\mid\! X = x\right]$. Here, we demonstrate that this conditional expectation is equivalent to the true CATE:
    \begin{align} 
        \label{proof-sutva}
        \ex_{A, Y}\left[\frac{A (Y - m(X))}{\pi_{A}(X)}\Bigg|{X}\right]  
        &= \ex_{A, Y(-1), Y(+1)}\left[\frac{A (Y(A) - m({X}))}{\pi_{A}(X)}\Bigg|{X}\right]   
        \\ \nonumber
        &= \ex_A \ex_{Y(-1), Y(+1) | A }\left[\frac{ A (Y(A) - m({X}))}{\pi_{A}(X)} \Bigg|{X} \right]
        \\ \nonumber
        &= \sum_{a \in \{-1, +1\}} \pi_{a}(X) \ex_{Y(-1), Y(+1) | A = a} \left[\frac{ a (Y(a) - m({X}))}{\pi_{a}(X)} \Bigg|{X} \right]
        \\ \nonumber
        &= \sum_{a \in \{-1, +1\}}  \ex_{Y(-1), Y(+1) | A = a} \left[ a (Y(a) - m({X})) \bigg|{X} \right]
        \\ \label{proof-ignorability}
        &=  \ex \left[ (Y(+1)\big|{X} \right] - m(\mathbf{X})   - \ex \left[ (Y(-1)\big|{X} \right] +  m({X})
        \\ \nonumber
        &= \tau({X}),
\end{align}
where we used SUTVA in \eqref{proof-sutva} and conditional ignorability in \eqref{proof-ignorability}.
\end{proof}

\vspace{-24pt}
\subsection{Proof of Theorem \ref{theo1-var-reduce}} 
\label{proof-theo1-var-reduce}
The variance of ${\tau}_{m}$ is: $\var({\tau}_{m}(x,A,Y)\mid X=x) = \ex({\tau}^2_{m}(x,A,Y)\mid X=x) - \ex^2({\tau}_{m}(x,A,Y)\mid X=x)$ where per Proposition \ref{prop1-cate-opt}, under the assumed conditions, the second term is equal to $\tau^2(x)$. So, only the first term, $\ex({\tau}^2_{m}(x,A,Y)\mid X=x)$ depends on $m(X)$. The first order optimality condition implies that, for any $x$, the optimal $m$ (calling it $m^*$) should satisfy $\ex_{A, Y}((Y-m^*(x))/\pi^2_A(x)) =0$, which we expand below:
\begin{align} \label{opt-cond}
    \ex_{A, Y}\left[\frac{Y-m^*(X)}{\pi^2_A(X)} \middle| X \right] 
    &=\ex_A \ex_{Y(-1), Y(+1) | A }\left[\frac{Y(A)-m^*(X)}{\pi^2_A(X)}\middle| X\right] 
    \\ \nonumber
    &= \sum_{a \in \{-1, +1\}} \pi_{a}(X) \ex_{Y(-1), Y(+1) | A = a} \left[\frac{Y(a)-m^*(X)}{\pi^2_a(X)}\middle| X\right] 
    \\ \nonumber
    &= \sum_{a \in \{-1, +1\}} \ex_{Y(-1), Y(+1) | A = a} \left[\frac{Y(a)-m^*(X)}{\pi_a(X)}\middle| X\right] 
    \\ \nonumber
    &= \ex_{Y(+1)} \left[\frac{Y(+1)-m^*(X)}{\pi_{+1}(X)}\middle| X\right] + \ex_{Y(-1)} \left[\frac{Y(-1)-m^*(X)}{\pi_{-1}(X)}\middle| X\right] 
    \\ \nonumber
    &= 0
\end{align}
Then for any $X = x$, the optimal function $m^*(x)$ should satisfy:
\begin{align*}
    &\ex_{A, Y}\left[\frac{Y-m^*(x)}{\pi^2_A(x)} \mid X=x \right] \\
    &=\frac{\ex\left[Y(+1)\mid X=x\right]}{\pi_{+1}(x)} - \frac{m^*(x)}{\pi_{+1}(x)}
    + \frac{\ex\left[Y(-1)\mid X=x\right]}{\pi_{-1}(x)} - \frac{m^*(x)}{\pi_{-1}(x)}
    \\ 
    &= \frac{\ex\left[Y(+1)\mid X=x\right]}{\pi_{+1}(x)}  
    +  \frac{\ex\left[Y(-1)\mid X=x\right]}{\pi_{-1}(x)} 
    -  m^*(x)\left(\frac{1}{\pi_{+1}(x)} + \frac{1}{\pi_{-1}(x)}\right)
    \\
    &= \frac{\ex\left[Y(+1)\mid X=x\right]}{\pi_{+1}(x)}  
    +  \frac{\ex\left[Y(-1)\mid X=x\right]}{\pi_{-1}(x)} 
    -  m^*(x)\left(\frac{1}{\pi_{+1}(x)\pi_{-1}(x)}\right) =0
\end{align*}
Solving for $m^*$ completes the proof: {$m^*(x) = \pi_{-1}(x)\ex[Y(+1)\mid X=x] + \pi_{+1}(x)\ex[Y(-1)\mid X=x] = \mu(x)$}. 

\subsection{More Context for Augmentation in CATE Estimation}
Here we provide a few more results that demonstrate the role of augmentation in estimation of the CATE using our pseuo-outcomes.
\begin{proposition} 
\label{prop2-decomp}  
The potential outcome mean can be decomposed into the counterfactual mean outcome (CMO) and the conditional average treatment effect (CATE) as follows: 
\begin{align} \nonumber
\mu_A(X) = \ex(Y(A) \!\mid\! X) = \mu(X) + w(X, A) \tau(X), 
\end{align} 
where the weight function $w(X, A) \equiv \frac{(1 + A) \pi_{+1}(X) - (1 - A) \pi_{-1}(X)}{2}$ is the only term that depends on $A$. 
\end{proposition}
\begin{proof}[Proof of Proposition \ref{prop2-decomp}] 
\label{proof-prop2-decomp}
Let's expand $\mu(X) + w(X, A) \tau(X)$:
\begin{align*}
\mu(X) + w(X, A) \tau(X) 
&= [\pi_{+1}(X) \mu_{-1}(X) + \pi_{-1}(X) \mu_{+1}(X)] \\ 
&+ \frac{(1 + A) \pi_{+1}(X) - (1 - A) \pi_{-1}(X)}{2} [\mu_{+1}(X)-\mu_{-1}(X)]
\end{align*}
Collecting terms for $\mu_{+1}(X)$ and $\mu_{-1}(X)$, the coefficients are:
\begin{align*}
  \mu_{+1}(X)&: \pi_{-1}(X) + \frac{(1 + A) \pi_{+1}(X) - (1 - A) \pi_{-1}(X)}{2} 
  = \begin{cases}
     \pi_{-1}(X) + \pi_{+1}(X) = 1 & A = +1 \\
     \pi_{-1}(X) - \pi_{-1}(X) = 0 & A = -1
  \end{cases} \\
  \mu_{-1}(X)&: \pi_{+1}(X) - \frac{(1 + A) \pi_{+1}(X) - (1 - A) \pi_{-1}(X)}{2} 
  = \begin{cases}
     \pi_{+1}(X) - \pi_{+1}(X) = 0 & A = +1 \\
     \pi_{+1}(X) + \pi_{-1}(X) = 1 & A = -1
  \end{cases}
\end{align*}
This verifies that $\mu(X) + w(X, A) \tau(X) = \mu_A(X)$ for both values of $A$.
\end{proof}

As a known special case, consider an RCT with equal treatment probabilities. Here, the decomposition simplifies to:
$\mu^r(X, A)=\mu_{A}^r(X)=\mu^r(X)+\frac{A}{2}\tau^r(X)$,
where the terms $\mu^r(X)$ and $\frac{A}{2}\tau^r(X)$ have previously been coined in the literature as the main effect of covariates and the treatment effect, respectively \citep{chen2017general}.

The following corollary combines Proposition \ref{prop1-cate-opt} (on bias) and Theorem \ref{theo1-var-reduce} (on variance):
\begin{corollary} \label{corollary1-useless}
Assume (A1)-(A3) hold and the true CMO $\mu(X)$ is known. For a given dataset at point $x$, $\{(x, A_i, Y_i)\}_{i=1}^n$, the estimator $\htau(x) = \frac{1}{n} \sum_{i=1}^n \tau_\mu(x, A_i, Y_i)$ is unbiased for the CATE at point $x$ and minimizes the population variance. 
\end{corollary}


\subsection{Proof of Theorem \ref{theo2-sandwich}} \label{proof-theo2-sandwich}
Per Proposition \ref{prop1-cate-opt}, under the assumed conditions, $\tau(X) = \ex({\tau}_{m}(X,A,Y)|{X})$ and therefore we can write the variance of CATE as $\var(\tau(X)) = \ex\left(\frac{(Y - m({X}))^2}{\pi_A^2({X})}|{X}\right) - \tau^2(X)$.
Only the first term of variance depends on $m(X)$. Our goal is to determine how much variance increases if $m(X):=\hmu(X)=\mu(X)+d(X)$ instead of $m(X):=\mu(X)$ which is the theoretical minimum. 
\begin{align*}
    &\var({\tau}_{\hmu}(X,A,Y)\mid X=x) \\ 
    &= \ex_{A, Y}\left(\left.\frac{(Y-\hmu(X))^2}{\pi^2_A(X)}\right| X\right) - \tau^2(X)
    \\ 
    &= \ex_{A, Y}\left(\left.\frac{(Y-\mu(X)-d(X))^2}{\pi^2_A(X)}\right| X\right) - \tau^2(X)
    \\ 
    &= \ex_{A, Y}\left(\left.\frac{(Y-\mu(X))^2 - 2(Y-\mu(X))d(X) + d^2(X)}{\pi^2_A(X)}\right| X\right) - \tau^2(X)
    \\ 
    &= \var({\tau}_{{\mu}}(X,A,Y)\mid X=x) 
    + \ex_{A, Y}\left(\left.\frac{- 2(Y-\mu(X))d(X) + d^2(X)}{\pi^2_A(X)}\right| X\right) 
    \\ 
    &= \var({\tau}_{{\mu}}(X,A,Y)\mid X=x) - 2d(x)\ex_{A, Y}\left(\left.\frac{(Y-\mu(X))}{\pi^2_A(X)}\right| X\right) 
    +d^2(x)\ex_{A}\left(\left.\frac{1}{\pi^2_A(X)}\right| X\right) 
\end{align*}
The middle term is zero because $\mu(X)$ is the minimizer of the variance and $\ex_{A, Y}\left(\left.\frac{(Y-\mu(X))}{\pi^2_A(X)}\right| X\right)$ is the exact optimality condition used in \eqref{opt-cond}. Moreover, $\ex_{A}\left(\left(\frac{1}{\pi^2_A(X)}\right| X=x\right)=\frac{1}{\pi_{+1}(x)\pi_{-1}(x)}$. Together, they result in:
\begin{align}
\nonumber
&\var({\tau}_{\hmu}(X,A,Y)\mid X=x) 
= \var({\tau}_{{\mu}}(X,A,Y)\mid X=x) + \frac{d^2(x)}{\pi_{+1}(x)\pi_{-1}(x)}
\\ \nonumber
&\Rightarrow \var({\tau}_{{\mu}}\mid X=x) + \frac{d^2(x)}{(1-\rho)^2} \leq \var({{\tau}}_{\hmu}\mid X=x) \leq \var({\tau}_{{\mu}}\mid X=x) + \frac{d^2(x)}{\rho^2}.
\end{align}
The last inequality holds because $\rho \leq \pi_{+1}(x) \leq 1 - \rho$ and $\pi_{-1}(x) = 1 - \pi_{+1}(x)$ and therefore $\frac{1}{(1-\rho)^2} \leq \frac{1}{\pi_{+1}(x)\pi_{-1}(x)} \leq \frac{1}{\rho^2}$. Taking expectation with respect to $X$ completes the proof for the population case.

\subsection{Proof of Theorem \ref{theo3-risk-bound}} \label{proof-theo3-risk-bound}
We begin by analyzing the risk of an arbitrary function:
\begin{lemma}[Decomposition of Risk]
    Let $\tau_m(X, A, Y)$ be an unbiased estimator of $\tau(X)$, i.e., $\ex_{Y,A\mid X}[\tau_m(X, A, Y)] = \tau(X)$. For any function $f: \mathcal{X} \to \mathbb{R}$, the risk
    \[
    R_m(f) \equiv \ex_X \ex_{Y,A\mid X} \left[\left(\tau_m(X, A, Y) - f(X)\right)^2\right]
    \]
    can be decomposed into irreducible error and estimation error:
    \[
    R_m(f) = \ex_X[\var(\tau_m\mid X)] + \ex_X\left[(f(X) - \tau(X))^2\right].
    \]
\end{lemma}

\begin{proof}
    We decompose the risk by expanding the squared difference:
    \begin{align*}
        R_m(f) 
        &= \ex_X \ex_{Y,A\mid X} \left[\left(\tau_m(X, A, Y) - f(X)\right)^2\right] \\
        &= \ex_X \ex_{Y,A\mid X} \left[\left(\tau_m(X, A, Y) - \tau(X) + \tau(X) - f(X)\right)^2\right] \\
        &= \ex_X \ex_{Y,A\mid X} \left[(\tau_m(X, A, Y) - \tau(X))^2 + (f(X) - \tau(X))^2 \right. \\ 
        &+ \left. 2(\tau_m(X, A, Y) - \tau(X))(\tau(X) - f(X))\right].
    \end{align*}
    
    The cross term vanishes because
    \begin{align*}
        &\ex_X \ex_{Y,A\mid X}\left[(\tau_m(X, A, Y) - \tau(X))(\tau(X) - f(X))\right] \\
        &= \ex_X\left[(\tau(X) - f(X))\ex_{Y,A\mid X}[\tau_m(X, A, Y) - \tau(X)]\right] = 0,
    \end{align*}
    where the last equality follows from the unbiasedness assumption.
    
    Therefore,
    \begin{align*}
        R_m(f) &= \ex_X \ex_{Y,A\mid X}\left[(\tau_m(X, A, Y) - \tau(X))^2\right] + \ex_X\left[(f(X) - \tau(X))^2\right] \\
        &= \ex_X[\var(\tau_m\mid X)] + \ex_X\left[(f(X) - \tau(X))^2\right],
    \end{align*}
    where we use the definition of conditional variance in the last step.
\end{proof}
\vspace{-12pt}
We are interested in three key functions:
\begin{itemize}
    \item True CATE: 
    $\tau(X) = \ex_{Y,A}[\tau_m(X, A, Y)\mid X]$    
    \item Best CATE estimator: The function in class $\cF$ that minimizes the mean squared error
    $\tau^* = \argmin_{f \in \cF} \ex_X\left[(f(X) - \tau(X))^2\right]$
    \item Empirical Risk Minimizer (ERM): $\htau = \argmin_{f \in \cF} \frac{1}{n} \sum_{i=1}^n \left[\tau_m(X_i, A_i, Y_i) - f(X_i)\right]^2$
\end{itemize}

Our primary objective is to bound the mean squared error of the ERM:$ex_X\left[(\htau(X) - \tau(X))^2\right].$ In statistical learning theory, this is achieved by bounding the excess risk (also known as approximation error or generalization gap):$R_m(\htau) - R_m(\tau^*),$ which quantifies the additional risk incurred by using the ERM instead of the best function in the function class. We leverage classic generalization bounds to establish an upper bound for the MSE of the CATE plug-in estimator.

\begin{lemma}[MSE Upper Bound of ERM]\label{lemma-ineq}
    Let $\tau(X)$ be the true CATE, $\htau$ be the ERM estimator in function class $\cF$, and $\tau^*$ be the best CATE estimator in $\cF$. Define the estimation error for any estimator $f$ as $\Delta_\tau^2(f) = \ex_X[(f(X) - \tau(X))^2]$, and the minimum estimation error in class $\cF$ as $\Delta_2^2(\cF, \tau) = \inf_{f \in \cF} \Delta_\tau^2(f)$. Then,
    \[
    \Delta_2^2(\htau, \tau) \leq 2G_m(\cF, \cD) + \Delta_2^2(\cF, \tau),
    \]
    where $G_m(\cF, \cD) \equiv \sup_{f \in \cF} |R_m(f) - \hat{R}_m(f)|$ is the supremum of the empirical risk gap over $\cF$ for training dataset $\cD=\{X_i, A_i, Y_i\}_{i=1}^n$.
\end{lemma}

\begin{proof}
    First, observe that the irreducible error cancels out in the excess risk:
    {\small
    \begin{align*}
        R_m(\htau) - R_m(\tau^*) 
        &= \ex_X[\var(\tau_m\mid X)] + \ex_X[(\htau(X) - \tau(X))^2]  - \ex_X[\var(\tau_m\mid X)] - \ex_X[(\tau^*(X) - \tau(X))^2] \\
        &= \Delta_2^2(\htau, \tau) - \Delta_2^2(\cF, \tau).
    \end{align*}}
    
    Therefore, the estimation error can be decomposed as:
    \begin{equation*}
        \Delta_2^2(\htau, \tau) = R_m(\htau) - R_m(\tau^*) + \Delta_2^2(\cF, \tau).
    \end{equation*}
    
    To bound the excess risk $R_m(\htau) - R_m(\tau^*)$, we decompose it as:
    \begin{align*}
        R_m(\htau) - R_m(\tau^*) 
        &= R_m(\htau) - \hat{R}_m(\htau) + \hat{R}_m(\htau) - \hat{R}_m(\tau^*) + \hat{R}_m(\tau^*) - R_m(\tau^*) \\
        &\leq R_m(\htau) - \hat{R}_m(\htau) + \hat{R}_m(\tau^*) - R_m(\tau^*) \\
        &\leq |R_m(\htau) - \hat{R}_m(\htau)| + |\hat{R}_m(\tau^*) - R_m(\tau^*)| \\
        &\leq 2\sup_{f \in \cF} |R_m(f) - \hat{R}_m(f)| = 2G_m(\cF, \cD),
    \end{align*}
    where the first inequality follows from $\hat{R}_m(\htau) - \hat{R}_m(\tau^*) \leq 0$ by definition of ERM.
    
    Combining these results yields the stated bound:
    \(
        \Delta_2^2(\htau, \tau) \leq 2G_m(\cF, \cD) + \Delta_2^2(\cF, \tau).
    \)
\end{proof}
Next, as a warm-up, we present a preliminary Theorem that presents our results under restrictive settings. 

\begin{theorem}[Generalization Bound via Rademacher Complexity] \label{theo-simple}
    Let $\cF$ be a function class, and $\htau$ be the ERM estimator in $\cF$. Assume the following:
    \begin{enumerate}
        \item The loss function $L(\cdot, \cdot)$ is bounded in $[0, l_\infty]$
        \item For any sample $(X_i, A_i, Y_i) \in \cD$ and functions $f \in \mathcal{F}$ and $m$: $|\tau_m(X_i, A_i, Y_i) - f(X_i)| \leq C$
    \end{enumerate}
    Then, with probability at least $1-\varepsilon_1$:
    \begin{equation*}
        \Delta_2^2(\htau, \tau) \leq \Delta_2^2(\cF, \tau) + 4C \cR_{n}(\cF) + \frac{l_\infty}{\sqrt{2n}}\sqrt{\log \frac{2}{\varepsilon_1}},
    \end{equation*}
    where $\cR_{n}(\cF)$ is the Rademacher complexity of function class $\cF$:
    \(
        \cR_{n}(\cF) \equiv \ex_{\epsilon, \cD}\left[\sup_{f \in \cF} \frac{1}{n} \sum_{i=1}^{n} \epsilon_i f(X_i)\right],
    \)
    with $\epsilon_i \in \{-1, +1\}$ being i.i.d. Rademacher random variables.
\end{theorem}

\begin{proof}
   The proof proceeds in three steps:
   \begin{enumerate}[leftmargin=12pt, itemsep=0pt]
   \item First, we bound the expected empirical risk gap using Rademacher complexity:
   \begin{equation}\label{eq-ex}
       \ex_{\cD}[G_m(\cF, \cD)] \leq \cR_{n}(\mathcal{H}_m) \leq 2C \cR_{n}(\cF),
   \end{equation}
   where $\mathcal{H}_m = \{h: (x,a,y) \mapsto L(\tau_m(x,a,y), f(x)), f \in \cF\}$. The second inequality follows from the fact that for any loss function $L(\cdot, \cdot)$ that is $G$-Lipschitz in its second argument (i.e., $|L(y_0, x') - L(y_0, x)| \leq G |x' - x|$), we have $\cR_{n}(\mathcal{H}_m) \leq G\cR_{n}(\cF)$. For the $\ell_2$ loss $L(y,x) = (y-x)^2$, under the assumption $|\tau_m(X_i, A_i, Y_i) - f(X_i)| \leq C$, we can show it is $2C$-Lipschitz in its second argument:
   \begin{align} \nonumber
       \left|L(y,x') - L(y,x)\right| 
       &= \left|(y-x')^2 - (y-x)^2\right| \\ \nonumber
       &= \left|(x-x')(2y-x - x')\right| \\ \nonumber
       &= \left|x-x'\right|\left|2y-x - x'\right| \\ \label{eq-lip-ineq}
       &= \left|x-x'\right|\left(\left|y-x\right|+\left|y-x'\right|\right) \\ \nonumber
       &\leq 2C\left|x'-x\right|
    \end{align}  
    
   \item By McDiarmid's inequality \citep{bach_learning_2024}, with probability at least $1-\varepsilon_1$:
   \begin{equation}\label{eq-prob}
      G_m(\cF, \cD) \leq 2\cR_{n}(\mathcal{H}_m) + \frac{l_\infty}{\sqrt{2n}}\sqrt{\log \frac{2}{\varepsilon_1}}.
   \end{equation}

    \item Combining bounds \eqref{eq-ex} and \eqref{eq-prob} with Lemma~\ref{lemma-ineq} yields the result.
   \end{enumerate}
\end{proof}
\vspace{-24pt}
\begin{remark}[On the Boundedness Assumption]
    The assumption $|\tau_m(X_i, A_i, Y_i) - f(X_i)| \leq C$ is crucial but potentially restrictive. Unlike typical learning theory settings where boundedness can be achieved by assuming bounded outcomes (i.e., $Y$ in $Y=f(X)+\epsilon$ setting) and function values ($f(X)$), here $\tau_m$ is an arbitrary transformation that depends on   function $m$.
    
    In what follows, we take a more practical approach by relaxing this assumption. Instead of requiring boundedness of arbitrary transformations $\tau_m$, we focus on an assumption about the optimal transformation $\tau_\mu$, which is more reasonable and limited. The final result explicitly connects the constants in our bound to the distance of $m(X)$ to the true CMO $\mu(X)$.
\end{remark}

To relax the boundedness assumption on the transformed outcome $\tau_m$, we first show in Lemma~\ref{lemma-uniform-var-bound} that, with high probability, the optimally transformed outcome $\tau_\mu$ remains uniformly close to the true CATE $\tau(X)$. As a direct consequence, in Lemma~\ref{lemma-loss-bound} we relax the loss-boundedness assumption used in Theorem~\ref{theo-simple}. 
Finally, in Lemma~\ref{lemma-lipschitz}, we derive a new Lipschitz constant for the $\ell_2$ loss to replace the constant $C$ in Theorem~\ref{theo-simple}. This new constant depends on how well the augmentation function $m(X)$ approximates the true CMO $\mu(X)$ under two reasonable assumptions: (1) the true treatment effect is bounded, i.e.\ $\lvert \tau(X) \rvert \leq B$, and (2) all functions in the class $\cF$ have bounded outputs, i.e.\ $\lvert f(X) \rvert \leq B$.

\begin{lemma} \label{lemma-uniform-var-bound}
    Let $X$ represent the covariates, $A \in \{-1, +1\}$ denote the treatment assignment, $Y$ be the observed outcome, and $\pi_A(X)$ represent the propensity score for treatment $A$. For observed variables $(X, A, Y)$, define the optimal transformed outcome $\tau_\mu(X, A, Y)$ as:
    \(
    \tau_\mu(X, A, Y) = \frac{A(Y - \mu(X))}{\pi_A(X)},
    \)
    where $\mu(X) \equiv \pi_{-1}(X) \mu_{+1}(X) + \pi_{+1}(X) \mu_{-1}(X)$ is the true Counterfactual Mean Outcome (CMO), and $\tau(X) \equiv \mu_{+1}(X) - \mu_{-1}(X)$  is the true CATE.
    
    Then, for any $\varepsilon_2 \in (0,1)$, with probability at least $1-\varepsilon_2$:
    \[
    \left|\tau_\mu(X, A, Y) - \tau(X)\right| \leq \frac{\sigma}{\sqrt{\varepsilon_2}},
    \]
    where $\sigma^2 = \sup_{x \in \mathcal{X}} \var(\tau_\mu\mid X=x)$ is the supremum of the conditional variance of the optimal transformed outcome over the domain of $X$.
\end{lemma}

\begin{proof}
    First, note that $\tau(X) = \ex_{A,Y}[\tau_\mu(X, A, Y)\mid X]$ by construction of $\tau_\mu$. 
    
    For any fixed $x \in \mathcal{X}$, applying Chebyshev's inequality to the random variable $\tau_\mu(x, A, Y)$ conditional on $X=x$:
    \[
    \mathbb{P}\left(\left|\tau_\mu(x, A, Y) - \tau(x)\right| \geq \frac{\sigma}{\sqrt{\varepsilon_2}}\mid X=x\right) \leq \frac{\var(\tau_\mu\mid X=x)}{(\sigma/\sqrt{\varepsilon_2})^2} \leq \varepsilon_2,
    \]
    where the last inequality follows from the definition of $\sigma^2$ as the supremum of the conditional variance.
    The result follows by the law of total probability over $X$.
\end{proof}

\vspace{-24pt}

\begin{remark}
The main assumption of Lemma~\ref{lemma-uniform-var-bound} is that the variance of the most efficient estimator remains bounded over the entire domain of covariates. This requirement is not restrictive. Indeed, from the definitions of $\tau_\mu(X, A, Y)$ and $\tau(X)$, one can see that it is equivalent to assuming
\[
\max\!\left(
\var\left(Y(+1) - \mu_{+1}(X)\mid X=x\right),
\var\left(Y(-1) - \mu_{-1}(X)\mid X=x\right)
\right)
< \infty,
\]
i.e., the per-arm conditional outcome variance is uniformly bounded. In the additive noise model $Y(a) = f_a(X) + \epsilon_a$, this condition simply means that the noise $\epsilon_a$ has bounded variance.
\end{remark}

One of the consequences of Lemma \ref{lemma-uniform-var-bound} is that now we can bound the loss function and relax the boundedness assumption of Theorem \ref{theo-simple}:  

\begin{lemma}[High-Probability Loss Bound]\label{lemma-loss-bound}
    Let $|\tau(X)| \leq B$ and $|f(X)| \leq B$ for all $f \in \cF$. Define $\sigma^2 = \sup_{x \in \mathcal{X}} \var(\tau_\mu\mid X=x)$ as the supremum of the conditional variance of the optimal transformed outcome over the domain of $X$. 
    Then, with probability at least $1-\varepsilon_1$, the squared loss is bounded as:
    \[
        \ell_2(\tau_m(X, A, Y), \, f(X)) = (\tau_m(X, A, Y) - f(X))^2 \leq l_\infty,
    \]
    where $l_\infty = \left(2B + \frac{\sigma}{\sqrt{\varepsilon_1}}+\frac{1}{\rho}\sup_{x\in\cX}\left|m(X) - \mu(X)\right|\right)^2$ and $\rho$ is a lower bound on the propensity scores.
\end{lemma}

\begin{proof}
    For any given sample $(X, A, Y)$ we have:
    \begin{align*}
        \left|\tau_m(X, A, Y) - f(X)\right| &= \left|\tau_m(X, A, Y) - \tau(X) + \tau(X) - f(X)\right| \\
        &\leq \left|\tau_m(X, A, Y) - \tau(X)\right| + \left|\tau(X) - f(X)\right| \\
        &\leq \left|\tau_m(X, A, Y) - \tau_\mu(X, A, Y) + \tau_\mu(X, A, Y) - \tau(X)\right| + 2B \\
        (\text{Lemma \ref{lemma-uniform-var-bound}})
        &\leq \left|\tau_m(X, A, Y) - \tau_\mu(X, A, Y)\right| + \frac{\sigma}{\sqrt{\varepsilon_1}} + 2B \\
        &\leq \frac{1}{\rho}\left|m(X) - \mu(X)\right| + \frac{\sigma}{\sqrt{\varepsilon_1}} + 2B \\
        &\leq \frac{1}{\rho}\sup_{x\in\cX}\left|m(X) - \mu(X)\right| + \frac{\sigma}{\sqrt{\varepsilon_1}} + 2B 
    \end{align*}
    where the third inequality onward hold with probability at least $1-\varepsilon_1$.
    
    Therefore, with the same probability:
    \begin{align*}
        \forall X, A, Y: \quad \ell_2(\tau_m(X, A, Y), \, f(X)) 
        &= (\tau_\mu(X, A, Y) - f(X))^2 \\
        &\leq \left(2B + \frac{\sigma}{\sqrt{\varepsilon_1}}+\frac{1}{\rho}\sup_{x\in\cX}\left|m(X) - \mu(X)\right|\right)^2 = l_\infty
    \end{align*}
\end{proof}
Finally, under relaxed assumptions, we derive a new Lipschitz constant to be used for Rademacher complexity in below Lemma. 

\begin{lemma}[Rademacher Complexity Bound for Loss Class] \label{lemma-lipschitz}
   Let $\cF$ be a function class where $|f(X)| \leq B$ for all $f \in \cF$, and assume $|\tau(X)| \leq B$. Let $\mathcal{H}_m = \{h: z \mapsto (\tau_m(z) - f(x))^2,\, z = (x,a,y),\, f \in \cF\}$ be the squared loss function class. Then, with probability at least $1-\varepsilon_2$:
   \[
       \cR_{n}(\mathcal{H}_m) \leq l_m \cR_{n}(\cF),
   \]
   where {\small $l_m = \left(4B + \frac{2\sigma}{\sqrt{\varepsilon_2}} + \frac{2}{\rho}\sup_{x\in\mathcal{X}}|\mu(x) - m(x)|\right)$} and $\rho$ is a lower bound on the propensity scores.
\end{lemma}

\begin{proof}
   For brevity, denote $Z_i = (X_i, A_i, Y_i)$ and let $\tau_m(Z_i) \equiv \tau_m(X_i, A_i, Y_i)$. We first examine the Lipschitz continuity condition of the $\ell_2$ loss from Equation \eqref{eq-lip-ineq}:
   \begin{align*}
       &\left|(\tau_m(Z_i) - f(X_i))^2 - (\tau_m(Z_i) - f(X_j))^2\right| \\
       &\leq \left|f(X_i)-f(X_j)\right| \left(\left|\tau_m(Z_i) - f(X_i)\right| + \left|\tau_m(Z_i) - f(X_j)\right|\right)
   \end{align*}
   
   Adding and subtracting the conditional expectation $\ex_{A,Y}[\tau_m\mid X_i] = \tau(X_i)$ and applying the triangle inequality:
   \begin{align*}
       &\left|(\tau_m(Z_i) - f(X_i))^2 - (\tau_m(Z_i) - f(X_j))^2\right| \\
       &\leq \left|f(X_i)-f(X_j)\right|\Big(\left|\tau_m(Z_i) - \ex_{A,Y}[\tau_m\mid X_i]\right| + \left|\tau(X_i) - f(X_j)\right| + \\ &\left|\tau_m(Z_i) - \ex_{A,Y}[\tau_m\mid X_i]\right| + \left|\tau(X_i) - f(X_i)\right|\Big) \\
       &\leq \left|f(X_i)-f(X_j)\right|\left(4B + 2\left|\tau_m(Z_i) - \ex_{A,Y}[\tau_m\mid X_i]\right|\right)
   \end{align*}
   where we used the boundedness assumptions.
   
   Decomposing via the optimal transformation $\tau_\mu$:
   \begin{align*}
       &\left|(\tau_m(Z_i) - f(X_i))^2 - (\tau_m(Z_i) - f(X_j))^2\right| \\ 
       &\leq \left|f(X_i)-f(X_j)\right|\Big(4B + 2\left|\tau_m(Z_i) - \tau_\mu(Z_i) + \tau_\mu(Z_i) - \tau(X_i)\right|\Big) \\
       &\leq \left|f(X_i)-f(X_j)\right|\Big(4B + \frac{2\sigma}{\sqrt{\varepsilon_2}} + 2\left|\tau_m(Z_i) - \tau_\mu(Z_i)\right|\Big)
    \end{align*}
   where the last inequality holds with probability $1-\varepsilon_2$ from Lemma \ref{lemma-uniform-var-bound}.
   
   Expanding the definition of transformation we have 
    \begin{align*}
        \left|\tau_m(Z_i) - \tau_\mu(Z_i)\right| 
        &= \left|\frac{A_i}{\pi_{A_i}(X_i)}\left[Y_i - m(X_i) - Y_i + \mu(X_i)\right]\right| \\
        &=\left|\frac{A_i}{\pi_{A_i}(X_i)}\right|\left|\mu(X_i) - m(X_i)\right| \\
        &\leq \frac{1}{\rho} \left|\mu(X_i) - m(X_i)\right|
    \end{align*}
    where we used the propensity score bound $\pi_{A_i}(X_i) \geq \rho$ in the last inequality. 

    The final form of the Lipschitz condition will be:
    {\small
   \begin{align*}
       \left|(\tau_m(Z_i) - f(X_i))^2 - (\tau_m(Z_i) - f(X_j))^2\right| 
       \leq \left|f(X_i)-f(X_j)\right|\left(4B + \frac{2\sigma}{\sqrt{\varepsilon_2}} + \frac{2}{\rho}\sup_{x\in\mathcal{X}}\left|\mu(x) - m(x)\right|\right)
   \end{align*}}
   
   Therefore, with probability at least $1-\varepsilon_2$, the squared loss class $\mathcal{H}_m$ is $l_m$-Lipschitz, and:
   \begin{align*}
       \cR_{n}(\mathcal{H}_m) 
       = \ex_{\epsilon, \cD} \left[\sup_{h \in \mathcal{H}} \frac{1}{n} \sum_{i=1}^{n} \epsilon_i h(Z_i)\right] 
       \leq l_m \ex_{\epsilon, \cD} \left[\sup_{f \in \cF} \frac{1}{n} \sum_{i=1}^{n} \epsilon_i f(X_i)\right] 
       = l_m \cR_{n}(\cF)
   \end{align*}
\end{proof}
\vspace{-24pt}
\begin{remark}
   While $m(X)$ is estimated from data (to approximate $\mu(X)$), we assume it is computed using an independent dataset through sample splitting. For instance, with $n$ samples, one can use half to learn $m(X)$ and the other half to learn $\tau(X)$. Cross-fitting, which swaps the roles of the samples and averages the learned parameters, can utilize the full dataset. This ensures independence of $L_m$ from dataset $\cD$, and therefore, we can take it out of the expectation. 
\end{remark}

\begin{theorem}[CATE Estimation Error Bound]
   Let $\cF$ be a function class and $\htau$ be the ERM estimator in $\cF$. Assume:
   \begin{enumerate}[leftmargin=12pt, itemsep=0pt]
       \item The true CATE and functions in $\cF$ are bounded: $|\tau(X)|, |f(X)| \leq B$ for all $f \in \cF$
       \item The propensity scores are bounded away from zero: $\pi_A(X) \geq \rho > 0$
       \item The conditional variance of the optimal transformed outcome is bounded: $\sigma^2 \equiv \sup_{x \in \mathcal{X}} \var(\tau_\mu\mid X=x) < \infty$
   \end{enumerate}
   Then, with probability at least $1-2\varepsilon$:
   \begin{equation*}
       \Delta_2^2(\htau, \tau) \leq \Delta_2^2(\cF, \tau) 
       + 2C(m, \varepsilon) \cR_{n}(\cF) 
       + C^2(m, \varepsilon) \sqrt{\frac{\log(2/\varepsilon)}{n}},
   \end{equation*}
   where {\small $C(m, \varepsilon) = \left(2B + \frac{\sigma}{\sqrt{\varepsilon}} + \frac{1}{\rho}\sup_{x\in\mathcal{X}}|\mu(x) - m(x)|\right)$} is the Lipschitz constant of the loss function.
\end{theorem}

\begin{proof}
   The proof combines four key results:
   \begin{enumerate}[leftmargin=12pt, itemsep=0pt]
        \item From Lemma~\ref{lemma-uniform-var-bound}, with probability at least $1-\varepsilon_1$, the optimal transformed outcome concentrates around the true CATE with radius $\sigma/\sqrt{\varepsilon_1}$.
        \item Consequently, from Lemma \ref{lemma-loss-bound}, with probability at least $1-\varepsilon_1$, the loss is bounded with $C^2(m, \varepsilon)$.
        \item From Lemma~\ref{lemma-lipschitz}, with probability at least $1-\varepsilon_2$, the squared loss class $\mathcal{H}_m$ is $C(m, \varepsilon)$-Lipschitz with respect to $\cF$, implying:
        $\cR_{n}(\mathcal{H}_m) \leq 2C(m, \varepsilon) \cR_{n}(\cF)$
        \item Applying McDiarmid's inequality with the bounded loss and the new Rademacher complexity while setting $\varepsilon_1 = \varepsilon_2 = \varepsilon$ completes the proof.         
   \end{enumerate}
\end{proof}
\vspace{-24pt}
\subsection{Rademacher Complexity Bounds for Common Function Classes} 
\label{r-complexity-bounds}

\renewcommand{\arraystretch}{1.5}


\begin{table}[ht]
\caption{\small Rademacher Complexity Bounds for Linear and Convex Models}
\label{tab:lin_convex}
{\small 
\begin{tabular}{|p{6cm}|p{4cm}|p{5cm}|}
\hline
\textbf{Function Class} & \textbf{Rademacher Complexity Bound} & \textbf{Key Assumptions} \\
\hline
\parbox[t]{6cm}{Linear, $\ell_2$-constraint:\\[3pt]
$\displaystyle \cF=\{x\mapsto w^\top x:\|w\|_2\le B\}$}
& $\displaystyle \cR_n(\cF)\le B\cdot\sqrt{\tr(\Sigma)/n}$
& Data bounded in $\ell_2$ or covariance $\Sigma=\E[xx^\top]$ bounded.
\\
\hline
\parbox[t]{6cm}{Linear, $\ell_1$-constraint:\\[3pt]
$\displaystyle \cF=\{x\mapsto w^\top x:\|w\|_1\le B\}$}
& $\displaystyle \cR_n(\cF)\le B\cdot\max_i\|x_i\|_\infty\sqrt{\ln d/n}$
& Samples $x_i$ have bounded $\|\cdot\|_\infty$ norm.
\\
\hline
\parbox[t]{6cm}{Sparse linear ($s$-sparse):\\[3pt]
$\displaystyle \cF=\{x\mapsto w^\top x:\|w\|_0\le s,\;\|w\|_2\le B\}$}
& $\displaystyle \cR_n(\cF)\le B\sqrt{s\ln(d/s)/n}$
& Bounded features; $d=\dim(x)$.
\\
\hline
\parbox[t]{6cm}{Convex Lipschitz:\\[3pt]
$\displaystyle \cF=\{f:\X\to\R,\;f\text{ convex},\;\|f\|_{\text{Lip}}\le L\}$}
& $\displaystyle \cR_n(\cF)\le L\cdot\text{diam}(\X)/\sqrt{n}$
& Domain $\X$ has diameter $\text{diam}(\X)$.
\\
\hline
\end{tabular}}
\end{table}

\begin{table}[h]
\caption{\small Neural Networks}
\label{tab:nn}
{\small
\begin{tabular}{|p{6cm}|p{4cm}|p{5cm}|}
\hline
\textbf{Function Class} & \textbf{Rademacher Complexity Bound} & \textbf{Key Assumptions} \\
\hline
\parbox[t]{6cm}{One-hidden-layer, $H$ units:\\[3pt]
$\displaystyle \cF = \left\{x \mapsto \sum_{j=1}^H v_j\,\sigma(w_j^\top x) : \|w_j\|_2 \le B_w,\; \|v\|_1 \le B_v \right\}$}
& $\displaystyle \cR_n(\cF) \le B_v B_w \|\X\|_F / \sqrt{n}$
& $\sigma$ is 1-Lipschitz; $\|\X\|_F$ is the Frobenius norm.
\\
\hline
\parbox[t]{6cm}{Deep nets, depth $L$:\\[3pt]
$\displaystyle \cF = \{f : \|W_l\|_2 \le B \}$}
& $\displaystyle \cR_n(\cF) \le B^L \|\X\|_F \prod_{l=1}^{L-1} \|W_l\|_2 / \sqrt{n}$
& Spectral-norm constraints on weight matrices.
\\
\hline
\parbox[t]{6cm}{ReLU nets, depth $L$, width $H$:\\[3pt]
$\displaystyle \cF = \{f : \|W_l\|_F \le B \}$}
& $\displaystyle \cR_n(\cF) \le B^L \sqrt{H} \|\X\|_F / \sqrt{n}$
& Frobenius norm bound on each weight matrix. Scales with $\sqrt{H}$ and $B^L$.
\\
\hline
\parbox[t]{6cm}{Conv nets, bounded filters:\\[3pt]
$\displaystyle \cF = \{f : \|W\|_F \le B \}$}
& $\displaystyle \cR_n(\cF) \le B \sqrt{d \ln d / n}$
& $d = $ total number of parameters.
\\
\hline
\end{tabular}}
\end{table}

\begin{table}[ht]
\caption{\small Other Function Classes}
\label{tab:other}
{\small
\begin{tabular}{|p{6cm}|p{4cm}|p{5cm}|}
\hline
\textbf{Function Class} & \textbf{Rademacher Complexity Bound} & \textbf{Key Assumptions} \\
\hline
\parbox[t]{6cm}{$\ell_p$-norm ball:\\[3pt]
$\displaystyle \cF = \{x \mapsto w^\top x : \|w\|_p \le r\}$}
& $\displaystyle \cR_n(\cF) \le r \|\X\|_{p^*} / \sqrt{n}$
& $p^*$ is the dual of $p$; $\|X\|_{p^*}$ is the dual norm of data matrix.
\\
\hline
\parbox[t]{6cm}{Depth-$D$ decision trees:}
& $\displaystyle \cR_n(\cF) \le \sqrt{D \ln(2n)/n}$
& Assumes roughly balanced splits.
\\
\hline
\parbox[t]{6cm}{$k$-nearest neighbors:}
& $\displaystyle \cR_n(\cF) \le \sqrt{k \ln n / n}$
& Assumes bounded outputs.
\\
\hline
\parbox[t]{6cm}{Smooth functions:\\[3pt]
$\displaystyle \cF = \{f : \|f\|_{\text{Lip}} \le L,\; \|f\|_\infty \le B\}$}
& $\displaystyle \cR_n(\cF) \le L \cdot \text{diam}(\X) / \sqrt{n}$
& Domain $\X$ of diameter $\text{diam}(\X)$.
\\
\hline
\end{tabular}}
\end{table}

\begin{table}[ht]
\caption{\small Non-parametric Methods}
\label{tab:nonparam}
{\small
\begin{tabular}{|p{6cm}|p{4cm}|p{5cm}|}
\hline
\textbf{Function Class} & \textbf{Rademacher Complexity Bound} & \textbf{Key Assumptions} \\
\hline
\parbox[t]{6cm}{Hölder smooth functions:\\[3pt]
$\displaystyle \cF = \{f \in C^{\alpha}(\X) : \|f\|_{C^{\alpha}} \le B\}$}
& $\displaystyle \cR_n(\cF) \le B \cdot (1/n)^{\alpha/(2\alpha + d)}$
& Functions with $\alpha$ derivatives, input dimension $d$. Optimal rate for nonparametric regression.
\\
\hline
\parbox[t]{6cm}{Sparse additive models (SpAM):\\[3pt]
$\displaystyle \cF = \left\{f(x) = \sum_{j \in S} f_j(x_j) : |S| \le s,\; \|f_j\|_\infty \le B \right\}$}
& $\displaystyle \cR_n(\cF) \le B \sqrt{s \ln(d)/n}$
& At most $s$ active features out of $d$ total features. Each component function bounded.
\\
\hline
\parbox[t]{6cm}{Besov spaces $B_{p,q}^\alpha$:\\[3pt]
$\displaystyle \cF = \{f \in B_{p,q}^\alpha : \|f\|_{B_{p,q}^\alpha} \le B\}$}
& $\displaystyle \cR_n(\cF) \le B \cdot (1/n)^{\alpha/(2\alpha + d(1/p - 1/2)_+)}$
& Wavelet-based smoothness spaces. Generalize Hölder and Sobolev spaces.
\\
\hline
\parbox[t]{6cm}{Sobolev spaces:\\[3pt]
$\displaystyle \cF = \{f \in W^{k,p}(\X) : \|f\|_{W^{k,p}} \le B\}$}
& $\displaystyle \cR_n(\cF) \le B \cdot (1/n)^{k/(2k + d)}$
& Functions with $k$ weak derivatives in $L^p$. Domain $\X \subset \mathbb{R}^d$.
\\
\hline
\end{tabular}}
\end{table}

\subsection{Proof of Proposition \ref{prop4-cmo-convergence}} 
\label{proof-prop4-cmo-convergence}
\begin{proof}
We prove parts (i) and (ii) separately.

\textbf{Part (i): MSE Convergence.}  
By definition, for any $x\in\cX$, we have
\[
\hat{\mu}(x)-\mu(x)
=
\pi_{-1}(x)\Bigl[\hat{\mu}_{+1}(x)-\mu_{+1}(x)\Bigr]
+
\pi_{+1}(x)\Bigl[\hat{\mu}_{-1}(x)-\mu_{-1}(x)\Bigr].
\]
Squaring both sides yields
\[
\begin{aligned}
\Bigl(\hat{\mu}(x)-\mu(x)\Bigr)^2 
&=
\pi_{-1}(x)^2\Bigl[\hat{\mu}_{+1}(x)-\mu_{+1}(x)\Bigr]^2
+
\pi_{+1}(x)^2\Bigl[\hat{\mu}_{-1}(x)-\mu_{-1}(x)\Bigr]^2\\[1mm]
&\quad
+\, 2\,\pi_{-1}(x)\pi_{+1}(x)
\Bigl|\hat{\mu}_{+1}(x)-\mu_{+1}(x)\Bigr|
\Bigl|\hat{\mu}_{-1}(x)-\mu_{-1}(x)\Bigr|.
\end{aligned}
\]
Since the propensity weights satisfy $0\le \pi_a(x)\le 1$, it follows that
\begin{align}
\Bigl(\hat{\mu}(x)-\mu(x)\Bigr)^2
&\le
\Bigl[\hat{\mu}_{+1}(x)-\mu_{+1}(x)\Bigr]^2
+
\Bigl[\hat{\mu}_{-1}(x)-\mu_{-1}(x)\Bigr]^2 \\
&+
2\,\Bigl|\hat{\mu}_{+1}(x)-\mu_{+1}(x)\Bigr|
\Bigl|\hat{\mu}_{-1}(x)-\mu_{-1}(x)\Bigr|.
\end{align}
Taking expectation with respect to $X$ and applying the Cauchy–Schwarz inequality to the cross term, we obtain
\[
\begin{aligned}
\mathbb{E}\Bigl[\Bigl(\hat{\mu}(X)-\mu(X)\Bigr)^2\Bigr]
&\le 
\mathbb{E}\Bigl[\Bigl(\hat{\mu}_{+1}(X)-\mu_{+1}(X)\Bigr)^2\Bigr]
+
\mathbb{E}\Bigl[\Bigl(\hat{\mu}_{-1}(X)-\mu_{-1}(X)\Bigr)^2\Bigr]\\[1mm]
&\quad\quad
+\, 2\,\sqrt{
\mathbb{E}\Bigl[\Bigl(\hat{\mu}_{+1}(X)-\mu_{+1}(X)\Bigr)^2\Bigr]
\cdot
\mathbb{E}\Bigl[\Bigl(\hat{\mu}_{-1}(X)-\mu_{-1}(X)\Bigr)^2\Bigr]
}.
\end{aligned}
\]
By the per-arm assumption,
\[
\mathbb{E}\Bigl[\Bigl(\hat{\mu}_{a}(X)-\mu_{a}(X)\Bigr)^2\Bigr]
=\cO_p\Bigl(r^2(n'_a)\Bigr)
\quad\text{for } a\in\{+1,-1\}.
\]
Since $n'_a\ge\eta n'$ for both arms, we have $r(n'_a)=\cO\bigl(r(n')\bigr)$. Consequently, there exists a constant $C>0$ (depending on $\eta$) such that
\[
\mathbb{E}\Bigl[\Bigl(\hat{\mu}(X)-\mu(X)\Bigr)^2\Bigr]
\le
C\Biggl(
r^2(n')
+
r(n')\,r(n')
\Biggr)
=
\cO_p\Bigl(r^2(n')\Bigr).
\]
This proves the MSE convergence claim.

\vspace{1mm}
\textbf{Part (ii): Uniform Convergence.}  
We need the following lemma ---a standard result in functional analysis--- to connect the two modes of convergence.
\begin{lemma}[MSE Implies Uniform Convergence]
\label{lemma-convergence}
Let $\cX\subset \mathbb{R}^d$ be compact, and suppose $\mu, \hmu_{n'}: \cX \to \mathbb{R}$ are both $L$-Lipschitz functions. Then
\begin{equation*}
\begin{minipage}{\textwidth}
\raggedright
$\Delta_{2}^2\!\left(\hat\mu_{n'}, \mu\right)
= \mathbb{E}\!\left[\left(\hat\mu_{n'}(X) - \mu(X)\right)^2\right]
= \mathcal{O}_p\left(r^2(n')\right)$
\par\noindent
\hfill
$\Longleftrightarrow$\quad
$\Delta_{\infty}\!\left(\hat\mu_{n'}, \mu\right)
= \sup_{x \in \mathcal{X}} \left| \hat\mu_{n'}(x) - \mu(x) \right|
= \mathcal{O}_p\left(r(n')\right)$
\end{minipage}
\end{equation*}

\end{lemma}

\begin{proof}
We show both directions:

\textbf{1) ($\Longrightarrow$) MSE $\implies$ Uniform Convergence.}
Assume
$\mathbb{E}\!\left[\left(\mu(X) - m(X)\right)^2\right] \le \xi$
for some small $\xi > 0$. Our goal is to prove $\sup_{x\in X} \left|\mu(x) - m(x)\right| \le C \sqrt{\xi},$
where $C$ depends on $L$ and the diameter of $X$. 

\begin{enumerate}[leftmargin=12pt, itemsep=0pt]
\item
By Lipschitz continuity of both $\mu$ and $m$, for any $x, y \in \cX$:
\[
\left|\left(m(x) - \mu(x)\right) - \left(m(y) - \mu(y)\right)\right|
\le
\left|m(x) - m(y)\right| + \left|\mu(x) - \mu(y)\right|
\le
2L \|x - y\|.
\]
Hence, $m(x) - \mu(x)$ is at most $2L$-Lipschitz.

\item
By compactness of $\cX$, for any $\epsilon > 0$ there exists a finite $\epsilon$-cover $x_1,\ldots,x_N \in \cX$ such that for every $x \in \cX$ there is an $x_{i(x)}$ with $\| x - x_{i(x)}\| \le \epsilon$. The integer $n$ depends on $\epsilon$ and the dimension $d$.

\item
For any $x\in \cX$, pick $i(x)$ such that $\| x - x_{i(x)}\|\le \epsilon$. Then
\[
\left|\mu(x) - m(x)\right|
\le
\left|\mu(x_{i(x)}) - m(x_{i(x)})\right|
+
2L\| x - x_{i(x)}\|
\le
\left|\mu(x_{i(x)}) - m(x_{i(x)})\right|
+
2L\epsilon.
\]
Taking the supremum over $x\in \cX$ gives
\[
\sup_{x\in \cX} \left|\mu(x) - m(x)\right|
\le
\max_{1\le i \le N} \left|\mu(x_i) - m(x_i)\right|
+
2L\epsilon.
\]

\item
By Markov's inequality, for any $t > 0$,
\(
P\!\left(\left|\mu(X) - m(X)\right|\right)
\le
\frac{\xi}{t^2}.
\)

Applying a union bound across the $n$ cover points,
\(
P\!\left(\max_{1\le i \le N}\left|\mu(x_i) - m(x_i)\right|\right)
\le
\frac{N\xi}{t^2}.
\)

Choose $t = \sqrt{\frac{N\xi}{\varepsilon}}$ for $\varepsilon\in(0,1)$. Then with probability at least $1-\varepsilon$,
\[
\max_{1\le i \le N}\left|\mu(x_i) - m(x_i)\right|
\le
\sqrt{\tfrac{N\xi}{\varepsilon}}.
\]
Combining with the bound in Step 3,
\(
\sup_{x\in \cX}\left|\mu(x) - m(x)\right|
\le
\sqrt{\tfrac{N\xi}{\varepsilon}}
+
2L\epsilon.
\)

\item
Setting $\epsilon = \frac{\sqrt{\xi}}{2L}$ yields:
$
\sup_{x\in X}\left|\mu(x) - m(x)\right|
\le
\sqrt{\tfrac{N\xi}{\varepsilon}}
+
2L\frac{\sqrt{\xi}}{2L}
=
\sqrt{\tfrac{N}{\varepsilon}}\sqrt{\xi}
+
\sqrt{\xi}
=
C\sqrt{\xi},
$
where $C = \sqrt{\tfrac{N}{\varepsilon}} + 1$ depends on the cover size $n$ (hence on $\dim(\cX)$ and the diameter of $\cX$).  

Thus, small MSE at rate $\xi$ implies $\sup_{x\in \cX}\left|\mu(x)-m(x)\right|$ is $\cO(\!\sqrt{\xi})$.  
\end{enumerate}

\textbf{2) ($\Longleftarrow$) Uniform Convergence $\implies$ MSE.}
Now, for the other direction, assume
\[
\sup_{x\in X}\left|\mu(x) - m(x)\right|
\le
r(n')
\quad
\text{with high probability.}
\]
Then pointwise we have
\(
\left|\mu(X) - m(X)\right|^2
\le
r(n')^2.
\)
Hence,
\(
\mathbb{E}\!\left[\left(\mu(X) - m(X)\right)^2\right]
\le
r(n')^2,
\)
so in probability we get 
$\Delta_{2}^2(m,\mu) = \cO_p\left(r^2(n')\right)$.  

Putting both directions together, we see that under the proposed assumptions,
\[
\Delta_{2}^2(m,\mu)=\cO_p\left(r^2(n')\right)
\Longleftrightarrow
\Delta_{\infty}(m,\mu)=\cO_p\left(r(n')\right).
\]
This completes the proof.
\end{proof}

Now, let's move back to the proof part (ii) of Proposition \ref{prop4-cmo-convergence}. 

Assume in addition that for each $a\in\{+1,-1\}$ the functions $\mu_a$ and their estimators $\hat{\mu}_a$ are $L$-Lipschitz on the compact set $\cX$. Then, by the standard result stated in Lemma~\ref{lemma-convergence}, the per-arm mean squared error control,
\(
\mathbb{E}\Bigl[\Bigl(\hat{\mu}_{a}(X)-\mu_{a}(X)\Bigr)^2\Bigr]
=\cO_p\Bigl(r^2(n'_a)\Bigr),
\)
implies that
\(
\sup_{x\in \cX}\Bigl|\hat{\mu}_{a}(x)-\mu_{a}(x)\Bigr|
=\cO_p\Bigl(r(n'_a)\Bigr)
=\cO_p\Bigl(r(n')\Bigr).
\)

Recall that the CMO estimator is constructed as
\(
\hat{\mu}(x)
=
\pi_{-1}(x)\,\hat{\mu}_{+1}(x)
+
\pi_{+1}(x)\,\hat{\mu}_{-1}(x),
\)
and similarly for $\mu(x)$. Thus, for any $x\in\cX$,
\[
\begin{aligned}
\Bigl|\hat{\mu}(x)-\mu(x)\Bigr|
&=
\Bigl|\pi_{-1}(x)\Bigl[\hat{\mu}_{+1}(x)-\mu_{+1}(x)\Bigr]
+\pi_{+1}(x)\Bigl[\hat{\mu}_{-1}(x)-\mu_{-1}(x)\Bigr]\Bigr|\\[1mm]
&\le
\Bigl|\hat{\mu}_{+1}(x)-\mu_{+1}(x)\Bigr|
+
\Bigl|\hat{\mu}_{-1}(x)-\mu_{-1}(x)\Bigr|.
\end{aligned}
\]
Taking the supremum over $\cX$ yields
\[
\Delta_{\infty}\left(\hat{\mu}, \mu\right)
=\sup_{x\in \cX}\Bigl|\hat{\mu}(x)-\mu(x)\Bigr|
\le
\sup_{x\in \cX}\Bigl|\hat{\mu}_{+1}(x)-\mu_{+1}(x)\Bigr|
+
\sup_{x\in \cX}\Bigl|\hat{\mu}_{-1}(x)-\mu_{-1}(x)\Bigr|
=\cO_p\Bigl(r(n')\Bigr).
\]
This completes the proof of uniform convergence.

\vspace{1mm}
Combining the two parts, we conclude that the CMO estimator $\hat{\mu}(x)$ converges to $\mu(x)$ in MSE at the rate $\cO_p\bigl(r^2(n')\bigr)$ (without extra smoothness assumptions), and if the per-arm functions are additionally $L$-Lipschitz on a compact domain, then the uniform convergence rate is $\cO_p\bigl(r(n')\bigr)$.
\end{proof}

\vspace{-24pt}
\subsection{Proof of Proposition \ref{prop3-rct-est}}

\begin{proof}
By Theorem \ref{theo3-risk-bound}, with probability at least $1-2\varepsilon$:
$
\Delta_2^2(\htau_{n^r}, \tau)
   \leq
   \Delta_2^2(\cF, \tau) 
   + 2C(m,\varepsilon)\cR_{n^r}(\cF)
   +
   C^2(m,\varepsilon)
   \sqrt{\frac{\log\left(2/\varepsilon\right)}{n^r}}
$.
Under the assumption (A5), $\Delta_2^2(\cF, \tau) = 0$. For any well-behaved function class $\cF$, the Rademacher complexity term $\cR_n(\cF)$ converges to zero as $n^r \to \infty$. The final term, containing the deviation penalty $\sqrt{\log(2/\varepsilon)}$, also vanishes at rate $1/\sqrt{n^r}$. 
Therefore, $\Delta_\tau^2(\htau^r) \to 0$ as $n^r \to \infty$, establishing consistency.
\end{proof}
\vspace{-24pt}

\section{Supplementary analyses for the Greenlight Plus study}\label{app:gps-supp}
This appendix gives the analyses summarized in Section~\ref{sec:gps}. Each clinical site is treated as a small trial; unless stated otherwise, R-OSCAR-1arm borrows the pooled three-site EHR controls ($n^o=8{,}867$), and performance is measured by held-out RCT control-arm outcome loss. ``Covered'' sites are the three with EHR data (Duke, UNC, Vanderbilt); ``uncovered'' sites are the three trial sites absent from the EHR (Miami, NYU, Stanford).

\subsection{Small-trial control-arm diagnostic}\label{app:gps-controlarm}
For a target site, RACER and R-OSCAR-1arm control models are cross-fit over $K$ folds of the site's RCT controls and evaluated on each held-out control $i$ using the paired squared-error difference $D_i = (Y_i-\hat\mu^{r}_{-1}(X_i))^2 - (Y_i-\hat\mu^{o,t}_{-1}(X_i))^2$, where $\hat\mu^{o,t}_{-1}$ is the pooled-EHR control model calibrated on the training-fold RCT controls (positive $D_i$ favors borrowing). The analysis reports $\bar D$ with a paired bootstrap $95\%$ CI ($2000$ resamples) and certifies borrowing when the interval excludes zero (Algorithm~\ref{alg:gps-controlarm}, Table~\ref{tab:gps-controlarm}). Covered sites benefit, with gains increasing as the trial shrinks; the smallest, UNC ($n^r=34$), is the only certifiable single-site case. Uncovered sites show no reliable benefit even though NYU ($n^r=86$) has the most room to improve, so the benefit tracks covariate-support overlap rather than sample count.

\begin{algorithm}[h]
{\small
\caption{Small-trial control-arm diagnostic}\label{alg:gps-controlarm}
\begin{algorithmic}[1]
\State \textbf{Input:} target-site RCT controls $\{(X_i,Y_i)\}$; pooled EHR controls $\{(X^o_j,Y^o_j)\}$; folds $K$
\For{$k=1,\dots,K$}
  \State fit RACER control model $\hat\mu^r_{-1}$ on the training-fold RCT controls
  \State fit the pooled-EHR control model and calibrate it on the training-fold RCT controls $\to \hat\mu^{o,t}_{-1}$
  \State held-out control $i$: $D_i \gets (Y_i-\hat\mu^r_{-1}(X_i))^2-(Y_i-\hat\mu^{o,t}_{-1}(X_i))^2$
\EndFor
\State \textbf{Output:} $\bar D=\mathrm{mean}_i\, D_i$; paired bootstrap $95\%$ CI; certify borrowing iff CI excludes $0$
\end{algorithmic}}
\end{algorithm}

\begin{table}[h]
\centering
\caption{\textbf{Small-trial control-arm diagnostic (pooled EHR borrowed).} Held-out control-arm loss reduction $\bar D$ (positive favors R-OSCAR-1arm) with paired bootstrap $95\%$ CI. ``Covered'' marks trial sites present in the EHR.}
\label{tab:gps-controlarm}
\small
\begin{tabular*}{\columnwidth}{@{\extracolsep{\fill}}llccrc@{}}
\hline
Trial site & Covered? & $n^r$ & $\bar D$ (95\% CI) & Rel. & Certifiable \\
\hline
Vanderbilt & yes & $241$ & $0.065\ [-0.044,\, 0.175]$  & $+4.6\%$  & No  \\
Duke       & yes & $55$  & $0.191\ [-0.053,\, 0.407]$  & $+16.9\%$ & No  \\
UNC        & yes & $34$  & $0.344\ [0.102,\, 0.621]$   & $+28.4\%$ & Yes \\
NYU        & no  & $86$  & $-0.021\ [-0.208,\, 0.130]$ & $-1.2\%$  & No  \\
Miami      & no  & $28$  & $0.027\ [-0.111,\, 0.171]$  & $+5.3\%$  & No  \\
Stanford   & no  & $23$  & $0.130\ [-0.048,\, 0.310]$  & $+11.5\%$ & No  \\
\hline
\end{tabular*}
\end{table}

\subsection{Diagnostic threshold sensitivity}\label{app:gps-diagsens}
For each site, the cross-fitted diagnostic of Section~\ref{sec:diagnostic} is run once, borrowing the pooled EHR, and its per-subject weighted discrepancy is bootstrapped to obtain a one-sided certification level $p=\Pr^{*}(\bar D^{\mathrm{w}*}\le 0)$ (Algorithm~\ref{alg:gps-diagsens}); borrowing is certified at two-sided level $\alpha$ exactly when $p<\alpha/2$. Table~\ref{tab:gps-diagsens} reports $p$ per site: as $\alpha$ is relaxed, the diagnostic certifies exactly the covered sites, in order Duke, UNC, Vanderbilt, and never an uncovered site (covered $p\le0.12$, uncovered $p\ge0.29$). Under the appropriate one-sided test ($p<\alpha$), Duke and UNC are certified at $\alpha=0.05$.

\begin{algorithm}[h]
{\small
\caption{Diagnostic threshold sensitivity}\label{alg:gps-diagsens}
\begin{algorithmic}[1]
\State \textbf{Input:} target-site RCT; pooled EHR controls
\State run the cross-fitted diagnostic of Section~\ref{sec:diagnostic}; collect its per-subject weighted discrepancy $\{D^{\mathrm{w}}_i\}$
\State bootstrap $\bar D^{\mathrm{w}}$ ($5000$ resamples); one-sided level $p \gets \Pr^{*}(\bar D^{\mathrm{w}*}\le 0)$
\State \textbf{Output:} $p$; borrowing certified at two-sided level $\alpha$ iff $p<\alpha/2$
\end{algorithmic}}
\end{algorithm}

\begin{table}[h]
\centering
\caption{\textbf{Diagnostic certification level by site.} One-sided bootstrap level $p$ (lower $=$ certified at a stricter threshold). Loosening $\alpha$ certifies the covered sites in order and never an uncovered site.}
\label{tab:gps-diagsens}
\small
\begin{tabular*}{0.8\columnwidth}{@{\extracolsep{\fill}}llccc@{}}
\hline
Trial site & Covered? & $n^r$ & $p$ (one-sided) & Certified at \\
\hline
Duke       & yes & $55$  & $0.000$ & $\alpha=0.05$ \\
UNC        & yes & $34$  & $0.036$ & $\alpha=0.10$ \\
Vanderbilt & yes & $241$ & $0.115$ & $\alpha=0.32$ \\
Stanford   & no  & $23$  & $0.294$ & --- \\
Miami      & no  & $28$  & $0.540$ & --- \\
NYU        & no  & $86$  & $0.652$ & --- \\
\hline
\end{tabular*}
\end{table}

\subsection{Outcome-permuted negative control}\label{app:gps-negctrl}
To isolate sample volume from outcome signal, the UNC small-trial analysis is repeated with all sample sizes fixed after randomly permuting the pooled-EHR control outcomes. The permutation destroys the covariate-outcome relationship while preserving the marginal outcome distribution, and results are averaged over $30$ permutations (Algorithm~\ref{alg:gps-negctrl}). The control-arm gain falls from $\bar D=0.34$ on the real data to $-0.02$ (SD $0.18$), and the diagnostic certifies borrowing in only $7\%$ of permutations. The benefit therefore reflects observational outcome signal rather than added sample volume.

\begin{algorithm}[h]
{\small
\caption{Outcome-permuted negative control}\label{alg:gps-negctrl}
\begin{algorithmic}[1]
\State \textbf{Input:} UNC RCT controls; pooled EHR controls; permutations $B$
\For{$b=1,\dots,B$}
  \State permute the EHR control outcomes $Y^o$ (break $X^o$--$Y^o$, preserve the marginal)
  \State recompute $\bar D$ (Algorithm~\ref{alg:gps-controlarm}) and the diagnostic decision with the permuted EHR
\EndFor
\State \textbf{Output:} mean $\bar D$ over permutations; diagnostic certification rate
\end{algorithmic}}
\end{algorithm}

\subsection{Ordered individual effects}\label{app:gps-caterpillar}
On the pooled data, the linear CATE for each estimator is fit by ordinary least squares on the lasso-selected support (the full six-covariate set here, the lasso selecting no sparser model) and a $95\%$ interval is formed for each participant's linear predictor, with participants ordered by point estimate (Algorithm~\ref{alg:gps-caterpillar}, Figure~\ref{fig:gps-caterpillar}). The two estimators recover the same monotone spread of individual effects (roughly $-0.8$ to $+0.7$ WFLz) with near-identical interval widths (mean $1.51$ for both), and no interval excludes zero. These working-model intervals are descriptive: they condition on the fitted linear model and do not propagate selection or borrowed-model uncertainty.

\begin{algorithm}[h]
{\small
\caption{Ordered individual CATE intervals}\label{alg:gps-caterpillar}
\begin{algorithmic}[1]
\For{each estimator (RACER, R-OSCAR-1arm)}
  \State write the CATE as linear $\hat\tau(x)=x^\top\beta$; select the support by lasso, refit by OLS
  \State form the $95\%$ interval for $\hat\tau(x_i)$ at each participant $i$ from the OLS fit
\EndFor
\State order participants by $\hat\tau(x_i)$ on a shared rank axis
\State \textbf{Output:} ordered point estimates with intervals; mean width; number of intervals excluding $0$
\end{algorithmic}}
\end{algorithm}

\begin{figure}[h]
    \centering
    \includegraphics[width=0.82\textwidth]{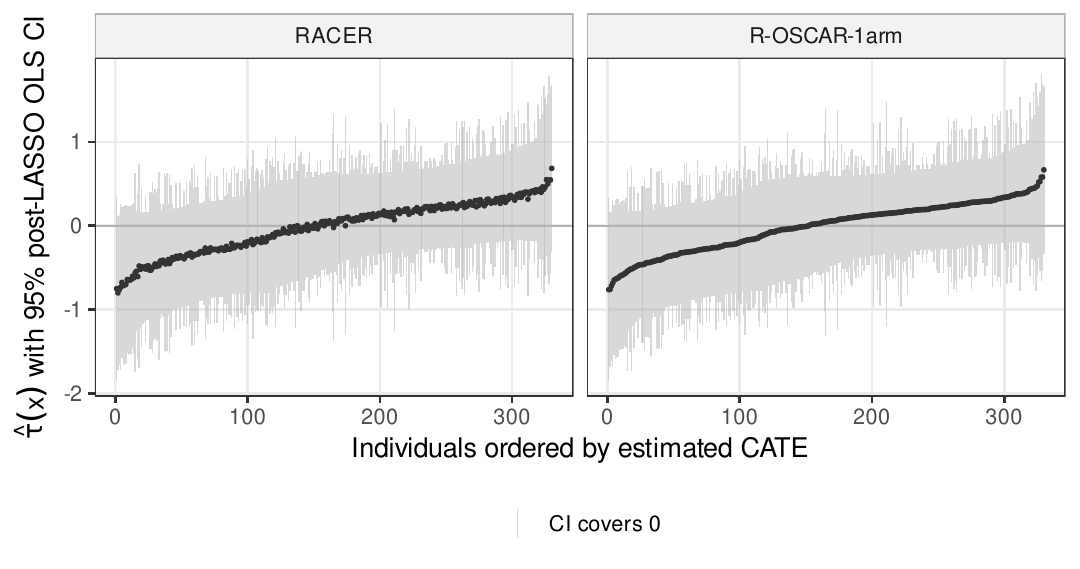}
    \caption{\small Ordered per-individual $\hat\tau(x)$ with $95\%$ OLS working-model intervals. RACER and R-OSCAR-1arm recover the same monotone gradient and near-identical widths; no interval excludes zero at $n^r=330$.}
    \label{fig:gps-caterpillar}
\end{figure}

\subsection{Control- versus treated-arm variance decomposition}\label{app:gps-vardecomp}
With the linear CATE written as $\hat\tau(x)=\hat\mu_{+1}(x)-\hat\mu_{-1}(x)$ and the arm models fit on disjoint RCT subsamples, the per-individual interval variance decomposes additively into treated- and control-arm contributions (Algorithm~\ref{alg:gps-vardecomp}). Averaged over participants, the two components are essentially equal (treated $0.085$, control $0.083$; treated share $0.51$), so neither arm drives the interval width. This explains why the certified control-arm gain of Appendix~\ref{app:gps-controlarm} does not narrow the intervals in Figure~\ref{fig:gps-caterpillar}: both estimators use the same covariates and OLS interval construction, whose width is set by the trial design and residual variance rather than by the borrowed control model's accuracy. Expressing the gain as a narrower CATE interval would require uncertainty quantification that propagates the low variance of the EHR-fit control model.

\begin{algorithm}[h]
{\small
\caption{Control- vs treated-arm variance decomposition}\label{alg:gps-vardecomp}
\begin{algorithmic}[1]
\State \textbf{Input:} OLS arm models $\hat\mu_{+1},\hat\mu_{-1}$ (disjoint RCT arms)
\State for each participant $i$: treated SE $s_{t,i}$, control SE $s_{c,i}$ (OLS prediction SEs)
\State $\widehat{\mathrm{Var}}\,\hat\tau(x_i) \gets s_{t,i}^2+s_{c,i}^2$;\quad treated share $\gets s_{t,i}^2/(s_{t,i}^2+s_{c,i}^2)$
\State \textbf{Output:} mean treated and control variance; mean treated share
\end{algorithmic}}
\end{algorithm}

\clearpage
\newpage
\bibliography{sample}

\end{document}